\documentclass[twocolumn,showpacs,preprintnumbers,amsmath,amssymb,fleqn]{revtex4}

\usepackage{graphicx}
\usepackage{bm}
\usepackage{subfigure}

\begin{document}

\title{The magnetic ground state of Sr$_2$IrO$_4$ and implications for second-harmonic generation}
\author{S. Di Matteo}
\affiliation{D\'epartement Mat\'eriaux Nanosciences, Institut de Physique de Rennes UMR UR1-CNRS 6251, Universit\'e de Rennes 1, F-35042 Rennes Cedex, France}
\author{M. R. Norman}
\affiliation{Materials Science Division, Argonne National Laboratory, Argonne, IL  60439, USA}

\date{\today}
\begin{abstract}
The currently accepted magnetic ground state of Sr$_2$IrO$_4$ (the $-++-$ state) preserves inversion symmetry.  This is at odds,
though, with recent experiments that indicate a magnetoelectric ground state, leading to the speculation that orbital
currents or more exotic magnetic multipoles might exist in this material.  Here, we analyze various magnetic
configurations and demonstrate that two of them, the magnetoelectric $-+-+$ state and the non-magnetoelectric $++++$ state, 
can explain these recent second-harmonic generation (SHG) experiments, obviating the need to invoke orbital currents.
The SHG-probed magnetic order parameter has the symmetry of a parity-breaking multipole in the $-+-+$ state and of a parity-preserving multipole in the $++++$ state.
We speculate that either might have been created by the laser pump used in the experiments. 
An alternative is that the observed magnetic SHG signal is a surface effect.
We suggest experiments that could be performed to test these various possibilities,
and also address the important issue of the suppression of the RXS intensity at the L$_2$ edge.
\end{abstract}

\maketitle

\section{Introduction}

The physical properties of layered iridates, in particular Sr$_2$IrO$_4$, have been thoroughly investigated since the seminal paper of B.~J.~Kim and collaborators \cite{kim1} and their suggested analogy with the physics of cuprate superconductors. The formation of a half-filled $t_{2g}$ doublet by the strong Ir spin-orbit interaction, that is then gapped by correlations, mimics what is seen in the cuprates, making Sr$_2$IrO$_4$ an insulator, despite its Ir$^{4+}$ ionic configuration with five occupied $t_{2g}$ electrons \cite{kim1,jackeli}. More recent experiments on doped iridates point to the emergence of a pseudogap \cite{kim3} and at low temperatures a d-wave gap \cite{kim4}, therefore strengthening the analogy with cuprates. Most recently, a new experiment based on second-harmonic generation (SHG) \cite{shg} claimed the detection of an odd-parity, magnetic hidden order in Sr$_2$IrO$_4$, suggesting the presence of orbital currents as proposed by Varma for cuprates \cite{varma}.
This followed an earlier bulk property study indicating a giant magnetoelectric effect in Sr$_2$IrO$_4$ \cite{chikara}.

Despite these analogies, there are also significant differences between Sr$_2$IrO$_4$ and La$_2$CuO$_4$.  First, the insulating gap has a different character, spin-orbit plus Mott versus charge transfer, therefore doped holes in Sr$_2$IrO$_4$ go into the Ir $5d$ states, and not in the oxygen $2p$ ones as in La$_2$CuO$_4$. Second, the $5d$ states of Ir are much more spatially extended than the $3d$ states of Cu, making the on-site Coulomb and exchange terms significantly weaker.
Therefore, the physical motivation for orbital currents, based as it is on the near degeneracy of the transition metal $d$ and oxygen $p$ states \cite{varma}, seems unlikely in the iridate case, where the oxygen $2p$ bands lie more than 3 eV below the Ir $t_{2g}$ doublet \cite{moon}.

The existence of an SHG signal \cite{shg} points to a reduction of the magnetic space group symmetry 2/m1$^\prime$ previously indicated by neutrons and resonant x-ray measurements. Whether this reduction is due to orbital currents or another mechanism remains to be seen. If we analyze the relative stacking along the $c$-axis of the ferromagnetic (FM) in-plane component of the moment in each of the IrO$_2$ planes, we find that three inequivalent configurations are possible. They can be labeled as $-++-$, $++++$ and $-+-+$  (Fig.~\ref{fig1}), where $\pm$ refer to the projection of the FM component in each plane along the $b$-axis, with the first configuration being that
identified in Sr$_2$IrO$_4$. As detailed in Section III, both $++++$ and $-+-+$ lead to a symmetry reduction of 2/m1$^\prime$ and, for different reasons, can explain the SHG results. The former was found by resonant x-rays in Ref.~\onlinecite{kim2} in a 0.3 T magnetic field and has also been identified by both neutrons and resonant x-rays upon doping with Rh \cite{clancy,ye15,footye}.
Since the inter-plane spin exchange has been estimated to be as small as 1 $\mu$eV \cite{fuji}, one could speculate that the inter-plane magnetic pattern might be disrupted by a laser pulse of 1mJ/cm$^2$ fluence as used in Ref.~\onlinecite{shg}. We argue that any such pattern breaking should in general lead to an SHG signal because of the resulting symmetry reduction. 
Another possibility is that SHG arises from a magnetized surface, which also has the desired symmetry.
For these reasons, we believe the SHG signal can in principle be explained by magnetism.

The aim of the present paper is to critically revisit several aspects of Ref.~\onlinecite{shg}, looking for alternative explanations of the SHG signal, and as a byproduct address some important issues on the suppression of the resonant x-ray scattering (RXS) intensity at the Ir L$_2$ edge. To reach our goals, the present article is organized as follows: in Section II, we review the details of the crystal and magnetic symmetries of Sr$_2$IrO$_4$, and show how RXS data on Rh-doped samples below $T_N$ might be explained in terms of the $-+-+$ state as well as the previously suggested $++++$ state \cite{clancy,ye15,footye}.
We propose further RXS and neutron experiments to clearly identify the actual magnetic pattern. Section III is devoted to an analysis of the SHG experiment from the quantum-mechanical microscopic expressions of the tensors involved. This allows us to show that only two magnetic space groups are consistent with the SHG experiment. The first is 2$^\prime$/m, advocated in Ref.~\onlinecite{shg}, which is also the magnetic space group of the $-+-+$ magnetic pattern. 
The second is 2$^\prime$/m$^\prime$, which is the magnetic group corresponding to the $++++$ pattern. We characterize the multipole ranks of the order parameters identified by the SHG experiment for each magnetic group. 
In particular, for the 2$^\prime$/m magnetic space group of the $-+-+$ state, the allowed order parameters have the symmetry of inversion-odd magnetic multipoles of rank one, two and three: toroidal dipole, magnetic quadrupole and toroidal octupole (the magnetic quadrupole, though, does not contribute in an SS polarization geometry). Instead, for the 2$^\prime$/m$^\prime$ magnetic space group of the $++++$ state, the allowed order parameters have the symmetry of inversion-even magnetic multipoles up to rank three: magnetic toroidal monopole, magnetic dipole, magnetic toroidal quadrupole and magnetic octupole.
In this Section, we also speculate on whether the
SHG signal is induced by the laser pump, or rather that it is a surface effect (the magnetic point group of the surface being 2$^\prime$).
Several experiments are suggested to test these possibilities.
In Section IV, we  address the important issue of the suppression of the RXS intensity at the L$_2$ edge and the results obtained in the literature on the doublet $J_{\rm eff}=1/2$ of Sr$_2$IrO$_4$. We discuss some of the critical aspects of this doublet and clarify its connection to the RXS experiments.
Finally, in Section V, ab initio simulations for some key x-ray absorption spectroscopy (XAS) experiments are presented, with the dual goal of confirming (or modifying) the 
$J_{\rm eff}=1/2$ doublet picture, and also to test whether the two proposed magnetic symmetries we suggest could explain the SHG experiment \cite{shg}.
Some conclusions are offered in Section VI.

\section{Crystal and magnetic symmetries in ${\rm Sr_2IrO_4}$: analysis of resonant structure factors}

The original analysis of the crystal space group of Sr$_2$IrO$_4$ suggested I4$_1$/acd \cite{original}, with the 8 Ir ions in this unit cell related by the symmetry operations of the 8a site of I4$_1$/acd, as detailed in Table I. For future reference, we note that, though the point symmetry of the Ir site is $\overline{4}$, the further reduction from the 4/m point symmetry, and therefore the breaking of inversion symmetry, is determined only by the oxygens in the IrO$_2$ planes above (below) that of the Ir, and these are quite distant ($>$ 6.5 \AA). For this reason, the effect of inversion-breaking at an Ir site, though non-zero, is extremely small.
Of course, inversion symmetry in the unit cell is restored for the global symmetry I4$_1$/acd.

Very recently, the crystal structure has been resolved by neutron scattering to be rather I4$_1$/a \cite{ye15}, explaining the observation of forbidden Bragg peaks in earlier experiments \cite{ye13,dhital} and in keeping with recent SHG data \cite{torch}. Within this space group, the 8 Ir atoms in the unit cell split into two nonequivalent groups of 4 atoms, each with the same point group $\overline{4}$ (sites 4a and 4b). The same remark concerning the inversion breaking made above still applies, as the symmetry reduction from I4$_1$/acd to I4$_1$/a, leading to the loss of the two glide planes containing the $c$-axis, is just determined by a tiny displacement of the planar oxygen atoms ($<$ 0.1\%). As all the symmetry analysis in the literature up to now has been performed with the I4$_1$/acd space group, in what follows, we shall use both the I4$_1$/acd and I4$_1$/a space groups, highlighting any differences between the two.

\subsection{Resonant structure factors for I4$_1$/acd and I4$_1$/a}

\begin{figure*}[ht!]
\includegraphics[width=1.0\textwidth]{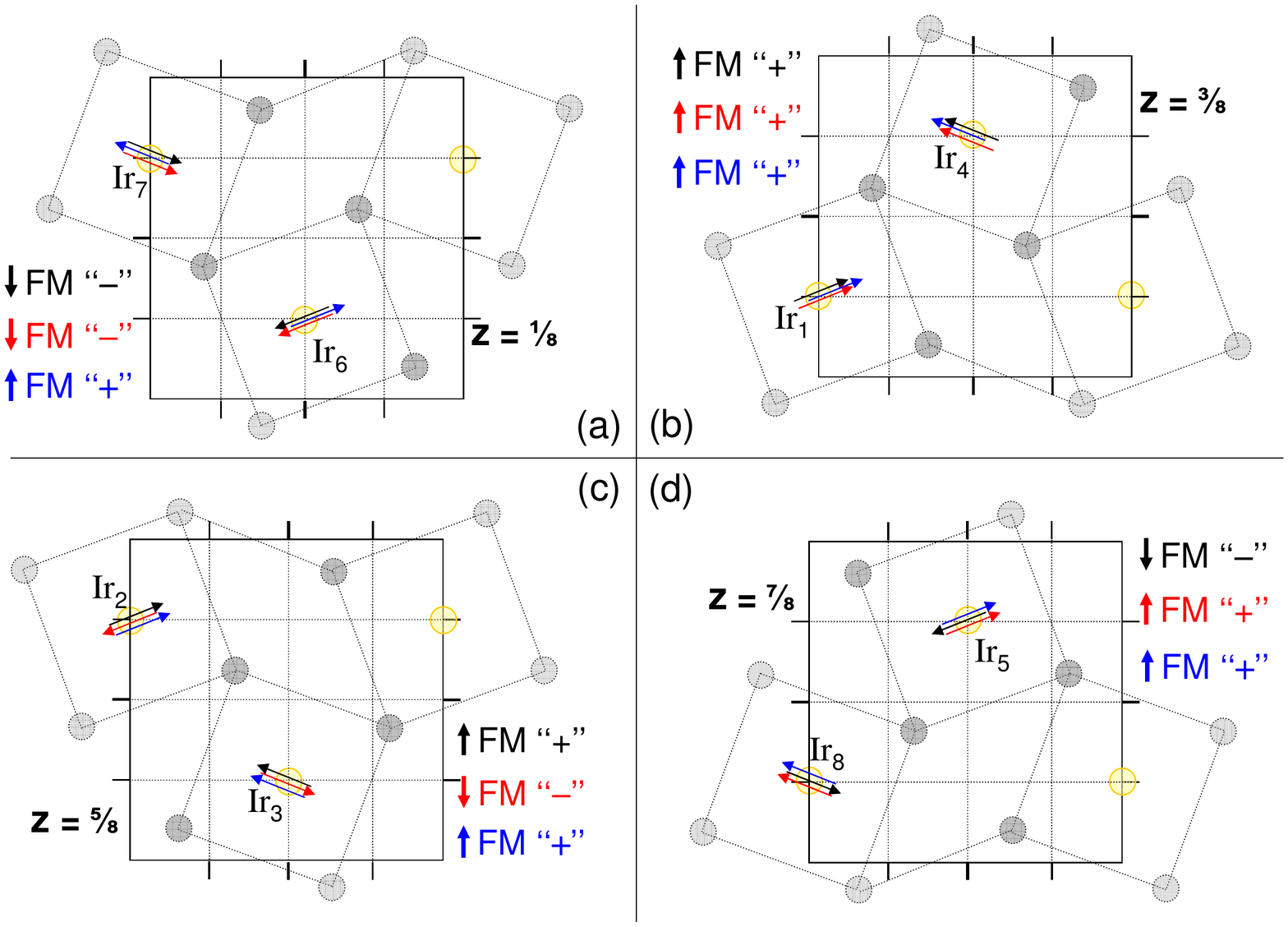}
\vspace{-30pt}
\caption {(Color online) Crystal and magnetic symmetries of Sr$_2$IrO$_4$ in the 4 IrO$_2$ planes of the unit cell: (a) $z=\frac{1}{8}$; (b) $z=\frac{3}{8}$; (c) $z=\frac{5}{8}$; (d) $z=\frac{7}{8}$. The three magnetic patterns of interest, $-++-$ (black arrows), $++++$ (blue arrows) and $-+-+$ (red arrows), are shown. They differ by the order of the ferromagnetic in-plane component along the $b$-axis. Iridium atoms are labelled as in Table I.}
\label{fig1}
\end{figure*}

In the I4$_1$/acd space group (settings 2), the 8 Ir atoms occupy the positions shown in Table I, in fractional units (with $a$=$b$=5.4846 \AA, $c$=25.804 \AA) \cite{original}. They are characterized by a surrounding distorted oxygen octahedron, as shown in Fig.~\ref{fig1} for the basal ($ab$) planes, with apical oxygens along the $c$-axis at 2.057 \AA~and planar oxygens at 1.979 \AA, with a tetragonal distortion of 4\% \cite{original}. The planar oxygens are rotated by about 12$^\circ$ around the $c$-axis: this rotation is the basis of the loss of the 
I4/mmm space-group symmetry that characterizes the analogous compound, Ba$_2$IrO$_4$. Below $T_N \simeq 230$ K \cite{fuji}, an antiferromagnetic state develops, characterized by magnetic moments lying in the basal plane and forming an angle of about 12$^\circ$ with the $a$-axis, as shown in Fig.~\ref{fig1}. The in-plane magnetic pattern is such as to have an antiferromagnetic order parameter along the $a$-axis and a ferromagnetic component along the $b$-axis (smaller by the ratio $\sin 12^\circ/\cos 12^\circ$), leading to the loss of tetragonal symmetry. The ferromagnetic component is however compensated when summed up over the four IrO$_2$ layers of the unit cell. In the undoped compound, the ferromagnetic component along the $b$-axis has the pattern $-++-$ \cite{kim2}, as shown in Fig.~\ref{fig1}.

In order to write down the resonant x-ray structure factor at the Ir L$_3$ edge, we need to know the magnetic symmetry relations among the 8 Ir atoms of the 8a site. In order to allow for comparison with previous work, we list in Table I the Ir atoms with the $-++-$ pattern and the symmetry operations that need to be applied to the first Ir atom in order to obtain the others (this column, for the first atom, corresponds to its point-group symmetries). For future convenience, we list also two other patterns of the ferromagnetic component, the $++++$ and the $-+-+$ patterns, as shown in Fig. \ref{fig1}. The $++++$ structure has been identified as the magnetic structure of Sr$_2$IrO$_4$ in a magnetic field ($H \ge 0.3$ T) directed in the $ab$-plane \cite{kim2,boseggia} and also suggested as the magnetic structure in Rh-doped samples
\cite{clancy,ye15}. The $-+-+$ magnetic structure has not been suggested in the literature up to now, but we claim that it can describe some of the experimental results on Rh-doped samples, as detailed below \cite{footye}.
Both patterns can explain the SHG experiment, as we shall see.

\begin{table}[ht!]
	\caption{Fractional positions in the I4$_1$/acd crystal space group and magnetic symmetries relative to Ir$_1$. The orientation of the magnetic moment for the $-++-$, $++++$ and $-+-+$ patterns relative to the $a$-axis is also given, with $\phi$ the rotation angle.
Note that certain operations are accompanied by glide, screw, or body-centered translations.}
	\centering
	\begin{ruledtabular}
		\begin{tabular}{c|c|c|c|c|c}
			Atom & frac.~pos. & sym$_{-++-}$ & $\phi_{-++-}$  & $\phi_{++++}$ & $\phi_{-+-+}$\\
			\colrule
			Ir$_1$ & (0,1/4,3/8) & ${\hat{E}}$, ${\hat{T}}{\hat{C}}_{2z}$ & 12$^o$  & 12$^o$  & 12$^o$  \\
			Ir$_2$ & (0,3/4,5/8) & ${\hat{I}}$, ${\hat{T}}{\hat{m}}_{z}$ & 12$^o$ & 12$^o$  & 192$^o$  \\
			Ir$_3$ & (1/2,1/4,5/8) & ${\hat{C}}_{2y}$, ${\hat{T}}{\hat{C}}_{2x}$  & 168$^o$ & 168$^o$  & 348$^o$ \\
			Ir$_4$ & (1/2,3/4,3/8) & ${\hat{m}}_{y}$, ${\hat{T}}{\hat{m}}_{x}$ & 168$^o$  & 168$^o$ & 168$^o$  \\
			Ir$_5$ & (1/2,3/4,7/8) & ${\hat{T}}$, ${\hat{C}}_{2z}$ & 192$^o$ & 12$^o$  & 12$^o$  \\
			Ir$_6$ & (1/2,1/4,1/8) & ${\hat{m}}_{z}$, ${\hat{T}}{\hat{I}}$ & 192$^o$ & 12$^o$  & 192$^o$  \\
			Ir$_7$ & (0,3/4,1/8) & ${\hat{C}}_{2x}$, ${\hat{T}}{\hat{C}}_{2y}$ & 348$^o$ & 168$^o$  & 348$^o$  \\
			Ir$_8$ & (0,1/4,7/8) & ${\hat{m}}_{x}$, ${\hat{T}}{\hat{m}}_{y}$ & 348$^o$ & 168$^o$  & 168$^o$  \\
		\end{tabular}
	\end{ruledtabular}
	\label{sym1}
\end{table} 
Here ${\hat{T}}$, ${\hat{I}}$ and ${\hat{E}}$ are the time-reversal operator, the inversion and the identity, and ${\hat{m}}_{i}$ and 
${\hat{C}}_{2i}$ are the mirror symmetry and the two-fold rotation around the axis $i$ ($x,y,z$ parallel to $a,b,c$). Taking into account these symmetries, the resonant x-ray structure factor for the $-++-$ pattern, summed over the 8 Ir atoms, can be written as: 

\begin{align}
F_{hkl}^{-++-} \propto & (1+{\hat{T}}(-1)^{h+k+l})(1+{\hat{I}}(i)^{2k+l}) \nonumber \\
 & (1+{\hat{T}}{\hat{m}}_{x}(-1)^{h+k})f_1
\label{strfac1}
\end{align}
where $f_1$ is the resonant atomic scattering amplitude for Ir atom 1 (see, e.g., Ref.~\onlinecite{jphysD}). Below, we write, for future use, the structure factors for the two other patterns, $++++$ and $-+-+$.  
It is important to notice that both the $++++$ and $-+-+$ magnetic structures are characterized by the same in-plane ferromagnetic component for IrO$_2$ layers $z=\frac{3}{8}$ and $z=\frac{7}{8}$, as shown in Fig.~\ref{fig1}. This implies that the time-reversal symmetry relating Ir$_1$ and Ir$_5$ in Eq.~(\ref{strfac1}) is replaced by the identity (see Eqs.~(\ref{strfac2}) and (\ref{strfac3})). The difference between the two structures lies in the way (for example) layer $z=\frac{7}{8}$ is related to layer $z=\frac{5}{8}$. For the $++++$ structure, Ir$_1$ and Ir$_2$ atoms are related by inversion, as for the $-++-$ structure, whereas for the $-+-+$ structure, Ir$_1$ and Ir$_2$ atoms are related by ${\hat{T}}{\hat{I}}$. This is the second factor in the right-hand side of Eqs.~(\ref{strfac2}) and (\ref{strfac3}). Overall, the corresponding structure factors are:

\begin{align}
F_{hkl}^{++++} \propto & (1+(-1)^{h+k+l})(1+{\hat{I}}(i)^{2k+l}) \nonumber \\
& (1+{\hat{T}}{\hat{m}}_{x}(-1)^{h+k})f_1  
\label{strfac2}
\end{align}

\begin{align}
F_{hkl}^{-+-+} \propto & (1+(-1)^{h+k+l})(1+{\hat{T}}{\hat{I}}(i)^{2k+l}) \nonumber \\
& (1+{\hat{T}}{\hat{m}}_{x}(-1)^{h+k})f_1  
\label{strfac3}
\end{align}

We remark that the last factor $(1+{\hat{T}}{\hat{m}}_{x}(-1)^{h+k})$ is unchanged in all three patterns because it relates the two in-plane Ir atoms, whose relative behavior is not affected by the overall stacking along the $c$-axis. This term is however responsible for changes in the structure factor when the symmetry is reduced to I4$_1$/a. In fact, such a reduction is determined by the breaking of the ${\hat{m}}_{x}$ (and ${\hat{m}}_{y}$) symmetry. This implies that (for example) Ir$_1$ and Ir$_4$ atoms in Table I now belong to two inequivalent sites (4a and 4b). This in turn leads to the altered structure factors:

\begin{align}
\tilde{F}_{hkl}^{-++-} \propto & (1+{\hat{T}}(-1)^{h+k+l})(1+{\hat{I}}(i)^{2k+l}) \nonumber \\
& (f_1 +(-1)^{h+k} f_4)
\label{strfac4}
\end{align}

\begin{align}
\tilde{F}_{hkl}^{++++} \propto & (1+(-1)^{h+k+l})(1+{\hat{I}}(i)^{2k+l}) \nonumber \\
& (f_1 +(-1)^{h+k} f_4)  
\label{strfac5}
\end{align}

\begin{align}
\tilde{F}_{hkl}^{-+-+} \propto & (1+(-1)^{h+k+l})(1+{\hat{T}}{\hat{I}}(i)^{2k+l}) \nonumber \\
& (f_1 +(-1)^{h+k} f_4)  
\label{strfac6}
\end{align}

We notice for the next section on SHG interpretation that all the previous equations are equally valid for the SHG experiment by putting $h=k=l=0$ (in the optical regime, only the zone center is involved). 
Interestingly, this shows that optical reflections in the $-++-$ pattern are only sensitive to time reversal and parity-even observables, otherwise the $(1+{\hat{T}})(1+{\hat{I}})$ prefactor of Eq.~(\ref{strfac4}) would be zero. Likewise, optical reflections in the $++++$ pattern are only sensitive to parity-even quantities (both magnetic and non-magnetic, see Eq.~(\ref{strfac5})) and in the $-+-+$ pattern they are sensitive to ${\hat{T}}{\hat{I}}$-even observables (i.e., either magnetic, parity-odd multipoles, or non-magnetic parity-even ones).

\begin{table}[ht!]
	\caption{Magnetic symmetries relative to Ir$_1$ in the I4$_1$/a crystal space group for the $-++-$ (2/m1$^\prime$ group), $++++$ (2$^\prime$/m$^\prime$) and $-+-+$ (2$^\prime$/m) patterns.
Note that certain operations are accompanied by glide, screw, or body-centered translations. Only the four equivalent Ir$_1$, Ir$_2$, Ir$_5$ and Ir$_6$ are shown.}
	\centering
	\begin{ruledtabular}
		\begin{tabular}{cccc}
			Atom & sym$_{-++-}$ & sym$_{++++}$ & sym$_{-+-+}$  \\
			\colrule
Ir$_1$ & ${\hat{E}}$, ${\hat{T}}{\hat{C}}_{2z}$ & ${\hat{E}}$, ${\hat{T}}{\hat{C}}_{2z}$  & ${\hat{E}}$, ${\hat{T}}{\hat{C}}_{2z}$  \\
Ir$_2$ &  ${\hat{I}}$, ${\hat{T}}{\hat{m}}_{z}$ & ${\hat{I}}$, ${\hat{T}}{\hat{m}}_{z}$ & ${\hat{m}}_{z}$, ${\hat{T}}{\hat{I}}$  \\
Ir$_5$ &  ${\hat{T}}$, ${\hat{C}}_{2z}$ & ${\hat{E}}$, ${\hat{T}}{\hat{C}}_{2z}$ & ${\hat{E}}$, ${\hat{T}}{\hat{C}}_{2z}$  \\
Ir$_6$ &  ${\hat{m}}_{z}$, ${\hat{T}}{\hat{I}}$ & ${\hat{I}}$, ${\hat{T}}{\hat{m}}_{z}$ & ${\hat{m}}_{z}$, ${\hat{T}}{\hat{I}}$  \\
		\end{tabular}
	\end{ruledtabular}
	\label{sym2}
\end{table} 

\subsection{Analysis of key x-ray reflections}

There are two groups of resonant x-ray reflections that have been studied in the literature at the Ir L$_3$ edge and deliver two independent pieces of information. 

In the first group, we have those reflections which served to identify the $-++-$ magnetic space group of the stoichiometric compound. They are of the kind (1,0,4n), (0,1,4n+2) \cite{twin} and (0,0,2n+1), that we analyze within the I4$_1$/acd space group to compare with the existing literature. Later, we shall highlight the differences induced by the reduction to I4$_1$/a.
From Eq.~(\ref{strfac1}), as $h+k+l$ is odd, these three reflections must be magnetic in the $-++-$ state, and they vanish for the other two patterns. Their presence is therefore a signature of the $-++-$ state. Notice also that for SP scattering \cite{notesp}, $f_1$ is proportional to $\vec{k}_o\cdot\vec{m}\equiv m_{k_o}$ (with $\vec{k}_o$ the outgoing scattering vector, $\vec{m}$ the magnetic moment). The selection rule imposed by the term $(1+{\hat{T}}{\hat{m}}_{x}(-1)^{h+k})$ gives a signal proportional to $(1+{\hat{m}}_{x}) m_{k_o}\propto m_a$ for both $F_{1,0,4n}^{-++-}$ and $F_{0,1,4n+2}^{-++-}$. Here $m_a$ is the projection of the magnetic moment along the $a$-axis. Instead, for $F_{0,0,2n+1}^{-++-}$, the signal is proportional to $(1-{\hat{m}}_{x}) m_{k_o}\propto m_b$, as noted earlier \cite{boseggia2}. This method allows one to obtain the direction of the magnetic moment (the rotation angle of $\simeq 12^\circ$ reported in the literature) by means of the ratio of the $m_a$ to the $m_b$ components. However, this assumes the I4$_1$/acd space group. The reduction to I4$_1$/a implies the breakdown of the $(1+{\hat{T}}{\hat{m}}_{x}(-1)^{h+k})$ selection rule due to the inequivalence of Ir$_1$ and Ir$_4$. Though this reduction is small, as stated above, it could lead to changes in the rotation angle, and it might be worthwhile to repeat this analysis for I4$_1$/a, as now the magnetic moments of the two in-plane Ir atoms become inequivalent (though the argument of Ref.~\onlinecite{jackeli} should still impose a locking with the rotation of the oxygen octahedra). As a last remark on the I4$_1$/acd to I4$_1$/a symmetry reduction, we emphasize that it does not play any role for the coupling along the $c$-axis, as the symmetry that it breaks is the one that links the two  Ir atoms in the same plane.

A second group of reflections more interesting for our work, as they are key to interpreting the SHG experiment, are of the kind (1,0,4n$\pm$1)  (similarly (0,1,4n$\pm$1)). As clear from Eq.~(\ref{strfac4}), these reflections cannot have magnetic origin and their appearance below $T_N$ indicates an alteration of the $-++-$ configuration. They were found either by applying a small magnetic field in the stoichiometric material \cite{kim2} or by Rh doping 
\cite{clancy,ye15}. In both cases they were suggested as being signatures of the $++++$ state. We show below that this is not necessarily the case, as the $-+-+$ configuration can give rise to the same magnetic or charge reflections, and further investigation is
needed to disentangle the two patterns, at least in the case of Rh doping \cite{footye}. 
As before, we start our analysis with the I4$_1$/acd space group. Though $h+k+l$ is even, reflections (1,0,4n$\pm$1) are Bragg forbidden for the $-++-$ configuration, since $h+k$ is odd (Eq.~(\ref{strfac1}). They can however be explained as magnetic in origin (but can also be non-magnetic in I4$_1$/a space group, see below) if the system undergoes a phase transition to either the $++++$ configuration or the $-+-+$ configuration. If this is the case, the structure factors become: 
\begin{equation}
F_{1,0,4n\pm 1}^{++++}\propto (1\pm i)m_a
\end{equation}
and 
\begin{equation}
F_{1,0,4n\pm 1}^{-+-+}\propto (1\mp i)m_a
\end{equation}
They lead to the same intensity, so from a purely magnetic analysis, they cannot be differentiated. We should notice here that care is needed in identifying the orthorhombic $a$ and $b$ axes, as reversing them might lead to an incorrect pattern identification. In fact, switching $h$ and $k$ Miller indices corresponds to switching the $++++$ and $-+-+$ patterns: 

\begin{equation}
F_{0,1,4n\pm 1}^{++++}\propto (1\mp i)m_a
\end{equation}
and 
\begin{equation}
F_{0,1,4n\pm 1}^{-+-+}\propto (1\pm i)m_a.
\end{equation}

In all these expressions, $m_a$ is the $a$-axis component of the magnetic moment at sites Ir$_1$ and Ir$_4$ (difference of $f_1$ and $f_4$). 

\begin{figure}
\includegraphics[width=0.35\textwidth]{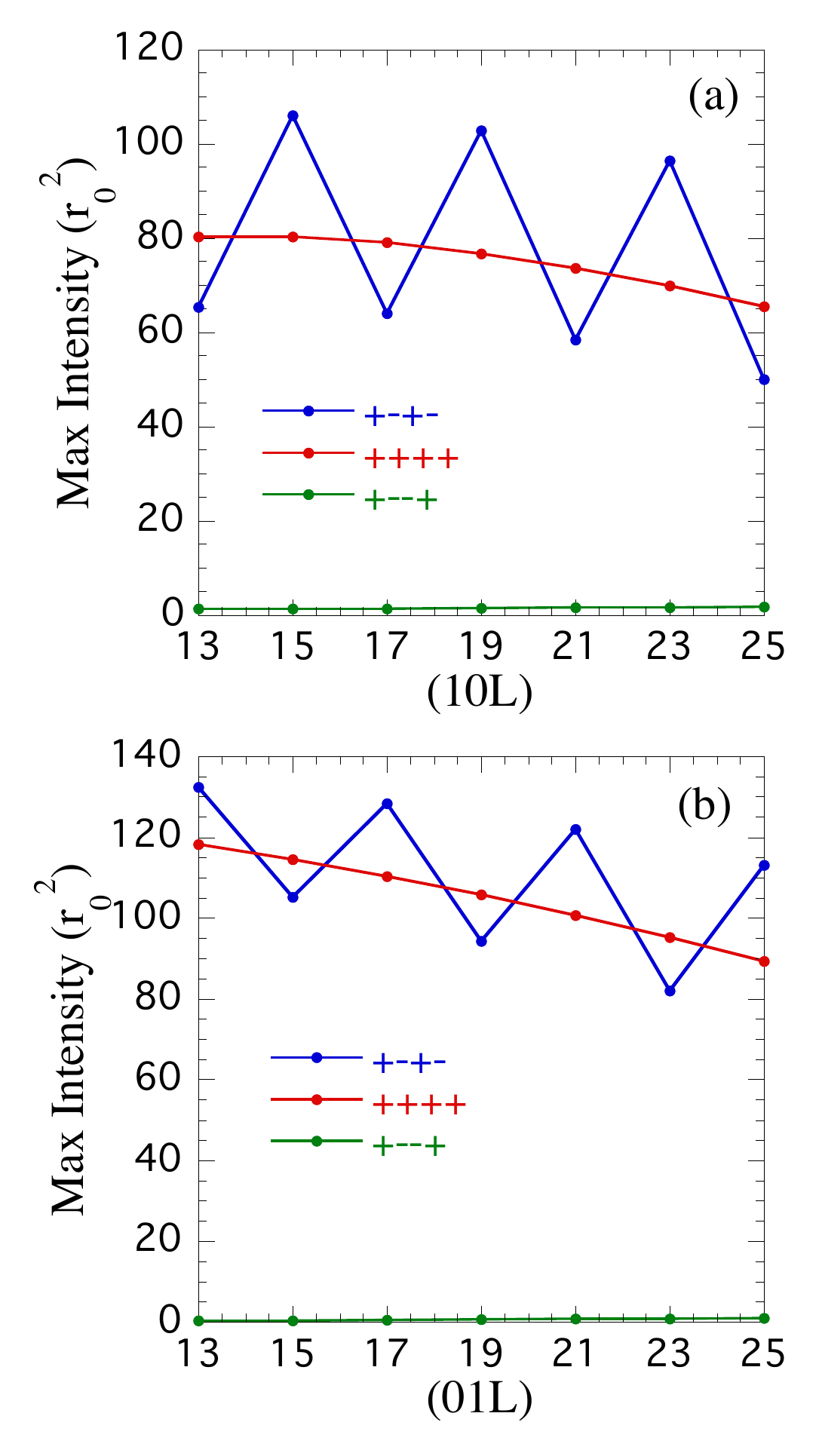}
\caption {(Color online) The alternating intensities (at the RXS maximum for L$_3$) for magnetic pattern $-+-+$ as opposed to $++++$, of (a) the (1,0,4n-1) and (1,0,4n+1) reflections, and (b) the (0,1,4n-1) and (0,1,4n+1) reflections.
These simulations were performed with the FDMNES code for a cluster radius of 6.5 \AA~and a Hubbard $U$ on the Ir sites of 2 eV.  For (10L), the azimuthal angle was 0$^\circ$ and for (01L), it was 90$^\circ$.  In all cases, the polarization geometry was SP, with an assumed core-hole lifetime of 5.25 eV.
Similar oscillations are found for a cluster radius of 3 \AA, though they are less pronounced.
}
\label{fig2}
\end{figure}

In order to disentangle the two magnetic patterns, we need to play on the differences between the ${\hat{T}}{\hat{I}}$ and ${\hat{I}}$ operators that relate f$_1$ to f$_2$. This can only be done by allowing an interference with the charge scattering. In fact, the charge scattering in resonant conditions is not only given by the (scalar) Bragg scattering, but also by the anisotropic scattering that does not obey the extinction rule $(1+{\hat{m}}_{x}(-1)^{h+k})=0$, because the mirror symmetry is not necessarily +1. The intensity of the charge scattering at these (1,0,4n$\pm$ 1) reflections will be increased by the symmetry reduction to the I4$_1$/a space group. As stated above, such a group breaks the ${\hat{m}}_{x}$-symmetry, violating the above extinction rule, even at the level of the scalar charge scattering, and in fact the existence of these reflections in neutron scattering above $T_N$ have been taken to be a signature of this space group \cite{ye13}.
For RXS, magnetic and non-magnetic terms are out of phase by $\pi/2$. This implies that, writing the non-magnetic atomic scattering factor as f$^c$ (in the I4$_1$/a space group, from Eqs.~(\ref{strfac5}) and (\ref{strfac6}), it is f$_1^c-$f$_4^c$) and the magnetic one as f$^m$ (= f$_1^m-$f$_4^m$), apart from an overall phase factor, the structure factors can be written as:

\begin{equation}
\tilde{F}_{1,0,4n\pm 1}^{++++}\propto (1 \pm i)f^c \mp (1 \mp i)f^m =(f^c \mp f^m) \pm i (f^c \pm f^m)
\end{equation}
and
\begin{equation}
\tilde{F}_{1,0,4n\pm 1}^{-+-+}\propto (1 \pm i)(f^c \pm f^m)
\end{equation}
From these expressions, we see that the interference of magnetic and non-magnetic terms allows differentiating the two patterns. In the case of the $++++$ pattern, the signals at (1,0,4n+1) and (1,0,4n$-$1) are identical (apart from the different geometrical factors due to the different $\vec{Q}$). But, they are different for the $-+-+$ pattern due to the constructive/destructive interference seen in these expressions. A numerical simulation of these findings by the FDMNES code (see Section V) is reported in Fig.~\ref{fig2}.

We should remark here on an important warning for this analysis. In the previous six equations, we have treated the atomic scattering factors f$^c$ and f$^m$ as if they were real quantities, whereas, close to an edge, they are complex, because of the energy denominator $(\hbar\omega - \Delta E +i\Gamma)$, where $\Delta E$ measures the energy difference of the two levels related by the photon transition and $\Gamma$ measures the inverse of the core-hole lifetime. If, however, we are very close to a resonance, then $|\hbar\omega - \Delta E| \ll \Gamma$, and the denominator becomes purely imaginary and all the previous discussion on interferences keeps its validity (Fig.~\ref{fig2}, being at the L$_3$ maximum, corresponds to this case). Of course, the same is true in the opposite, non-resonant case: $|\hbar\omega - \Delta E| \gg \Gamma$. It is not, however, true when $|\hbar\omega - \Delta E| \sim \Gamma$, or if two resonances are sufficiently close to one another.

Even though magnetic neutron scattering differs from magnetic RXS (in particular, no resonant denominator), in Ref.~\onlinecite{ye15}, the authors find a different signal at (101) and at (103) in the spin-flip versus non-spin-flip ratios.
In the light of our previous analysis, this would seem to point to the $-+-+$ configuration.
In RXS, this shows us as an alternation of the (10L) (L odd) intensity versus L for $-+-+$ (due to the above mentioned interference) as opposed to the smooth dependence for the $++++$ configuration, which, as we show in Fig.~\ref{fig2}, is quite pronounced in ab initio calculations.  
A smooth behavior versus L was seen by Clancy {\it et al.}~for their Rh-doped samples \cite{clancy} which is probably why they advocated the $++++$ state over the $-+-+$ state.  One issue is that their azimuthal plots indicate multi-domain effects. Another is that the energy at which the measurement was performed might have induced extra interference due to the energy denominator, as discussed above. We suggest that additional RXS and neutron experiments on single domains be performed to
check which pattern ($-+-+$ or $++++$) is actually induced by Rh doping.
Regardless, both $-+-+$ and $++++$ configurations can explain the SHG experiment \cite{shg}, without the need to invoke exotic magnetic symmetries. As shown in the Table II, the $-+-+$ state has the magnetic space group 2$^\prime$/m, and the $++++$ state the 2$^\prime$/m$^\prime$ one. As detailed in the next Section, both are compatible with the SHG result.

\section{A reanalysis of the symmetries of the SHG experiment}

In this Section, we reanalyze the SHG experiment \cite{shg}, listing all the symmetries that allow one to address the experimental results.  
In subsection A, we discuss the two susceptibility tensors, $\chi^{(e)}$ and $\chi^{(m)}$ (defined in Appendix A.1), possibly involved in the interference pattern with the high-temperature signal, determined by $\chi^{(q)}$ \cite{torch}. Both tensors should be investigated on the same footing because it is known, e.g., for Cr$_2$O$_3$, that the magnetic part \cite{nota2} of $\chi^{(e)}$ and $\chi^{(m)}$ are of the same order of magnitude \cite{muto}.
We also identify the symmetry of the order parameters associated with $\chi^{(e)}$ and $\chi^{(m)}$ for linear polarizations and their multipolar ranks. Then, in subsection B, we evaluate the different azimuthal dependences of $\chi^{(e)}$ and $\chi^{(m)}$. This, together with a full analysis of the magnetic symmetries of Sr$_2$IrO$_4$ and the findings of subsection A, allows us to point towards the following interpretation of the SHG experiment \cite{shg}: only two magnetic space groups can explain the interference pattern of the SHG experiment. The former is 2$^\prime$/m (as already suggested in Ref.~\onlinecite{shg}). 
The associated order parameters are all inversion and time-reversal odd, with the symmetry of either a toroidal dipole, a magnetic quadrupole or a toroidal octupole (the magnetic quadrupole cannot be observed in SS geometry).
The latter is 2$^\prime$/m$^\prime$, which is characterized by an order parameter with the symmetry of either a magnetic toroidal monopole, a magnetic dipole, a magnetic toroidal quadrupole or a magnetic octupole. 
In either case, we suggest that the observed symmetry reduction is not determined by exotic magnetic patterns, but by a transition to the $-+-+$ state (2$^\prime$/m group), or to the $++++$ state (2$^\prime$/m$^\prime$ group). How this transition could happen is discussed at the end of this Section, along with an alternate explanation that what is observed is surface magnetic SHG.

\subsection{Order parameters associated with SHG tensors}

Second harmonic generation is a third-order process in the matter-radiation interaction, determined by two absorptions of a photon $\hbar\omega$ and the emission of a photon $2\hbar\omega$ \cite{fiebig}, as pictorially described in Fig.~\ref{fig3}. The full cross-section and its explicit derivation are reported in Appendix A.1. 
The total scattering amplitude, $A_{SHG}$, is obtained, as for RXS \cite{jphysD}, by a scalar coupling of tensors representing the properties of the sample with the corresponding tensors describing the electromagnetic field. The full SHG amplitude, reported in Eq.~(\ref{scatSHG2}), is then composed of terms of the kind: $A_{SHG} \sim \chi^{(e)}_{\alpha\beta\gamma}\epsilon_{\alpha}^*\epsilon_{\beta}\epsilon_{\gamma}+ \chi^{(m)}_{\alpha\beta\gamma} (\vec{\epsilon}^*\times\vec{k})_{\alpha}\epsilon_{\beta} \epsilon_{\gamma} + \chi^{(q)}_{\alpha\beta\gamma\delta}\epsilon_{\alpha}^*k_{\beta}\epsilon_{\gamma}\epsilon_{\delta}$, where $\vec{\epsilon}$ and $\vec{k}$ are the polarization and wave-vectors of the electromagnetic field.

\begin{figure*}[ht!]
\includegraphics[width=1.0\textwidth]{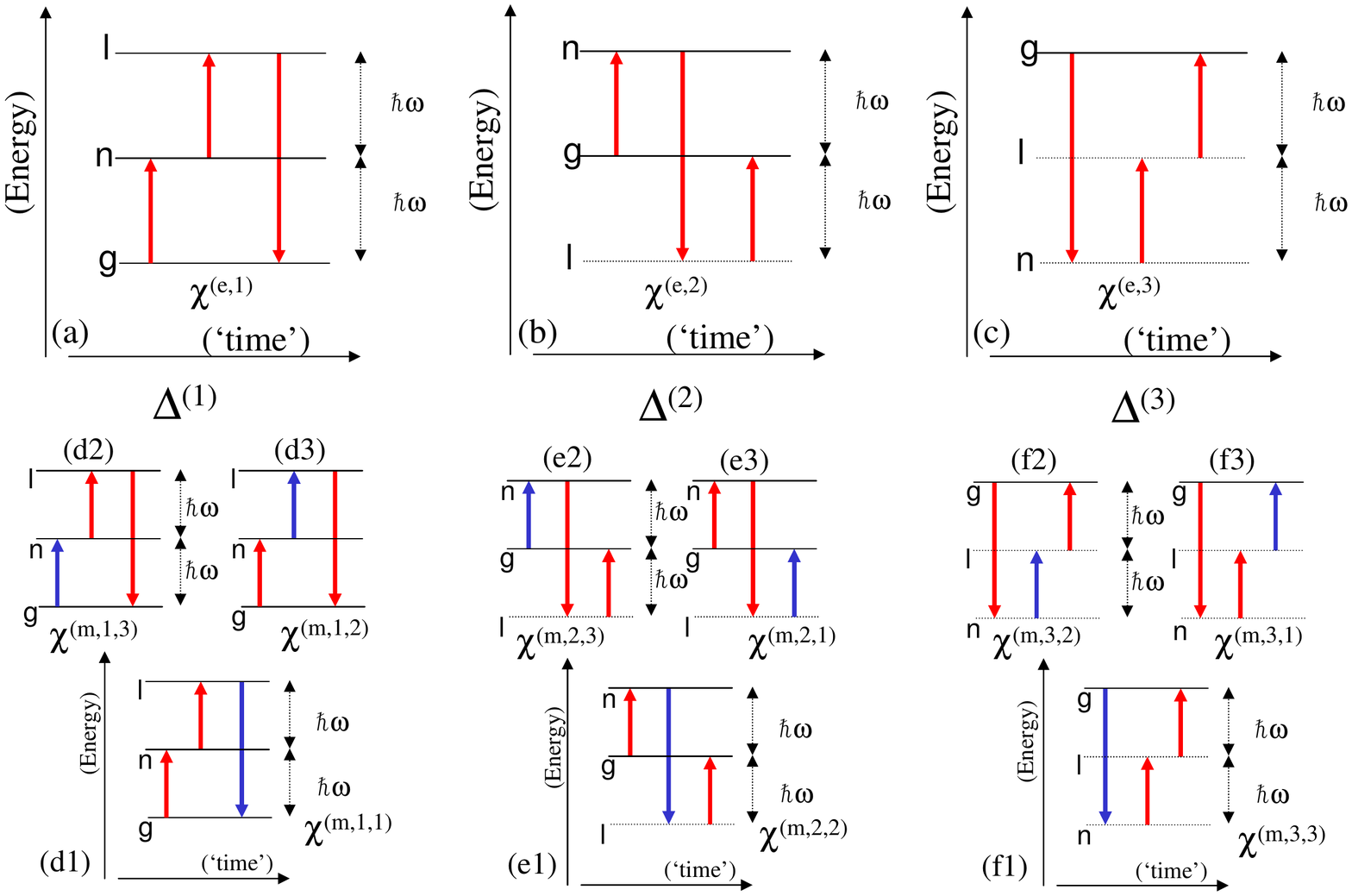}
\vspace{-60pt}
\caption {(Color online) Schematic representation of the SHG process. E1 transitions are represented in red and M1 transitions in blue. The three processes leading to $\chi^{(e)}$ are shown in (a), (b), (c). The nine processes leading to $\chi^{(m)}$ are shown in (d), (e), (f). All these processes are expressed in Appendix A through Eqs.~(\ref{polmate}) to (\ref{polmatm}). Each label $\chi^{(m,j,i)}$ is explicitly defined in one of these equations. The non-resonant processes are highlighted by an energy level represented by a dotted line. The denominators $\Delta^{(i)}$, defined in the text, correspond to doubly-resonant, singly-resonant and non-resonant processes.}
\label{fig3}
\end{figure*}

As clear from the definitions in Appendix A.1, $\chi^{(e)}$ is characterized by all three transitions (the two at frequency $\omega$ and the one at frequency $2\omega$) of electric-dipole character (therefore inversion odd), and $\chi^{(m)}$ is characterized by one magnetic-dipole and two electric-dipole transitions (therefore inversion even). They can both interfere with the electric-quadrupole tensor $\chi^{(q)}$ (characterized by one electric-quadrupole and two electric-dipole transitions), responsible for the SHG signal in the high-temperature phase of Sr$_2$IrO$_4$ \cite{torch}. 
From Eqs.~(\ref{chie}), (\ref{chim}) and (\ref{chiq}) in Appendix A, $\eta^{(q)}$, the transition-matrix element associated to the susceptibility $\chi^{(q)}$, contains an extra factor of $i$ coming from the expansion of $e^{i\vec{k}\cdot\vec{r}}$, and this determines the phase shift of $\pi/2$ of these matrix elements with respect to those of the $\eta^{(e)}$ term. Following the analysis reported in Appendix A.2, we have therefore that the time-reversal even part of $\eta^{(q)}$ is imaginary (and the time-reversal odd part of $\eta^{(q)}$ real); the time-reversal even part of $\eta^{(e)}$ is real (and its time-reversal odd part is imaginary); finally, the time-reversal even part of $\eta^{(m)}$ is imaginary (and its time-reversal odd part is real). This implies that, in the non-resonant regime where denominators are real, the time-reversal even part of $\chi^{(q)}$ can interfere with both the time-reversal odd part of $\chi^{(e)}$ and with the time-reversal even part of $\chi^{(m)}$, but not with the time-reversal even part of $\chi^{(e)}$ and with the time-reversal odd part of $\chi^{(m)}$. However, the presence of the imaginary damping factors $i\Gamma_n$ in the resonant denominators scrambles up this analysis, as detailed in Appendix A.2, so that all terms can interfere among themselves. Yet, this analysis in terms of $\eta^{(e,m,q)}_{\alpha\beta\gamma;ln}$ without the resonant denominator is fundamental, as in the case of RXS, in order to identify the time-reversal properties of the order parameters associated with this process, that are determined just by the matrix elements. In fact, the physical origin of the complex numerators and denominators in second-order (RXS), or third-order (SHG) expressions is profoundly different. The imaginary unit in the numerator is a consequence of magnetism: in its absence, eigenstates $\Phi_j$ of Eq.~(\ref{scatSHG}) can be chosen as real and all matrix elements $\eta^{(e,m,q)}_{\alpha\beta\gamma;ln}$ (see Appendix A.2) would be real as well. Therefore, the time-reversal symmetry of the matrix elements $\eta^{(e,m,q)}_{\alpha\beta\gamma;ln}$ is related to their real and imaginary parts. The imaginary unit in the denominator is, instead, a consequence of damping due to spontaneous emission \cite{sakurai}: that sign cannot be changed, because damping is irreversible. 

The previous classification reminds one of the simpler RXS case \cite{jphysD,prbvarma}. There are however several differences, because of the intrinsic asymmetry of the SHG amplitude (two photons $\omega$ in; one photon $2\omega$ out). 
For example, it is well known that the electric dipole-electric dipole (E1-E1) approximation in RXS leads to a time-reversal odd, imaginary part of the matter tensor, which is proportional to the magnetic dipole and is scalarly coupled to $\epsilon^*_{\rm out} \times \epsilon_{\rm in}$. This in turn implies that for E1-E1, no magnetic signal occurs in the SS channel.  This is not the case for SHG, because of the asymmetry of the denominators $\Delta_{l,n}^{(i)}$ with respect to the exchange $n \leftrightarrow l$, as shown in detail in Appendix A.1 and A.2: indeed, a magnetic $SS$ signal is quite common in SHG \cite{fiebig}.

The general classification of the order parameters associated with the transition matrix elements in the SHG susceptibilities is quite lengthy and will be treated in a future publication. As already reported in Ref.~\onlinecite{marri} for second-order susceptibilities, the order parameters in the optical regime, differently from the x-ray regime, are correlation functions, which are much harder to analyze. Here we focus on the SHG experiment of Ref.~\onlinecite{shg} and, in particular, consider just the symmetry of the order parameter associated with the $\chi^{(e)}$ and $\chi^{(m)}$ tensors when the incoming and outgoing electric fields are linearly polarized, and therefore real.

Consider for example $\chi^{(e)}$ and Eq.~(\ref{ordpar_eee}). $A_{SHG}^{(e)}$ is a scalar quantity. We can take advantage of this property to decompose the susceptibilities $\chi^{(e)}$ in spherical tensors, as each of them must be scalarly coupled to an equivalent spherical tensor representing the polarization properties. Each spherical tensor derived from $\chi^{(e,m,q)}_{\alpha\beta\gamma}$ is an irreducible representation of the rotation group whose symmetry can be identified with that of a given multipole. We can rewrite Eq.~(\ref{ordpar_eee}) as:  

\begin{align} \label{shgspher}
A_{SHG}^{(e)} & \propto  {\tilde \chi}^{(e)}_{\alpha\beta\gamma} \epsilon^o_{\alpha}\epsilon^i_{\beta}\epsilon^i_{\gamma}  \\
& = \frac{1}{2} ({\tilde \chi}^{(e)}_{\alpha\beta\gamma} \epsilon^o_{\alpha}\epsilon^i_{\beta}\epsilon^i_{\gamma} + {\tilde \chi}^{(e)}_{\alpha\gamma\beta} \epsilon^o_{\alpha}\epsilon^i_{\gamma}\epsilon^i_{\beta}) \nonumber \\  
& =\frac{1}{4} ({\tilde \chi}^{(e)}_{\alpha\beta\gamma} +  {\tilde \chi}^{(e)}_{\alpha\gamma\beta} ) \epsilon^o_{\alpha}(\epsilon^i_{\beta}\epsilon^i_{\gamma} +\epsilon^i_{\gamma}\epsilon^i_{\beta}) \nonumber  
\end{align}

Though there are several ways to couple three vectors in irreducible components, Eq.~(\ref{shgspher}) suggests the most natural one: 
the symmetry of the $\epsilon^i_{\gamma} \epsilon^i_{\beta}$ part implies that only 6 out of 9 cartesian components contribute in this coupling: they form a scalar tensor ($T^{(0)}$) and a second-rank spherical tensor ($T^{(2)}_m$, $m=-2$ to $2$). In turn, these two spherical tensors couple to the remaining outgoing polarization vector $\vec{\epsilon}^o$. The coupling of the scalar (order-zero spherical tensor) with the vector $\vec{\epsilon}^o$ (first-rank spherical tensor), gives a first-rank spherical tensor ($\overline{O}^{(1)}$). The coupling of the second-rank spherical tensor $T^{(2)}_m$ with the vector $\vec{\epsilon}^o$ leads to three spherical tensors: $\tilde{O}^{(1)}$, $\tilde{O}^{(2)}$ and $\tilde{O}^{(3)}$ (as in the usual coupling of angular momenta).
The explicit expression for all these tensors is given in Appendix A.3.

The case of $\chi^{(m)}$ is less straightforward because of the substitution of $\epsilon^{i,o}_{\alpha}$ with $(\vec{\epsilon}^{i,o}\times\vec{k}^{i,o})_{\alpha}$ and the associated symmetrization over all three terms, as reported in Appendix A.3. However, {\it mutatis mutandis}, the order parameters are in this case spherical tensors of rank $i=$0, 1, 2, 3: $\tilde{\tilde{O}}^{(i)}$.
In this case, more tensors of the same order can appear, as detailed in Appendix A.3. 
Of course, the $\tilde{O}^{(i)}$ are all inversion-odd and the $\tilde{\tilde{O}}^{(i)}$ are all inversion-even. The former are associated with order parameters with the symmetry of a toroidal dipole, a magnetic quadrupole and a toroidal octupole (for the time-reversal odd part) and with the symmetry of an electric dipole, an axial toroidal quadrupole and an electric octupole (for the time-reversal even part). The latter are associated with order parameters with the symmetry of a magnetic toroidal monopole, a magnetic dipole, a magnetic toroidal quadrupole and a magnetic octupole (for the time-reversal odd part) and with the symmetry of an electric charge, an axial toroidal dipole, an electric quadrupole and an axial toroidal octupole (for the time-reversal even part). As stated above, we can just speak of an order parameter `with the symmetry of', because in the optical regime, differently from the x-ray regime, all involved states are band-like and the ``order parameters" are rather many-body correlation functions \cite{marri,benoist}.

In the following subsection, we specialize this analysis to the magnetic symmetries of Sr$_2$IrO$_4$.

\subsection{SHG symmetry analysis applied to Sr$_2$IrO$_4$}

The magnetic space group of the $-++-$ state of Sr$_2$IrO$_4$ associated with the I4$_1$/a crystal symmetry is 2/m1$^\prime$, as clear from Table \ref{sym2}, when only the 4 equivalent Ir atoms of the I4$_1$/a group are considered (e.g., Ir$_1$, Ir$_2$, Ir$_5$, Ir$_6$). The behavior of the two tensors $\chi^{(e)}$ and $\chi^{(m)}$ under the magnetic symmetry group 2/m1$^\prime$ and subgroups is analyzed here and in Appendix A.4, the aim being to find out what magnetic subgroups allow for both the interference with the time-reversal even $\chi^{(q)}$ signal of the non-magnetic phase (found in Ref.~\onlinecite{torch}) and the odd $\psi$-dependence seen in the experimental data (Figs.~1 and 3 of Ref.~\onlinecite{shg}). As shown in Eq.~(\ref{azim}) below, the key feature for having an odd $\psi$-dependence is to have allowed cartesian tensors with an even dependence on $z$ (which means an odd dependence on $x$ and $y$, as both $\chi^{(e)}$ and $\chi^{(m)}$ are third-rank cartesian tensors). This statement comes from the following expressions for the electromagnetic field (in = incoming, frequency $\omega$; out = outgoing, frequency $2\omega$; $\theta$ is the angle between the outgoing beam and the $c$-axis; $\psi$ is the azimuthal angle around the $c$-axis, with $\psi =0$ when the in-plane projection of the incoming wave-vector is along the $a$-axis):
\begin{align} \label{azim}
E_S^{in} & =  (-\sin\psi,\cos\psi,0)=-H_P^{in} \\
E_P^{in} & =  (\cos\theta \cos\psi,\cos\theta \sin\psi,\sin\theta)=H_S^{in} \nonumber \\
E_P^{out} & =  (-\cos\theta \cos\psi,-\cos\theta \sin\psi,\sin\theta)=H_S^{out} \nonumber
\end{align}
noting that $E_S^{in}=E_S^{out}$ and $H_P^{in}=H_P^{out}$. For future use, we also write $k^{in}=(\sin\theta \cos\psi,\sin\theta \sin\psi,-\cos\theta)$ and $k^{out}=(\sin\theta \cos\psi,\sin\theta \sin\psi,\cos\theta)$.

For the eight magnetic groups discussed below, those allowing for a third-rank tensor with an odd $\psi$-dependence can easily be picked out from Tables 4 and 7 of Birss \cite{birss}.
For the time-odd case of interest, there are only two possibilities: a polar tensor for 2$^\prime$/m (that is, $\chi^{(e)}$) and an axial tensor for 2$^\prime$/m$^\prime$
(that is, $\chi^{(m)}$).  For the unlikely time-even case, there is only a polar tensor for m1$^\prime$ and an axial tensor for $\bar{1}1^\prime$. A detailed demonstration of these properties is provided in Appendix A.4. 
We remark that the surface electric dipole contribution for the relevant point groups, as listed in Ref.~\onlinecite{torch} (supplemental material), have an even dependence on $\psi$ and can be excluded for this reason.

It is interesting to identify the allowed components of the order parameters for the two space groups 2$^\prime$/m and 2$^\prime$/m$^\prime$. For the latter, the magnetic-dipole order parameter is the in-plane ferromagnetic component along the $b$-axis, noting that the SHG signal is actually determined by higher-order correlation functions with the same symmetry (Appendix A.3) as opposed to those determined
from core-hole spectroscopies. For the 2$^\prime$/m space group, the calculation of the toroidal dipole and magnetic quadrupole is slightly more complex: we can explicitly calculate their values, for the $-+-+$ pattern, taking, respectively, the antisymmetric and the symmetric traceless parts of the following cartesian tensor: ${\cal M}_{ij} \equiv \sum_{n=1}^8 r_i^{(n)} m_j^{(n)}$. Here the sum is performed over the 8 Ir atoms in the unit cell, $r_i^{(n)}$ is the $i^{\rm th}$ cartesian component ($i=x,y,z$) at the $n^{\rm th}$ Ir site of the position vector and $m_j^{(n)}$ the $j^{\rm th}$ cartesian component of the magnetic moment at the $n^{\rm th}$ Ir site (here, we consider magnetic patterns where $m_z=0$ for all sites). A direct calculation using Table I shows that the only components different from zero for the $-+-+$ pattern are the toroidal dipole $\Omega_x$ (antisymmetric) and the magnetic quadrupole $M_{yz}$ (symmetric, traceless). All other components are zero. The absolute value of both $\Omega_x$ and $M_{yz}$ is $\frac{1}{2}|m| |c| \sin\phi$, where $|m|$ is the value of the magnetic moment at the Ir sites, $|c|$ is the value of the $c$-axis length (25.804 \AA) and $\phi \sim 12^\circ$ is the angle of the magnetic moment with the $a$-axis. Again, we remind that the SHG signal is actually associated with higher-order correlation functions with the same symmetry as $\Omega_x$ and $M_{yz}$ (Appendix A.3).

\begin{figure}
\includegraphics[width=0.35\textwidth]{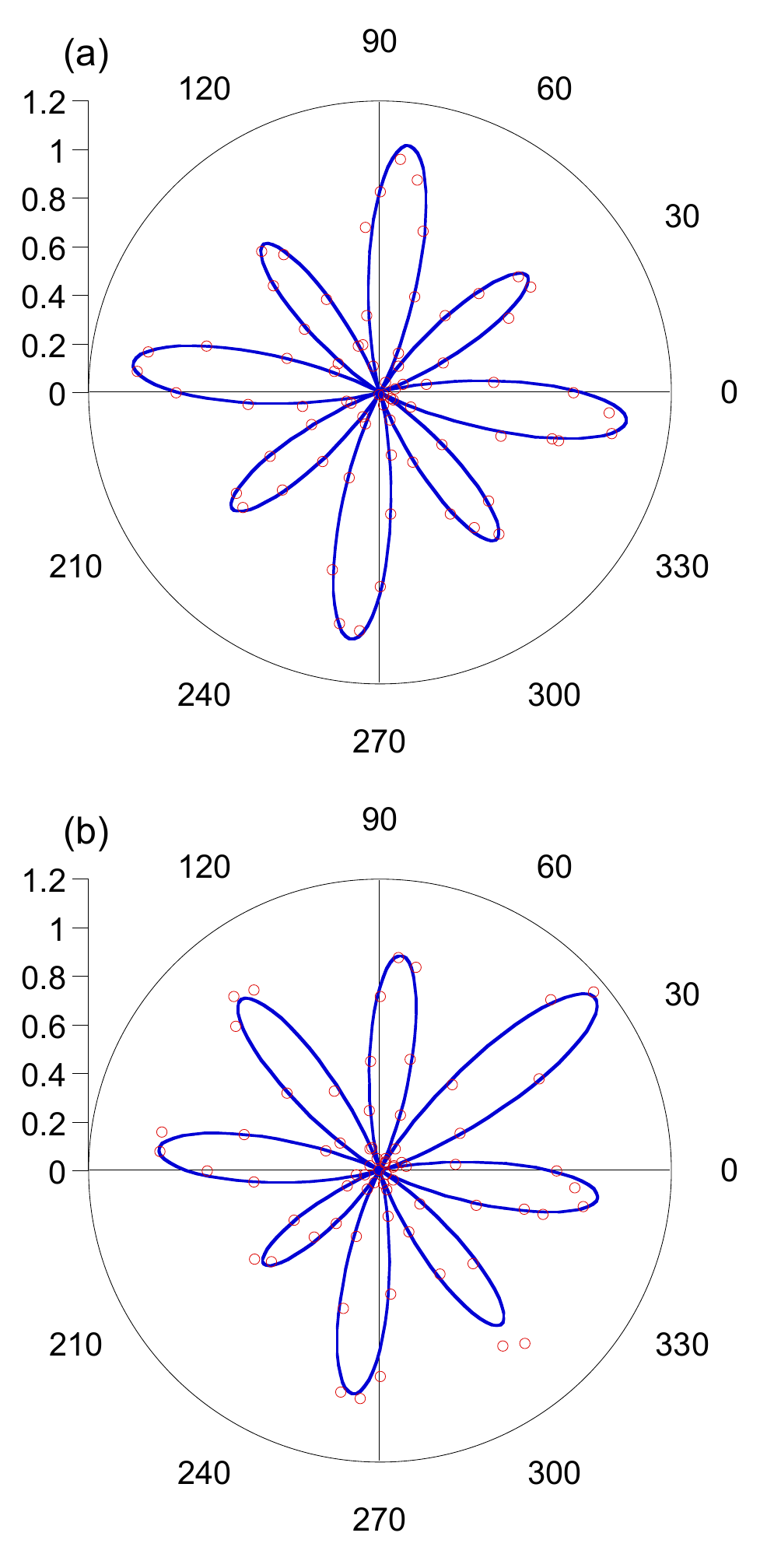}
\caption {(Color online) Second harmonic signal in SS geometry at (a) 295K and (b) 175K \cite{shg}.  The zero of the angle
is along the $a$-axis.  The solid curves are fits to angular functions as
tabulated in Ref.~\onlinecite{torch}.  For (a), the curve is the square of $0.91 - 0.42\sin(4\psi) - 1.69\sin^2(2\psi)$.
Note that the smallest coefficient is for the only allowed term in 4/mmm (the rest only occur for 4/m).  For (b), the curve is the square of
$0.86 - 0.40\sin(4\psi) - 1.70\sin^2(2\psi) - 0.03\cos^3(\psi) - 0.25\cos^2(\psi)\sin(\psi) - 0.11\cos(\psi)\sin^2(\psi) + 0.005\sin^3(\psi)$.
Note the small changes in the first three coefficients relative to (a), and the fact that the last four coefficients indicate
the significance of the octupole terms.}
\label{fig4}
\end{figure}

Some brief comments on the azimuthal dependence of the experimental data are in order.  The high temperature data originates from a $\chi^{(q)}$ signal, since $\chi^{(e)}$ is not allowed and $\chi^{(m)}$ does not have the correct angular dependence \cite{birss,torch}.  These data cannot be fit by functions of just 4/mmm symmetry \cite{torch}.  In fact, the coefficients of the 4/m terms not in 4/mmm are larger than the 4/mmm term, despite the weak nature of the 4/m symmetry breaking (Fig.~\ref{fig4}(a)).  Similar observations apply as well to the third harmonic signal \cite{torch}.  Whether this is a real effect, or due to other factors not taken into account in the analysis, remains to be seen.  In the low temperature phase, where C$_4$ symmetry is broken down to C$_1$ \cite{shg}, we find that dipole terms alone are not sufficient to fit the change in the azimuthal dependence (in SS geometry, quadrupole terms do not enter, and the dipole term goes as $\sin(\psi)$).  This means that octupole terms play a significant role (see caption of Fig.~\ref{fig4}(b)), as often found as well in RXS.

To summarize, the 2$^\prime$/m and 2$^\prime$/m$^\prime$ magnetic groups are the only ones that can explain the SHG data with time-reversal odd order parameters. Two other groups, $\overline{\rm 1}$1$^\prime$ and m1$^\prime$, could have been compatible with the interference pattern, but are characterized by non-magnetic order parameters that make them implausible. 
Both 2$^\prime$/m, corresponding to the $-+-+$ magnetic pattern, and 2$^\prime$/m$^\prime$, corresponding to the $++++$ magnetic pattern, can therefore explain the experimental SHG data without the need to invoke any exotic, higher-order, multipolar magnetic order or orbital currents.

However, the reason why the $-+-+$ or $++++$ magnetic patterns replaces the $-++-$ pattern remains to be found. Here we advance some hypotheses and propose new experiments to check them.  We begin with the nature of the SHG process in Sr$_2$IrO$_4$.

\begin{figure}
\vspace{-3cm}
\includegraphics[width=0.6\textwidth]{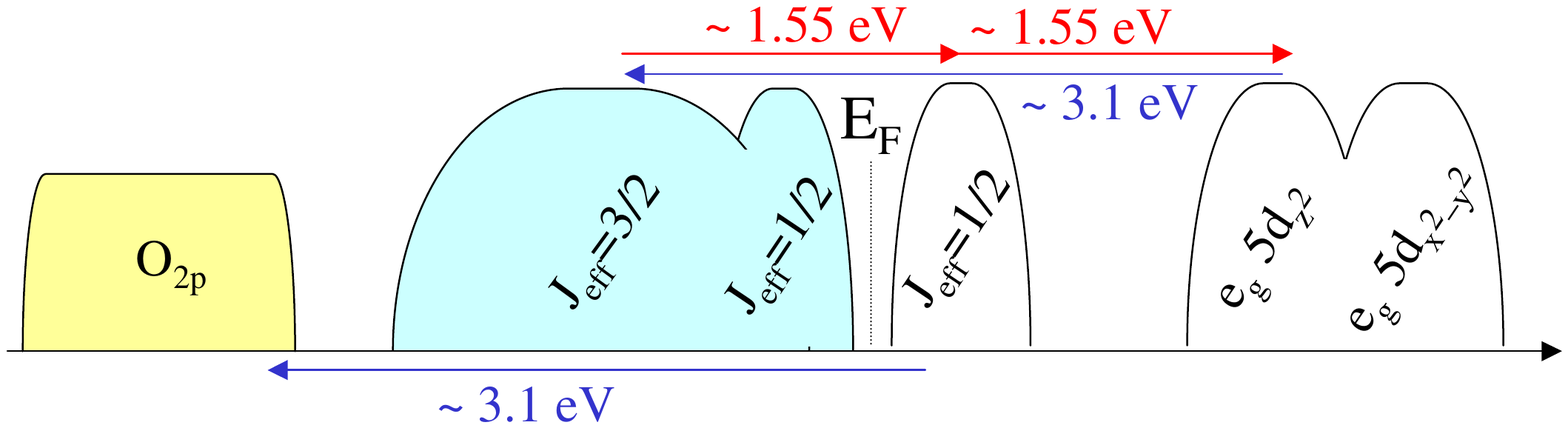}
\vspace{-0.5cm}
\caption {(Color online) Sketch of the density of states of Sr$_2$IrO$_4$, with the possible transitions leading to the SHG signal: a doubly resonant one, involving $J_{\rm eff}=3/2$ electrons, and a non-resonant one, involving O $2p$ electrons.}
\label{fig5}
\end{figure}

Zhao {\it et al.}~\cite{shg} propose a doubly non-resonant virtual transition for the incoming 1.55 eV photon ($\lambda = 800$ nm) from the O $2p$ band since it is more than 3 eV below the Fermi energy (Fig.~\ref{fig5}). We propose instead that the incoming 1.55 photons undergo a doubly resonant transition from the filled $J_{\rm eff}=3/2$ band to the empty partner of the $J_{\rm eff}=1/2$ doublet and then to the lower part of the $e_g$ band (Fig.~\ref{fig5}). The existence of the first resonance is demonstrated by a previous pump-probe experiment \cite{tsieh} and the position of the second one can be inferred from O K-edge measurements in x-ray absorption \cite{moon} and x-ray inelastic scattering \cite{moretti2}.
This level scheme has been advocated by a recent optics measurement as well \cite{sohn}.
The intensity of the doubly resonant path will be enhanced by a factor $\sim 500$ (for a typical damping $\Gamma \sim 0.1$ eV), due to the resonant denominator of Eq.~(\ref{scatSHG2}).
Whereas in reflection geometry, as in Ref.~\onlinecite{shg}, both processes generate an SHG signal, measurements in a transmission geometry would unambiguously identify one or the other \cite{bloem}: the strong damping due to real absorption of the doubly-resonant mechanism would deplete it, contrary to the the non-resonant process involving O $2p$ states.

We also speculate that the laser pump itself might modify the SHG signal.  After all, only a 0.3 T field or a few \% Rh doping are
needed to stabilize the $++++$ state (or possibly the $-+-+$ state for Rh doping).  In support of our conjecture, 
it is known that laser-induced non-thermal changes in the magnetism can occur that follows the time profile of a short laser pulse (48 fs) as demonstrated in Ref.~\onlinecite{bigot},
where the effect was attributed to both the coupling of the electric and magnetic fields of the laser to the spins,
and the alteration of the electric field of the ions due to the photodoped carriers \cite{fluence}.
In FeBO$_3$, this leads to a change in the anisotropy of the probe polarization that has C$_1$ symmetry \cite{kalashnikova}.
The laser can also generate changes in the symmetry of the lattice on this time scale, again due to the photodoped carriers,
as demonstrated recently for Cr$_2$O$_3$ \cite{sala}.
There, the symmetry corresponded to an even parity mode, but coupling of excitations to an odd parity mode was recently suggested in Sr$_2$IrO$_4$ \cite{sohn}
for an energy corresponding to the pump energy of Zhao {\it et al.}~\cite{shg}, which would again lead to a C$_1$ distortion of the SHG signal.
Moreover, a non-thermal transition from an antiferromagnetic state to a ferromagnetic state was generated on the time scale of the laser pulse in
a manganite \cite{li}, though this required a critical fluence and also an external field to align the ferromagnetic moments.
In this context, Dean {\it et al.}~\cite{dean} monitored the (-3,-2,28) magnetic Bragg peak associated with the $-++-$ ground state and found that it was strongly suppressed with a laser fluence of 1 mJ/cm$^2$.  
They speculate that this was due to destruction of the inter-plane magnetic correlations due to their weak nature, though thermal-induced demagnetization is the
most likely cause due to laser heating.  Regardless, if they are right, even if the resulting 3D pattern were
random in nature, we would expect an SHG signal to be present from the symmetry reduction (noting that even for the ordered ground states, only $-++-$ has a null SHG signal).  
Such an excited state might also be less sensitive to Rh doping, which would then explain why $T_{\Omega}$ is less sensitive to Rh doping than T$_N$.
But in order to obtain the measured azimuthal scan (Fig.~\ref{fig4}), a given laser pulse should not induce a random orientation of the magnetic domains below T$_{\Omega}$  since the measurement averages over many pulses.  This is doubtful, since Zhao {\it et al.}~\cite{shg} found the domains to change on each successive cool down.
Moreover, if the pump polarization oriented the domains, then four domains would not have been seen as they did.
And any effect would have to occur on the time scale of the emitted SHG photons.
Still, these speculations could be tested by future experiments: for instance, 
monitoring in parallel magnetic Bragg peaks associated with $++++$ or $-+-+$,
measuring the SHG signal as a function of laser fluence, or in the presence of a magnetic field, or at different photon energies
(some of the effects mentioned above are very sensitive to the photon energy), or performing actual pump-probe SHG experiments.

Because of potential issues with a laser-induced interpretation, one might wonder if there is an alternate explanation.  There is.
Magnetic SHG can also originate 
from the surface, as identified in a number of past studies \cite{kirilyuk}.  The magnetic point group of the surface of Sr$_2$IrO$_4$ is 2$^\prime$.  This allows for both the eee and eem SHG signals discussed above.  Although it might seem unusual to have interference between a bulk eeq signal and a magnetic surface eee/eem signal, this has been
observed in Fe/Au superlattices \cite{sato}, with the origin of the interference attributed to the complex nature of the energy denominator as we discussed above.
This seems to us a realistic possibility, and could be tested experimentally. The SHG energy of Zhao {\it et al.}~\cite{shg} corresponds to a maximum in the
conductivity \cite{sohn}.  This implies that many of the SHG photons are absorbed, meaning the resulting signal is dominated by the surface.  On the other hand,
there is a pronounced minimum at 2 eV \cite{sohn}, meaning 
that experiments with a pump energy of 1 eV (SHG energy of 2 eV) should be less surface sensitive and the interference reduced.

We now turn to an analysis of the ground state of Sr$_2$IrO$_4$, which we shall use in Section V. 

\section{Kramers doublet ground state in ${\rm Sr_2IrO_4}$}

The breakthrough idea in the original paper of B.~J.~Kim {\it et al.}~\cite{kim1} was the identification of the ground state of Sr$_2$IrO$_4$ as an octahedral Kramers doublet. As shown in the last section, a proper knowledge of the ground and the excited states of Sr$_2$IrO$_4$ are necessary, in order to explain the SHG experiment.
As a consequence, we focus here on the theoretical description of this Kramers doublet and on the experimental evidence for its existence. We believe, in fact, that in spite of a number of papers published on the subject \cite{kim2,lovesey,boseggia,moretti}, there are some experimental consequences of this state that have not (or not correctly) been stated.
As authors in the iridate literature have sometimes followed different conventions for the definition of spherical orbitals, we list our own definitions in Appendix B. 

One of the key experimental evidences leading to the octahedral Kramers doublet was the absence of magnetic RXS intensity at the Ir L$_2$ edge for reflections that showed a big resonant intensity at the L$_3$ edge \cite{kim2,boseggia}. This feature has been confirmed in several other non-stoichiometric compounds \cite{boseggia,calderMn,calderRu}. Initially interpreted as definitive evidence for this Kramers doublet \cite{kim2}, the absence of a magnetic signal at the L$_2$ edge was considered as inconclusive, because an in-plane magnetic moment could lead to the same conclusion \cite{lovesey} even if the octahedral limit was not realized. However, the fact that for some doped samples the magnetic moments point out of the $xy$-plane and the magnetic RXS intensity at the Ir L$_2$ edge was still strongly depleted \cite{calderMn,calderRu} represent yet another hint towards the physical realization of an octahedral Kramers doublet. However, as this last mentioned experimental evidence referred to Ru-doped \cite{calderRu} and Mn-doped \cite{calderMn} samples, it is still an open question to find definitive experimental evidence of the octahedral Kramers doublet in Sr$_2$IrO$_4$.

\begin{figure*}[ht!]
\vspace{-6cm}
\includegraphics[width=1.0\textwidth]{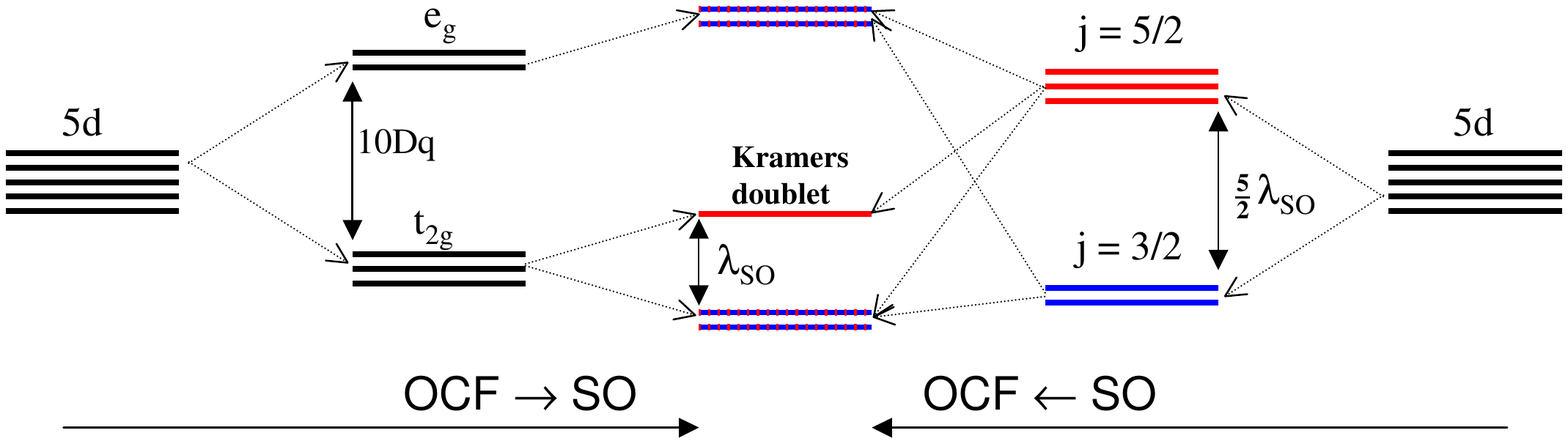}
\vspace{-2cm}
\caption {(Color online) A schematic representation of the Kramers doublet formation. When $jj$ coupling is considered (first spin-orbit (SO) and then octahedral crystal field (OCF), right-to-center path) it appears that the Kramers doublet is the only purely $j=5/2$ state (in red). All other states are a mixing of $j=3/2$ and $j=5/2$ (blue and red lines).}
\label{fig6}
\end{figure*}

We want to show here, by revisiting some of the calculations, that direct experimental evidence for the realization of this Kramers doublet in Sr$_2$IrO$_4$ is possible, looking at the Ir L$_2$ edge by X-ray absorption in partial yield in such a way as to reduce the core-hole lifetime to a value below 2 eV. The reason is due to the fact that two low-lying empty states are present in the spectrum: the empty partner of the Kramers doublet, within 1 eV of the Fermi level, and the $e_g$ states (around $\sim 3$ eV above the Fermi level). 
The two cannot be identified in usual XAS measurements, because of the L$_2$ core-hole width of around 5 eV. 
If an octahedral Kramers doublet existed, this would be a pure $j=5/2$ state (see Fig.~\ref{fig6} and calculations below) and no dipole transition at the L$_2$ edge would be allowed. Therefore in this case, the first, low lying peak would disappear and only the $e_g$ peak at higher energies would be present, as $e_g$ states are a mixture of both $j=5/2$ and $j=3/2$ states (Appendix B). Indeed, a hint towards the presence of the two peaks was highlighted by Boseggia {\it et al.}~\cite{boseggia}, who noticed two bumps in their magnetic spectra at the L$_2$ edge. We think that at least the higher energy peak corresponds to the $e_g$ states that, though just slightly magnetically polarized, can contribute to the magnetic intensity.

In what follows, we explain the details of our calculations. Note that as our main objective is to write down the L$_{2,3}$-edge cross-section, we shall not work with the effective angular momentum for the t$_{2g}$ states often employed in the literature, but with the real ones. We feel that this representation is more transparent if we analyze core-level transitions, as the expression of the Kramers doublet in terms of the $|j=\frac{5}{2},j_z\rangle$ states directly informs us about whether this transition is dipole-forbidden or not at the L$_2$ edge (we remind that dipole transitions are characterized by $\Delta j = \pm 1$ and therefore the $||j=1/2\rangle$ core states of the L$_2$ edge cannot be promoted to $|j=5/2\rangle$ states above the Fermi level). This would not be the case with $|J_{\rm eff} = 1/2 \rangle$. In Appendix B, we give the formulas to pass from one representation to the other, in order to compare with the existing literature. 

If we solve the crystal-field plus spin-orbit Hamiltonian at an Ir site, as done several times in the literature \cite{jackeli,perkins,abragam}, and as reported in detail in Appendix B, but then express the solution in terms of the spin-orbit coupled $j=5/2$ and $j=3/2$ states that can be associated with the $5d$ electrons of Ir, we obtain:
\begin{align} \label{kramers2a}
|\psi_{+}\rangle & = \frac{1}{\sqrt{5N}} \left[(2+\frac{R}{\sqrt{2}})|\frac{5}{2},\frac{3}{2}\rangle \right.  \\
			& \left. + (\sqrt{2}R-1)|\frac{3}{2},\frac{3}{2}\rangle- R\sqrt{\frac{5}{2}}|\frac{5}{2},-\frac{5}{2}\rangle\right] \nonumber
\end{align}
and
\begin{align} \label{kramers2b}
|\psi_{-}\rangle & = \frac{1}{\sqrt{5N}} \left[(2+\frac{R}{\sqrt{2}})|\frac{5}{2},-\frac{3}{2}\rangle \right.  \\
& \left. - (\sqrt{2}R-1)|\frac{3}{2},-\frac{3}{2}\rangle- R\sqrt{\frac{5}{2}}|\frac{5}{2},\frac{5}{2}\rangle\right] \nonumber
\end{align}
where the coefficients $R(\eta)$ and $N(\eta)$
depend solely on the ratio $\eta=\Delta_t/\lambda$ between the tetragonal crystal field, $\Delta_t$, and the spin-orbit coupling, $\lambda$:

$R(\eta)=-\frac{1}{\sqrt{2}}\left(1-\frac{1}{2}(1+2\eta+\sqrt{9-4\eta+4\eta^2}) \right)$

$N(\eta)=1+ \frac{1}{2}\left(1-\frac{1}{2}(1+2\eta+\sqrt{9-4\eta+4\eta^2})\right)^2$

\noindent Notice that in the octahedral limit, $\eta=0$, $R(\eta=0)=\frac{1}{\sqrt{2}}$.
From this expression, it is clear that if the octahedral limit is realized, then the coefficient of the $|\frac{3}{2},\pm\frac{3}{2}\rangle$ terms becomes zero and no signal at the L$_2$ edge can be detected. 

What should be underlined here is that the absence of intensity at the L$_2$ edge is {\it not} limited to the magnetic signal: {\it all} the signal at the L$_2$ edge associated with the empty part of the Kramers doublet would be zero if the octahedral limit is satisfied, even the non-magnetic absorption. In fact, in this case $|\psi_{\pm}\rangle$ would be a pure $|j=5/2\rangle$ state (see Fig. \ref{fig6}) and no dipole transition can occur between $2p_{1/2}$ and $5d_{5/2}$ ($\Delta j = 2$ is dipole-forbidden).

If we rewrite Eqs.~(\ref{kramers2a}) and (\ref{kramers2b}) using the cartesian representation, we obtain the form usually given in the literature for the half-filled Kramers doublet \cite{nota7}:

\begin{align}
|\psi_{+}\rangle & = \frac{1}{\sqrt{N}} \left(iR d_{xy\downarrow}-\frac{1}{\sqrt{2}}d_{xz\uparrow}-\frac{i}{\sqrt{2}}d_{yz\uparrow}\right)   \nonumber\\
|\psi_{-}\rangle & =\frac{1}{\sqrt{N}} \left(-iR d_{xy\uparrow}+\frac{1}{\sqrt{2}}d_{xz\downarrow}-\frac{i}{\sqrt{2}}d_{yz\downarrow}\right)
\label{kramers4}
\end{align}

This expression, with $R(\eta=0)=\frac{1}{\sqrt{2}}$, gives the usually quoted doublet with equal weights of t$_{2g}$ states. 
The expression above for the doublet corresponds to the case where the magnetic moment is along the $c$-axis, as $|\psi_{+}\rangle$ and $|\psi_{-}\rangle$ are eigenstates of both $L_z$ and $S_z$,  of eigenvalue $\pm 2/3$ and $\pm 1/6$, respectively, when $R=1/\sqrt{2}$. If we want it in {\it any} direction, we should make a linear combination of the two as follows:
\begin{align}
|\psi_{\rm any}\rangle & = \frac{1}{\sqrt{N}} \left[ \cos(\beta)\left(iR d_{xy\downarrow}-\frac{1}{\sqrt{2}}d_{xz\uparrow}-\frac{i}{\sqrt{2}}d_{yz\uparrow}\right) \right.   \nonumber\\
& \left. +\sin(\beta) e^{-i\gamma} \left(iR d_{xy\uparrow}-\frac{1}{\sqrt{2}}d_{xz\downarrow}+\frac{i}{\sqrt{2}}d_{yz\downarrow}\right)\right]
\label{psireal}
\end{align}

In this expression, $\beta=0$ and $\beta=\pi/2$ corresponds to the magnetic moment oriented along $\pm c$, whereas the magnetic moment in the $ab$-plane, as detailed in Appendix B.3, is obtained when $\cos(\beta)=\sin(\beta)$. In this case, $\gamma$ corresponds to the angle with respect to the local $x$-axis in the direction of a planar oxygen (e.g., for Ir$_1$ in Fig.~\ref{fig1}, $\beta=\gamma=\pi/4$). 

The experimental configuration, where the magnetic moment lies in the $ab$-plane, 45$^\circ$ from the local octahedron axes \cite{nota4}, leads to the following expressions for the Kramers doublets: 

\begin{align} \label{psirealpm}
|\psi_{\rm real}^+\rangle & = \frac{1}{\sqrt{2N}} \left[ \left(iR d_{xy\downarrow}-\frac{1}{\sqrt{2}}d_{xz\uparrow}-\frac{i}{\sqrt{2}}d_{yz\uparrow}\right) \right.   \\
& \left. + \frac{1}{\sqrt{2}} (1-i) \left(iR d_{xy\uparrow}-\frac{1}{\sqrt{2}}d_{xz\downarrow}+\frac{i}{\sqrt{2}}d_{yz\downarrow}\right)\right] \nonumber \\
|\psi_{\rm real}^-\rangle & = \frac{1}{\sqrt{2N}} \left[ \left(iR d_{xy\downarrow}-\frac{1}{\sqrt{2}}d_{xz\uparrow}-\frac{i}{\sqrt{2}}d_{yz\uparrow}\right) \right.   \nonumber\\
& \left. - \frac{1}{\sqrt{2}} (1-i) \left(iR d_{xy\uparrow}-\frac{1}{\sqrt{2}}d_{xz\downarrow}+\frac{i}{\sqrt{2}}d_{yz\downarrow}\right)\right] \nonumber
\end{align}

The same expressions in the $|j,j_z\rangle$ basis are reported in Appendix B.1.

Notice that the expression given in Ref.~\onlinecite{moretti} for the moment along $a$ does not appear to be correct. In their expression, that in the octahedral limit is $\frac{1}{\sqrt{3}} \left((d_{xy\uparrow}-id_{xy\downarrow})/\sqrt{2}+id_{xz\downarrow}+d_{yz\uparrow}\right)$, only $d_{yz\uparrow}$ and $d_{xz\downarrow}$ appear, violating the phase relation of the $J_{\rm eff}=1/2$ subspace. This state therefore brings in some admixture with components from the $J_{\rm eff}=3/2$ subspace \cite{nota3}. 

We can now evaluate the matrix elements that appear in both magnetic RXS and XAS. The details are shown in Appendix B.2. We draw here the conclusions of such calculations, Eqs.~(\ref{L2edge}) and (\ref{L3edge}). Where the L$_2$ edge is concerned, Eq.~(\ref{L2edge}) clearly shows that in the octahedral limit, $R(\eta=0)=\frac{1}{\sqrt{2}}$ and {\it all} matrix elements are zero. This is to be expected: in this limit, the Kramers doublet becomes a purely $|j=5/2\rangle$ state and no transition is possible for the L$_2$ edge. Eq.~(\ref{L2edge}) tells us also that when the magnetic moment is in the $ab$-plane, whatever its orientation, $\cos(2\beta)=0$ and off-diagonal elements are zero, independent of $R$. This implies, in particular, that the magnetic RXS signal is zero. We remind that this conclusion is valid because magnetic RXS is proportional to the antisymmetric part of the $L^{(2)}_{\alpha\beta}$ tensor (Eq.~(\ref{L2edge})) and that the latter is an irreducible tensor of rank one: if it is zero in one frame, it will be zero in any other rotated frame. The opposite also is true, valid at the L$_3$ edge, Eq.~(\ref{L3edge}): if at least one component is non-zero in a given frame, there will be at least one non-zero component in any other rotated frame. This is sufficient to affirm that there is always a magnetic RXS signal at the L$_3$ edge, whatever $R$ and $\beta$ are (this rule, of course, does not consider eventual extinctions due to the structure factor). At the L$_3$ edge, moreover, Eq.~(\ref{L3edge}) shows us that the absorption coefficient for the empty part of the Kramers doublet is always different from zero for any incoming polarization ($2R^2+\sqrt{2}R+1>0$, for any $R$) and for any direction of the magnetic moment. We finally notice that, contrary to what was stated in Ref.~\onlinecite{moretti}, the L$_2$ edge magnetic RXS in the $\pi-\pi$ channel is zero whatever $R$ is if the magnetic moment is confined within the $ab$-plane, as Eq.~(\ref{L2edge}) implies that no magnetic signal exists in this case, in any frame (any incoming and outgoing polarizations).

\section{Key experiments for ${\rm Sr_2IrO_4}$}

This Section is focused on the description of some key experiments with the double aim to find the fingerprint of a) the octahedral Kramers doublet in the stoichiometric material and b) the magnetic space groups, $++++$ and $-+-+$, of interest for the SHG experiment \cite{shg}.
As explained in the previous Section, one key experiment to confirm whether the $J_{\rm eff}=1/2$ doublet is realized in Sr$_2$IrO$_4$ is to compare XAS experiments at the L$_2$ and L$_3$ edges of Ir, using the high-resolution capabilities of partial fluorescence detection to reduce the value of the core-hole width at these edges. Typical values of the core-hole width for Ir at the L$_2$ and L$_3$ edges are 5.69 eV and 5.25 eV, respectively \cite{tables}. We remark in this respect that, as depicted in Fig.~\ref{fig6}, $e_g$ orbitals have both $j=3/2$ and $j=5/2$ character, as well as the $J_{\rm eff}=3/2$ submanifold of $t_{2g}$ \cite{notakim}. This is evident by a simple inspection of the transformation formulas given in Appendix B. In the octahedral limit, the only purely $j=5/2$ state is the Kramers doublet $J_{\rm eff}=1/2$. Therefore, dipole transitions can reach the empty $e_g$ states but not the empty partner of the Kramers doublet $J_{\rm eff}=1/2$. This is not the case for the L$_3$ absorption transitions that can reach both $e_g$ states and the empty partner of the Kramers doublet. A comparison of the two high-resolution XAS spectra could therefore provide a definitive confirmation of this issue for the stoichiometric compound.

\begin{figure}
\includegraphics[width=0.5\textwidth]{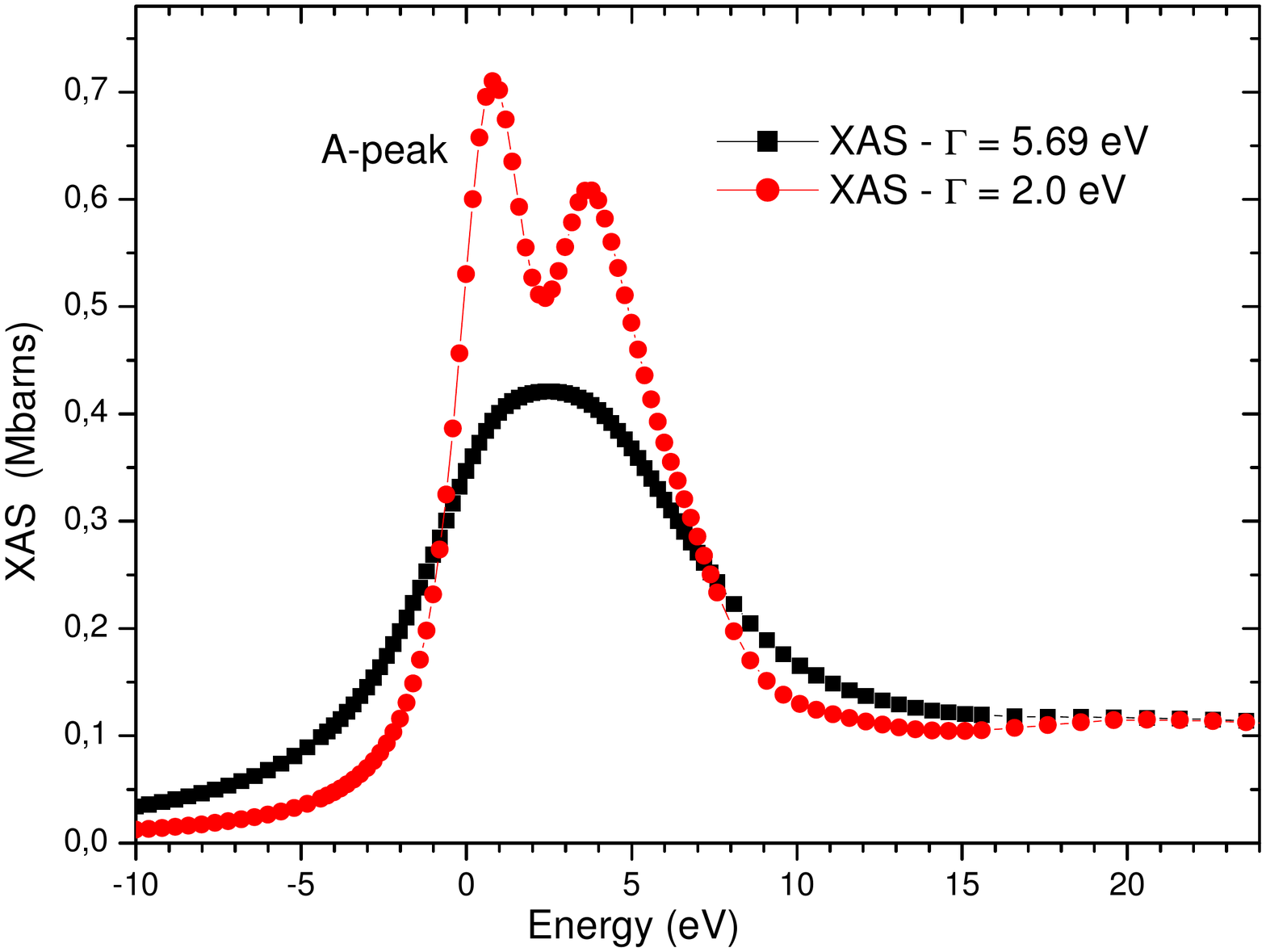}
\caption {(Color online) Ir L$_2$ edge XAS for two different core-hole widths. The core-hole width of total fluorescence yield, $\Gamma = 5.69$ eV, does not allow one to separate the underlying peak structure. The A-peak should be absent in the octahedral limit. The $-++-$ state was assumed (the $-++-$ and $++++$ states give indistinguishable results at this scale). The Fermi level sets the zero of the energy scale.}
\label{fig7}
\end{figure}

In Fig.~\ref{fig7} we show the results of a numerical simulation of L$_2$-edge XAS by means of the FDMNES code \cite{joly}.
We remind that the FDMNES code is based on a spin-polarized multiple-scattering approach including spin-orbit akin to LSDA+SO (local spin density approximation plus spin-orbit) calculations \cite{input}.
We performed two XAS calculations, one with a core-hole width of 5.69 eV and the other with a core-hole width of 2 eV. We used a cluster radius of 6.5 \AA, containing 85 atoms \cite{finite}.
With $\Gamma=2.0$ eV, the doublet structure of what appeared to be a single peak for $\Gamma=5.69$ eV clearly emerges.
The lower-lying structure, labelled as the A-peak in Fig.~\ref{fig7}, should not be there if the octahedral Kramers doublet is realized (i.e., $R=1/\sqrt{2}$). Its absence is a fingerprint of the octahedral Kramers doublet.
Unfortunately, the simulation does not reproduce all of the features of the data: for example, the energy splitting of the $e_g$ $5d_{3z^2-r^2}$ and $5d_{x^2-y^2}$ states is underestimated (only $\sim 0.6$ eV, as compared to 1.6 eV experimentally \cite{moretti2}), though it would take very high resolution to see this splitting in XAS. 

We performed also the same calculation with a Hubbard $U=2$ eV, but it did not change qualitatively the results of Fig.~\ref{fig7}, except for a shift of 0.5 eV of the higher-energy shoulder to still higher energies (however, this difference would be difficult to observe experimentally).
As shown in the literature in the case of actinides, it is possible to select photons emitted from decay channels characterized by longer lifetimes, i.e.~a smaller core-hole width, with resolutions of $\sim$2.0 eV at L edges, or even down to 1.2 eV for the M$_2$ edge in uranium compounds \cite{kvashnina}. These resolutions would clearly allow for the detection of the presence or absence of the $J_{\rm eff}=1/2$ peak at the L$_2$ edge, thereby providing the final word on this issue \cite{nota8}.

\begin{figure}
\includegraphics[width=0.5\textwidth]{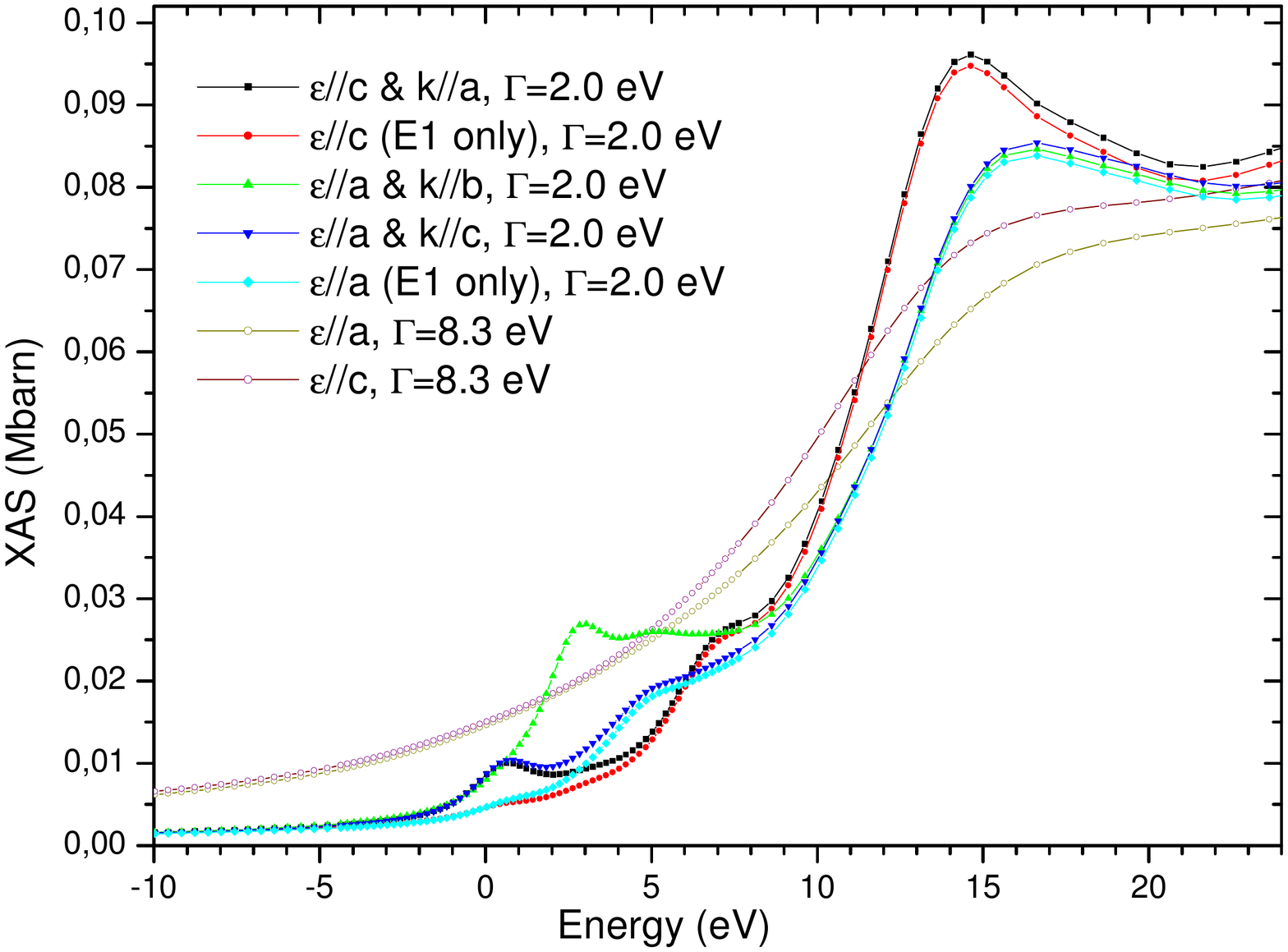}
\caption {(Color online) Iridium L$_1$ XAS, highlighting E2 transitions below 8 eV and the positions of the $t_{2g}$ and $e_g$ states (see the main text for explanations). The $-+-+$ state was assumed, and the Fermi level sets the zero of the energy scale.}
\label{fig8}
\end{figure}

A further experiment to double-check the behavior of the Kramers doublet is suggested by Eq.~(\ref{L2edge}). Here it is shown that when XAS is measured with incoming polarization along $z$, one should get a null L$_2$ signal for the Kramers doublet, regardless of whether one is in the octahedral limit or not.  That is, $z$ polarization only picks up the $e_g$ states. This would allow one to fix experimentally the energy level(s) of the $e_g$ states. If, starting from this configuration, we rotate the polarization towards, say, the $x$-direction, any signal that develops at lower energies would necessarily imply that we are filling in the unoccupied partner of the Kramers doublet and therefore we are not in the octahedral limit. The results of Moon {\it et al.}~\cite{moon} at the O K edge and our simulations by FDMNES at the Ir L$_1$ edge, shown in Fig.~\ref{fig8}, allow us to confirm that any signal of $x$ character developed below the lowest $z$ peak cannot be of $e_g$ origin (e.g., $5d_{x^2-y^2}$ orbitals), because the latter are higher in energy than the $5d_{3z^2-r^2}$ states. Here as well, the experiment would strongly rely on a high resolution to resolve the three peaks.

Further independent pieces of information on the nature of the low-lying energy states and their orbital distribution come from polarized XAS spectra at the Ir L$_1$ edge, where it is possible to play with both incoming polarization and wave-vector through electric-quadrupole (E2) transitions. The main results are shown in Fig.~\ref{fig8}. In particular, we show some results that would be measured by high-resolution XAS ($\Gamma \sim 2$ eV). The comparison with the curves characterized by $\Gamma=8.3$ eV (the typical core-hole width for L$_1$ XAS \cite{tables}) highlights the necessity for high-resolution XAS. For all curves with $\Gamma=2.0$ eV, the E2 signal is clearly visible in the region up to 8 eV above the Fermi energy (the zero of our energy scale). The E2 origin of the signal is demonstrated by the different behavior of the black and red curves in the region from 0 to 4 eV, and the behavior of the green and light-blue curves in the region from 0 to 8 eV. In particular, setting $\vec{\epsilon} \parallel c$ and $\vec{k} \parallel a$ (black curve) makes XAS blind to both the $5d_{x^2-y^2}$ and $5d_{3z^2-r^2}$ states and only sensitive to the $t_{2g}$ $5d_{xz}$ ones. This allows the identification of the lowest-lying peak, between 0 and 1 eV, as due to these $5d_{xz}$ states. We remark that the E2 nature of this peak is demonstrated by the difference with the purely E1 calculation (red curve). Experimentally, such a feature can be shown by rotating $\vec{k}$. When we rotate the polarization to the $a$ direction and the wave-vector to the $b$ direction (both in-plane), E2 transitions become allowed for both the $5d_{x^2-y^2}$ and $5d_{3z^2-r^2}$ states
(the latter, because of the $x^2+y^2$ part contained in the $r^2$ term). This is seen by the green curve in Fig.~\ref{fig8}. Finally, the dark-blue curve with $\vec{\epsilon} \parallel a$ and $\vec{k} \parallel c$ allows to double-check a) the E2 nature of the lowest-lying peak (because it is symmetric in the $a\leftrightarrow c$ exchange, as only E2 transitions are) and b) the partly E1 nature of the features around 4 eV (of course, the latter is evident also from the purely E1, light-blue curve). 

We remark also that a smooth extrapolation of the E1 peaks around 14 eV shows that the Ir $6p$ density of states is present down to the Fermi energy and, in correspondence with the E2 features, further $p$ density of states, probably due as well to the hybridization of the Ir $5d$ and O $2p$ states. The possibility that this inversion-odd density of states is magnetized, so as to be $TI$-invariant, as required in the 2$^\prime$/m group suggested by the SHG experiment, can be analyzed by non-reciprocal x-ray linear dichroism or magnetochiral (XM$\chi$D) dichroism \cite{gouloncr2o3,sdmcr2o3}. Such a technique would highlight the presence of a toroidal, magnetic signal of E1-E2 origin, therefore providing an independent confirmation of the presence of toroidal multipoles. By symmetry, this is only possible for the $-+-+$ state. 
Unfortunately, as the inversion breaking at the Ir sites is only determined by further oxygen neighbors (see Section II), E1-E2 radial matrix elements are extremely small, less than $10^{-4}$ of the E2-E2 radial matrix elements leading to the XAS pre-edge peak shown in Fig.~\ref{fig8}.
We have verified this by FDMNES, where we find an XM$\chi$D peak at the pre-edge for $k$ parallel to the $a$-axis that is of order 10$^{-4}$ of the L$_1$ XAS maximum (this requires a cluster radius of at least 7 \AA).
The same order of magnitude applies to the K edge. This implies that, were the toroidal nature of the SHG signal confirmed, SHG would be a much more sensitive tool to detect a tiny $\hat{T}\hat{I}$-breaking order parameter than dichroism in XAS.

In order to identify the actual magnetic state indicated by the SHG experiment, a more sensitive tool than XM$\chi$D is x-ray magnetic circular dichroism (XMCD), whose signal is zero for the $-+-+$ and $-++-$ states and non-zero for the $++++$ state.  An XMCD signal has been reported in Ref.~\onlinecite{haskel} in the presence of a magnetic field (therefore inducing the $++++$ state) with an intensity about $1\%$ of the XAS intensity \cite{xmcd}. 

As shown already in Section II in Fig.~\ref{fig2}, the two magnetic configurations $-+-+$ and $++++$ can be identified also by RXS experiments at the L$_3$ edge, with the aim of identifying whether the $-+-+$ state or the $++++$ state is realized in Sr$_2$IrO$_4$ in the case of Rh doping. Though challenging, it would be interesting to perform both the RXS and the XMCD experiments under the influence of a laser beam, so as to check whether the experimental conditions of the SHG experiment in
Ref.~\onlinecite{shg} could have possibly induced a phase transition from the $-++-$ state to the $++++$ or $-+-+$ states. We remind that a relatively small applied magnetic field ($H\simeq 0.3$ T) has been shown to induce the $++++$ state \cite{kim2}.
In fact, it would be very interesting to repeat the SHG experiments in the presence of a magnetic field that would induce the $++++$ state.

Finally, we have tried to highlight the difference between the $++++$ and $-+-+$ states also at the Ir L$_1$ edge, by interference of E1 and E2 signals at the pre-edge (experimentally, they could be disentangled with the technique of phase plates developed in Ref.~\onlinecite{wilkins}). Unfortunately, our FDMNES calculations find that charge scattering is the dominant term in this energy range, so no qualitative differences between the two states could be determined.

\section{Conclusions}

To conclude, we summarize here the main achievements of the present paper:

1) We have shown that the SHG experiment \cite{shg} can be explained by either the 2$^\prime$/m or the 2$^\prime$/m$^\prime$ magnetic space groups. The former was already identified in Ref.~\onlinecite{shg} where it was interpreted in terms of toroidal moments induced by orbital currents. However, an order parameter with the symmetry of a toroidal dipole is not sufficient to explain the azimuthal scan, as noted in Section III. The octupole is needed as well. 

2) We have demonstrated that it is not necessary to invoke exotic orbital currents to obtain an SHG signal: an induced transition to the $-+-+$ state or to the $++++$ state would have the same effect. The latter was explicitly excluded in Ref.~\onlinecite{shg}, probably because only $\chi^{(e)}$ was considered in that paper, not $\chi^{(m)}$. Two other magnetic space groups might explain the odd-$\psi$ azimuthal dependence of the interference SHG signal: m1$^\prime$ and $\overline{\rm 1}$1$^\prime$. The former was also identified in Ref.~\onlinecite{shg}. However, both of them are characterized by time-reversal even order parameters, and it appears implausible that they play a central role below the magnetic transition temperature. Finally, the $-+-+$ state has a magnetoelectric space group and could explain also the results of Ref.~\onlinecite{chikara}, that invokes the breaking of both spatial parity and time-reversal symmetry.
Either effect could also arise from surface magnetic SHG, and we have suggested that the photon energy could be changed to test for surface sensitivity of the SHG signal.

3) We have suggested new experiments to highlight the interplay of the three states, $-++-$, $++++$, or $-+-+$, characterized by different magnetic space groups. Though neutron and x-ray diffraction data clearly show that $-++-$ is the ground state for stoichiometric Sr$_2$IrO$_4$, the three are very close in energy (the inter-plane exchange $\sim$ $\mu$eV \cite{fuji}). A laser pulse of 1mJ/cm$^2$ fluence is known to suppress the $-++-$ state at fs timescales \cite{dean}.
Whether this is what produces the additional C$_1$ distortion of the SHG signal below T$_{\Omega}$ remains to be determined by future experiments, several of which were outlined in Section III.

4) From the calculations of Appendix B, we have highlighted a new experimental criterion, based on high-energy resolution XAS, to identify the octahedral nature of the Kramers doublet in stoichiometric Sr$_2$IrO$_4$. Such an XAS measurement is independent of the direction of the Ir magnetic moments (in-plane or out-of-plane), as it only relies on the purely $j=5/2$ nature of the Kramers doublet in the octahedral limit. It can provide, therefore, an independent confirmation of the realization of the octahedral Kramers doublet in the stoichiometric compound.

\begin{acknowledgments}
The authors would like to thank Dr.~Liuyan Zhao and Prof.~David Hsieh for providing the data plotted in Fig.~\ref{fig4} and for several
discussions about this paper.  They also thank Dr.~Feng Ye for clarification concerning the neutron scattering results for Rh-doped samples,
as well as Mark Dean for discussions about their pump-probe experiments.
Finally, we would like to thank one of the Referees for suggesting the possibility of surface magnetic SHG.
Work by MRN was supported by the Materials Sciences and Engineering
Division, Basic Energy Sciences, Office of Science, US DOE.
\end{acknowledgments}

\appendix
\section{Technical details on the SHG calculations}

\subsection{The total SHG amplitude} 

In the most general case, the transition probability per unit time from a state $\Phi_g$ to a state $\Phi_f$ can be written in term of the transition operator $T_I$ as follows: 

\begin{align}
W_{gf} = \frac{2\pi}{\hbar} \rho_f \big| \langle \Phi_f| T_I | \Phi_g \rangle \big| ^2  
\label{transop}
\end{align}

\noindent where $\rho_f$ is the density of the final states and $T_I=H_I + H_I G_{H_T} H_I$, with $H_I$ the matter-radiation interaction Hamiltonian and $G_{H_T}=(\Sigma_g - H_T +i\Gamma)^{-1}$ the Green's function related to the total Hamiltonian $H_T = H_0 + H_I$. $\Sigma_g$ is the total energy associated with $\Phi_g$. Here $H_0$ is the sum of the matter Hamiltonian and the radiation Hamiltonian, separately. With the usual Dyson expansion $G_{H_T} = G_{H_0} + G_{H_0} H_I G_{H_0}+ ...$, we can replace in the above expression for $T_I$ and rewrite Eq.~(\ref{transop}) up to any order in $H_I$. The scattering cross-section is obtained from the transition probability per unit time by dividing by the incoming flux
($c/V$ for photons, when the vector potential is normalized to one photon per unit volume, $V$ \cite{sakurai}). We can then pick out the third order for the scattering cross-section, of interest for SHG, $\sigma_{SHG} = \frac{2\pi V}{\hbar c} \rho_f \big| A_{SHG}\big| ^2$, where the amplitude reads (with $\Phi_f=\Phi_g$):  

\begin{align}
A_{SHG} = \sum_{l,n} \frac{\langle \Phi_g| H_I | \Phi_l \rangle \langle \Phi_l|  H_I| \Phi_n \rangle \langle \Phi_n|  H_I | \Phi_g \rangle}{(\Sigma_g-\Sigma_n)(\Sigma_g-\Sigma_l)}  
\label{scatSHG}
\end{align}

We can now separate in $H_I$ the radiation part from the matter part, as $\Sigma_i = E_i + m\hbar\omega$. Here $E_i$ is the energy of the material alone (without radiation) for state $i$ and $\hbar\omega$ measures the photon energy associated with a given matrix element. In particular, for SHG, $m=1$ in absorption and $m=2$ in emission, because of the absorption of two photons of energy $\hbar\omega$ and the emission of one photon of energy $2\hbar\omega$. Three different terms are possible, as represented in Fig.~\ref{fig3}, and they correspond to the three processes: 1) absorption ($\hbar\omega$) - absorption ($\hbar\omega$) - emission ($2\hbar\omega$): this is a doubly resonant process; 2) absorption ($\hbar\omega$) - emission ($2\hbar\omega$) - absorption ($\hbar\omega$): this is a singly resonant process; 3)  emission ($2\hbar\omega$) - absorption ($\hbar\omega$)  - absorption ($\hbar\omega$): this is a non-resonant process. Three different energy denominators are associated with each term, as shown below. For the interaction Hamiltonian, we suppose that we can perform a multipole expansion of the vector potential contained in $H_I$ \cite{jphysD} and consider electric-dipole (E1), magnetic-dipole (M1) and electric-quadrupole (E2) terms, only. With this hypothesis, Eq.~(\ref{scatSHG}) can be written as (a sum over repeated variables $\alpha,\beta,\gamma=x,y,z$ is employed): 

\begin{align}
& A_{SHG} = \sum_{l,n} \big(\Delta_{l,n}^{(1)}\big(\eta^{(e,1)}_{\alpha\beta\gamma;ln}O^{(e,1)}_{\alpha\beta\gamma} \nonumber \\
& +\sum_{i=1}^3\eta^{(m,1,i)}_{\alpha\beta\gamma;ln}O^{(m,1,i)}_{\alpha\beta\gamma} +\sum_{i=1}^3\eta^{(q,1,i)}_{\alpha\beta\gamma\delta;ln}O^{(q,1,i)}_{\alpha\beta\gamma\delta}\big) \nonumber \\
& +\Delta_{l,n}^{(2)}\big(\eta^{(e,2)}_{\alpha\beta\gamma;ln}O^{(e,2)}_{\alpha\beta\gamma} +\sum_{i=1}^3\eta^{(m,2,i)}_{\alpha\beta\gamma;ln}O^{(m,2,i)}_{\alpha\beta\gamma} \nonumber \\
& +\sum_{i=1}^3\eta^{(q,2,i)}_{\alpha\beta\gamma\delta;ln}O^{(q,2,i)}_{\alpha\beta\gamma\delta}\big) +\Delta_{l,n}^{(3)}\big(\eta^{(e,3)}_{\alpha\beta\gamma;ln}O^{(e,3)}_{\alpha\beta\gamma} \nonumber \\
& +\sum_{i=1}^3\eta^{(m,3,i)}_{\alpha\beta\gamma;ln}O^{(m,3,i)}_{\alpha\beta\gamma}+ \sum_{i=1}^3\eta^{(q,3,i)}_{\alpha\beta\gamma\delta;ln}O^{(q,3,i)}_{\alpha\beta\gamma\delta}\big)\big) \nonumber \\
& \equiv \chi^{(e,1)}_{\alpha\beta\gamma}O^{(e,1)}_{\alpha\beta\gamma} +\sum_{i=1}^3\big(\chi^{(m,1,i)}_{\alpha\beta\gamma}O^{(m,1,i)}_{\alpha\beta\gamma} +\chi^{(q,1,i)}_{\alpha\beta\gamma\delta}O^{(q,1,i)}_{\alpha\beta\gamma\delta}\big) \nonumber \\
& +\chi^{(e,2)}_{\alpha\beta\gamma}O^{(e,2)}_{\alpha\beta\gamma} +\sum_{i=1}^3\big(\chi^{(m,2,i)}_{\alpha\beta\gamma}O^{(m,2,i)}_{\alpha\beta\gamma} 
+\chi^{(q,2,i)}_{\alpha\beta\gamma\delta}O^{(q,2,i)}_{\alpha\beta\gamma\delta}\big) \nonumber \\
& +\chi^{(e,3)}_{\alpha\beta\gamma}O^{(e,3)}_{\alpha\beta\gamma} +\sum_{i=1}^3\big(\chi^{(m,3,i)}_{\alpha\beta\gamma}O^{(m,3,i)}_{\alpha\beta\gamma}+ \chi^{(q,3,i)}_{\alpha\beta\gamma\delta}O^{(q,3,i)}_{\alpha\beta\gamma\delta}\big)
\label{scatSHG2}
\end{align}

\noindent where the denominators are: $\Delta_{l,n}^{(1)}=((E_{lg}-2\hbar\omega-i\Gamma_l)(E_{ng}-\hbar\omega-i\Gamma_n))^{-1}$, $\Delta_{l,n}^{(2)}=((E_{lg}+\hbar\omega)(E_{ng}-\hbar\omega-i\Gamma_n))^{-1}$, and $\Delta_{l,n}^{(3)}=((E_{lg}+\hbar\omega)(E_{ng}+2\hbar\omega))^{-1}$. 

Cartesian tensors $\chi^{(e,m,q)}_{\alpha\beta\gamma}$ represent tensor properties of the matter and $O^{(e,m,q)}_{\alpha\beta\gamma}$ those of incoming and outgoing polarizations and wave-vectors. We remark that the quantities $\chi^{(e,i)}_{\alpha\beta\gamma}\equiv \sum_{l,n} \Delta_{l,n}^{(i)}\eta^{(e,i)}_{\alpha\beta\gamma;ln}$ correspond to the SHG susceptibilities usually defined in the literature (e.g., Ref.~\onlinecite{fiebig}) from the macroscopic relation: $\vec{P}^{(2\omega)} = \chi : \vec{E}^{(\omega)} \vec{E}^{(\omega)}$, and analogously for $\chi^{(m,i,j)}_{\alpha\beta\gamma}$ and $\chi^{(q,i,j)}_{\alpha\beta\gamma}$. The only difference is that in the quantum-mechanical approach, all the processes depicted in Fig.~\ref{fig3} for $\chi^{(e,i)}_{\alpha\beta\gamma}$ and $\chi^{(m,i,j)}_{\alpha\beta\gamma}$ must be considered. In Eq.~(\ref{scatSHG2}) we have chosen to introduce the tensors $\eta^{(e,m,q)}_{\alpha\beta\gamma;ln}$ (defined below), factorizing denominators explicitly because, as detailed in Sections III and A.2, this factorization allows a clearer identification of the irreversible part, related to the denominators, and of the transition matrix elements, related to the order parameters, where time-reversal can be investigated.   
The three denominators, $\Delta_{l,n}^{(1)}$, $\Delta_{l,n}^{(2)}$ and $\Delta_{l,n}^{(3)}$ weight the three quantum-mechanical transition amplitudes leading to the SHG signal (two resonant processes, only one resonant process, no resonant processes), as shown in Fig.~\ref{fig3}.

For simplicity, we call $\chi^{(e)}$ the tensors with three electric-dipole transitions (E1-E1-E1), often called in the optics community $\chi^{(eee)}$, $\chi^{(m)}$ the tensors with two electric-dipole and one magnetic-dipole transition (E1-E1-M1), often called $\chi^{(eem)}$, and $\chi^{(q)}$ the tensors with two electric-dipole and one electric-quadrupole transition (E1-E1-E2), often called $\chi^{(eeq)}$. For symmetry reasons, the M1 and E2 transitions can appear at any of the three positions and this is the origin of the extra index $i$ for $\chi^{(m)}$ and $\chi^{(q)}$. It is pictorially shown in Fig.~\ref{fig3} for $\chi^{(m)}$. Using the notation $E_{ij} = E_i - E_j$, the explicit expressions of the $\eta^{(e,m,q)}_{\alpha\beta\gamma;ln}$ tensors are:

\begin{align}
\eta^{(e,1)}_{\alpha\beta\gamma;ln} = i \frac{e^3}{\hbar^3} E_{gl}E_{ln}E_{ng} \langle \phi_g| r_{\alpha} | \phi_l \rangle \nonumber \\
\label{chie}
\langle \phi_l| r_{\beta}| \phi_n \rangle \langle \phi_n|  r_{\gamma} | \phi_g \rangle \\
\eta^{(e,2)}_{\alpha\beta\gamma;ln} =\eta^{(e,3)}_{\alpha\beta\gamma;ln} =\eta^{(e,1)}_{\alpha\beta\gamma;ln}
\end{align}

\begin{align}
\eta^{(m,1,1)}_{\alpha\beta\gamma;ln} = \frac{i e^3}{2 m_e \hbar^2} E_{ln}E_{ng} \langle \phi_g| (L+2S)_{\alpha} | \phi_l \rangle \nonumber \\ 
\label{chim}
\langle \phi_l| r_{\beta}| \phi_n \rangle \langle \phi_n|  r_{\gamma} | \phi_g \rangle \\
\eta^{(m,1,2)}_{\alpha\beta\gamma;ln} = \frac{i e^3}{2 m_e \hbar^2} E_{gl}E_{ng} \langle \phi_g| r_{\alpha} | \phi_l \rangle \nonumber \\ 
\langle \phi_l| (L+2S)_{\beta}| \phi_n \rangle \langle \phi_n|  r_{\gamma} | \phi_g \rangle \\
\eta^{(m,1,3)}_{\alpha\beta\gamma;ln} = \frac{i e^3}{2 m_e \hbar^2} E_{gl}E_{ln} \langle \phi_g| r_{\alpha} | \phi_l \rangle \nonumber \\ 
\langle \phi_l| r_{\beta}| \phi_n \rangle \langle \phi_n|  (L+2S)_{\gamma} | \phi_g \rangle  \\
\eta^{(m,2,1)}_{\alpha\beta\gamma;ln} = \eta^{(m,1,1)}_{\alpha\beta\gamma;ln} \\
\eta^{(m,2,2)}_{\alpha\beta\gamma;ln} = \eta^{(m,1,2)}_{\alpha\beta\gamma;ln} \\
\eta^{(m,2,3)}_{\alpha\beta\gamma;ln} = \eta^{(m,1,3)}_{\alpha\beta\gamma;ln} \\
\eta^{(m,3,1)}_{\alpha\beta\gamma;ln} = \eta^{(m,1,1)}_{\alpha\beta\gamma;ln} \\
\eta^{(m,3,2)}_{\alpha\beta\gamma;ln} = \eta^{(m,1,2)}_{\alpha\beta\gamma;ln} \\
\eta^{(m,3,3)}_{\alpha\beta\gamma;ln} = \eta^{(m,1,3)}_{\alpha\beta\gamma;ln} 
\end{align}

\begin{align}
\eta^{(q,1,1)}_{\alpha\beta\gamma\delta;ln} = -\frac{e^3}{2 \hbar^3} E_{gl}E_{ln}E_{ng} \langle \phi_g| r_{\alpha} r_{\beta}| \phi_l \rangle \nonumber \\
\label{chiq}
\langle \phi_l| r_{\gamma} | \phi_n \rangle \langle \phi_n|  r_{\delta} | \phi_g \rangle \\
\eta^{(q,1,2)}_{\alpha\beta\gamma\delta;ln} = -\frac{e^3}{2 \hbar^3}E_{gl}E_{ln}E_{ng}\langle \phi_g| r_{\alpha} | \phi_l \rangle \nonumber \\
\langle \phi_l| r_{\beta} r_{\gamma} | \phi_n \rangle \langle \phi_n|  r_{\delta} | \phi_g \rangle \\
\eta^{(q,1,3)}_{\alpha\beta\gamma\delta;ln} = -\frac{e^3}{2 \hbar^3} E_{gl}E_{ln}E_{ng} \langle \phi_g| r_{\alpha} | \phi_l \rangle \nonumber \\ 
\langle \phi_l| r_{\beta}| \phi_n \rangle \langle \phi_n|  r_{\gamma}r_{\delta} | \phi_g \rangle  \\
\eta^{(q,2,1)}_{\alpha\beta\gamma\delta;ln} = \eta^{(q,3,1)}_{\alpha\beta\gamma\delta;ln} = \eta^{(q,1,1)}_{\alpha\beta\gamma\delta;ln} \\
\eta^{(q,2,2)}_{\alpha\beta\gamma\delta;ln} = \eta^{(q,3,2)}_{\alpha\beta\gamma\delta;ln} = \eta^{(q,1,2)}_{\alpha\beta\gamma\delta;ln} \\
\eta^{(q,2,3)}_{\alpha\beta\gamma\delta;ln} = \eta^{(q,3,3)}_{\alpha\beta\gamma\delta;ln} = \eta^{(q,1,3)}_{\alpha\beta\gamma\delta;ln}
\end{align}
Here $\phi_j$ are the eigenstates of the matter alone from $H_0$.
It should be noted that $\eta^{(e,1)}_{\alpha\beta\gamma} = \eta^{(e,2)}_{\alpha\beta\gamma} = \eta^{(e,3)}_{\alpha\beta\gamma}\equiv \eta^{(e)}_{\alpha\beta\gamma}$ (we remove the $_{;ln}$ label when not explicitly needed, though the dependence on intermediate states should not be forgotten). In fact, even though each term in the product $(E_g-E_l)(E_l-E_n)(E_n-E_g)$ can be different in the three cases (as states $g$, $l$ and $n$ are different), the combined product is always the same.
Notice also that $\eta^{(m)}$ and $\eta^{(q)}$ ($\chi^{(m)}$ and $\chi^{(q)}$) are dimensionally homogeneous, whereas $\eta^{(e)}$ ($\chi^{(e)}$) should not be directly compared to $\eta^{(m)}$ and $\eta^{(q)}$ ($\chi^{(m)}$ and $\chi^{(q)}$) as they are not dimensionally homogeneous ($\chi^{(m)}/\chi^{(e)}\sim$ length $\sim \lambda/2\pi$ as $\chi^{(e)}$ does not multiply the wave-vector, as do the former two, see Eq.~(\ref{polmate}) and below).

The polarization matrices $O^{(e,m,q;1,2,3)}$ are (here $\epsilon^i$, $\epsilon^o$, $k^i$ and $k^o$ are incoming and outgoing polarizations and wave-vectors):

\begin{align}
\label{polmate}
O^{(e,1)}_{\alpha\beta\gamma} & = & \epsilon^{o*}_{\alpha}\epsilon^i_{\beta}\epsilon^i_{\gamma} \\
O^{(e,2)}_{\alpha\beta\gamma} & = & \epsilon^i_{\alpha}\epsilon^{o*}_{\beta}\epsilon^i_{\gamma} \\
O^{(e,3)}_{\alpha\beta\gamma} & = & \epsilon^i_{\alpha}\epsilon^i_{\beta}\epsilon^{o*}_{\gamma} 
\end{align}

\begin{align}
O^{(m,1,1)}_{\alpha\beta\gamma} & = & (\epsilon^{o*}\times k^o)_{\alpha}\epsilon^i_{\beta}\epsilon^i_{\gamma} \\
O^{(m,2,1)}_{\alpha\beta\gamma} & = & (\epsilon^i\times k^i)_{\alpha}\epsilon^{o*}_{\beta}\epsilon^i_{\gamma} \\
O^{(m,3,1)}_{\alpha\beta\gamma} & = & (\epsilon^i\times k^i)_{\alpha}\epsilon^i_{\beta}\epsilon^{o*}_{\gamma} \\
O^{(m,1,2)}_{\alpha\beta\gamma} & = & \epsilon^{o*}_{\alpha}(\epsilon^i\times k^i)_{\beta}\epsilon^i_{\gamma} \\
O^{(m,2,2)}_{\alpha\beta\gamma} & = & \epsilon^i_{\alpha}(\epsilon^{o*}\times k^o)_{\beta}\epsilon^i_{\gamma} \\
O^{(m,3,2)}_{\alpha\beta\gamma} & = & \epsilon^i_{\alpha}(\epsilon^i\times k^i)_{\beta}\epsilon^{o*}_{\gamma} \\
O^{(m,1,3)}_{\alpha\beta\gamma} & = & \epsilon^{o*}_{\alpha}\epsilon^i_{\beta}(\epsilon^i\times k^i)_{\gamma} \\
O^{(m,2,3)}_{\alpha\beta\gamma} & = & \epsilon^i_{\alpha}\epsilon^{o*}_{\beta}(\epsilon^i\times k^i)_{\gamma} \\
O^{(m,3,3)}_{\alpha\beta\gamma} & = & \epsilon^i_{\alpha}\epsilon^i_{\beta}(\epsilon^{o*}\times k^o)_{\gamma} 
\label{polmatm}
\end{align}

\begin{align}
O^{(q,1,1)}_{\alpha\beta\gamma\delta} & = & \epsilon^{o*}_{\alpha}k^o_{\beta}\epsilon^i_{\gamma}\epsilon^i_{\delta} \\
O^{(q,2,1)}_{\alpha\beta\gamma\delta} & = & \epsilon^i_{\alpha}k^i_{\beta}\epsilon^{o*}_{\gamma}\epsilon^i_{\delta} \\
O^{(q,3,1)}_{\alpha\beta\gamma\delta} & = & \epsilon^i_{\alpha}k^i_{\beta}\epsilon^i_{\gamma}\epsilon^{o*}_{\delta} \\
O^{(q,1,2)}_{\alpha\beta\gamma\delta} & = & \epsilon^{o*}_{\alpha}\epsilon^i_{\beta}k^i_{\gamma}\epsilon^i_{\delta} \\
O^{(q,2,2)}_{\alpha\beta\gamma\delta} & = & \epsilon^i_{\alpha}\epsilon^{o*}_{\beta}k^o_{\gamma}\epsilon^i_{\delta} \\
O^{(q,3,2)}_{\alpha\beta\gamma\delta} & = & \epsilon^i_{\alpha}\epsilon^i_{\beta}k^i_{\gamma}\epsilon^{o*}_{\delta} \\
O^{(q,1,3)}_{\alpha\beta\gamma\delta} & = & \epsilon^{o*}_{\alpha}\epsilon^i_{\beta}\epsilon^i_{\gamma} k^i_{\delta} \\
O^{(q,2,3)}_{\alpha\beta\gamma\delta} & = & \epsilon^i_{\alpha}\epsilon^{o*}_{\beta}\epsilon^i_{\gamma} k^i_{\delta} \\
O^{(q,3,3)}_{\alpha\beta\gamma\delta} & = & \epsilon^i_{\alpha}\epsilon^i_{\beta}\epsilon^{o*}_{\gamma} k^o_{\delta}
\label{polmatq}
\end{align}

\subsection{Time-reversal and interference}

Consider first the case of the E1-E1-E1 transition amplitude, $A_{SHG}^{(e)}$, associated with $\chi^{(e)}$. In this case, we get:

\begin{align}
A_{SHG}^{(e)} & = \sum_{l,n} \big(\Delta_{l,n}^{(1)}\eta^{(e,1)}_{\alpha\beta\gamma}O^{(e,1)}_{\alpha\beta\gamma} \nonumber \\
& +\Delta_{l,n}^{(2)}\eta^{(e,2)}_{\alpha\beta\gamma}O^{(e,2)}_{\alpha\beta\gamma} + \Delta_{l,n}^{(3)}\eta^{(e,3)}_{\alpha\beta\gamma}O^{(e,3)}_{\alpha\beta\gamma}\big) \nonumber \\
& = \sum_{l,n} \eta^{(e)}_{\alpha\beta\gamma} \big( \Delta_{l,n}^{(1)}O^{(e,1)}_{\alpha\beta\gamma} +\Delta_{l,n}^{(2)}O^{(e,2)}_{\alpha\beta\gamma} \nonumber \\
& +\Delta_{l,n}^{(3)}O^{(e,3)}_{\alpha\beta\gamma}\big)  = \epsilon^o_{\alpha} \epsilon^i_{\beta} \epsilon^i_{\gamma} {\tilde \chi}^{(e)}_{\alpha\beta\gamma}
\label{ordpar_eee}
\end{align}

In the last equality of the above expression, we defined ${\tilde \chi}^{(e)}_{\alpha\beta\gamma} = \sum_{l,n} \big(\Delta_{l,n}^{(1)} \eta^{(e)}_{\alpha\beta\gamma;ln} + \Delta_{l,n}^{(2)} \eta^{(e)}_{\gamma\alpha\beta;ln} + \Delta_{l,n}^{(3)} \eta^{(e)}_{\beta\gamma\alpha;ln}\big) $. We also considered the specific conditions of the SHG experiment \cite{shg}, where all polarizations are real.
 
An important element of our analysis is the recognition that, analogously to the case of RXS \cite{jphysD}, each $\eta^{(e,m,q)}$ tensor is characterized by a time-reversal odd and a time-reversal even part, due to the matrix elements and independent of the complex energy denominators. They can be found by looking for the real and imaginary parts of each tensor:
$\eta^{(e)}= \Re \eta^{(e)} + i\Im \eta^{(e)}$, $\eta^{(m)}= \Re \eta^{(m)} + i\Im \eta^{(m)}$ and $\eta^{(q)}= \Re \eta^{(q)} + i\Im \eta^{(q)}$. Notice that we did not consider in our analysis the common imaginary unit multiplying the E1-E1-E1 transition matrix elements in Eq.~(\ref{chie}). Of course, we factorized it also in Eq.~(\ref{chim}) and Eq.~(\ref{chiq}), so that $\eta^{(q)}$ is always phase shifted by $\pi/2$ compared to $\eta^{(e)}$ and $\eta^{(m)}$. We remark that this analysis corresponds to Birss' separation into $i$-tensors and $c$-tensors \cite{birss}.
Starting from Eq.~(\ref{ordpar_eee}), we report the full expression only for $\eta^{(e)}$ (in order not to overburden the notation, in the following we remove the label $^{(e)}$, superfluous as we just deal with $\eta^{(e)}$): 

\begin{align}
& A_{SHG}^{(e)} = \epsilon^o_{\alpha} \epsilon^i_{\beta} \epsilon^i_{\gamma} {\tilde \chi}^{(e)}_{\alpha\beta\gamma} \nonumber \\
& = \epsilon^o_{\alpha} \epsilon^i_{\beta} \epsilon^i_{\gamma} \big[ \sum_{ln} \big( \tilde{\Delta}^{(1+)}_{lng} \Re \tilde{\eta}_{\alpha\beta\gamma}^{(glng)}  + \nonumber \\
&   \tilde{\Delta}^{(2+)}_{lng} \Re \tilde{\eta}_{\gamma\alpha\beta}^{(glng)}  +\tilde{\Delta}^{(3+)}_{lng} \Re \tilde{\eta}_{\beta\gamma\alpha}^{(glng)} \big) \big] \nonumber \\
& +i \epsilon^o_{\alpha} \epsilon^i_{\beta} \epsilon^i_{\gamma} \big[ \sum_{ln} \big( \tilde{\Delta}^{(1-)}_{lng} \Im \tilde{\eta}_{\alpha\beta\gamma}^{(glng)}  + \nonumber \\
&   \tilde{\Delta}^{(2-)}_{lng} \Im \tilde{\eta}_{\gamma\alpha\beta}^{(glng)}  +\tilde{\Delta}^{(3-)}_{lng} \Im \tilde{\eta}_{\beta\gamma\alpha}^{(glng)} \big) \big] 
\label{ReIm_eee}
\end{align}
where we defined $\tilde{\Delta}^{(i-)}_{lng} = (\tilde{\Delta}^{(i)}_{lng} - \tilde{\Delta}^{(i)}_{\bar{l}\bar{n}\bar{g}})/2$, $\tilde{\Delta}^{(i+)}_{lng} = (\tilde{\Delta}^{(i)}_{lng} + \tilde{\Delta}^{(i)}_{\bar{l}\bar{n}\bar{g}})/2$, $\Re \tilde{\eta}_{\alpha\beta\gamma}^{(glng)} = (\tilde{\eta}_{\alpha\beta\gamma}^{(glng)} + \tilde{\eta}_{\gamma\beta\alpha}^{(gnlg)})/2$ and $\Im \tilde{\eta}_{\alpha\beta\gamma}^{(glng)} = (\tilde{\eta}_{\alpha\beta\gamma}^{(glng)} - \tilde{\eta}_{\gamma\beta\alpha}^{(gnlg)})/(2i)$. We have grouped in $\tilde{\Delta}^{(i)}_{lng}$ all coefficients of Eq.~(\ref{chie}), $i \frac{e^3}{\hbar^3} E_{gl}E_{ln}E_{ng}$, so as to leave in $\tilde{\eta}_{\beta\gamma\alpha}^{(glng)}$ only the matrix elements; so, for example, $\tilde{\Delta}^{(1)}_{lng} = i \frac{e^3}{\hbar^3} E_{gl}E_{ln}E_{ng} {\Delta}^{(1)}_{lng}$ and $\tilde{\eta}_{\alpha\beta\gamma}^{(glng)}= \langle \phi_g| r_{\alpha} | \phi_l \rangle \langle \phi_l| r_{\beta}| \phi_n \rangle \langle \phi_n|  r_{\gamma} | \phi_g \rangle$. Following the notation of Ref.~\onlinecite{prbvarma}, we labelled the energy spectrum of the time-reversed configuration as $E_{\bar{n}}$. So, if time-reversal is a symmetry for our system, then $E_{\bar{n}}=E_n$, so that $\tilde{\Delta}^{(i)}_{lng} = \tilde{\Delta}^{(i)}_{\bar{l}\bar{n}\bar{g}}$ and
$\tilde{\Delta}^{(i-)}_{lng}=0$.

We have also highlighted the order of appearance of intermediate states in the matrix elements through ${(glng)}$, which plays a fundamental role in the analysis. In fact, the complex conjugate of $\tilde{\eta}_{\alpha\beta\gamma}^{(glng)}$ is $\tilde{\eta}_{\gamma\beta\alpha}^{(gnlg)}$: this implies that we should not only reverse the order of the cartesian indices (which, alone, would have led to the antisymmetrization of the $\gamma$ and $\alpha$ labels), but also keep track of the order of the intermediate states $l,n$ (or $n,l$). The latter point is what makes the profound difference, mathematically, with the RXS case, where the imaginary part of the cartesian, third-rank tensor is antisymmetric in two labels $\alpha$ and $\gamma$ (see Ref.~\onlinecite{prbvarma} for the analysis of the analogous E1-E2 third-rank cartesian tensor in the case of RXS) and therefore only couples with the corresponding antisymmetric part of the polarization, leading to a powerful selection rule based on time-reversal. In SHG this is not possible, for this technical reason: $\Im \tilde{\eta}_{\alpha\beta\gamma}^{(glng)}$ is {\it not} antisymmetric in $\alpha \leftrightarrow \gamma$. Physically, this is related to the order of the absorption and emission processes (through the denominators $\tilde{\Delta}^{(i)}_{lng}$, which have no definite symmetry in the exchange $l\leftrightarrow n$), that does not allow time-reversal to be a symmetry of the SHG process, as it is of RXS.

In an analogous way, we can write similar expressions also for $\chi^{(m)}$ and $\chi^{(q)}$ and deduce that $\Re \tilde{\eta}^{(e)}$, $\Im \tilde{\eta}^{(m)}$ and $\Im \tilde{\eta}^{(q)}$ are time-reversal even tensors (non-magnetic or $i$-tensors in Birss' notation \cite{birss}), whereas $\Im \tilde{\eta}^{(e)}$, $\Re \tilde{\eta}^{(m)}$ and $\Re \tilde{\eta}^{(q)}$ are time-reversal odd (magnetic or $c$-tensors in Birss' notation). These cartesian tensors can then be analyzed in terms of the irreducible spherical decompositions, as shown in the next subsection and in Section III.

In the light of our findings, it turns out that, of our two candidates to explain the SHG signal, one (for the 2$^\prime$/m case) is the time-reversal odd part of $\eta^{(e)}$, which is imaginary and therefore can interfere with the time-reversal even part of $\eta^{(q)}$, which is imaginary as well. The other (for the 2$^\prime$/m$^\prime$ case) is the time-reversal odd part of $\eta^{(m)}$, which is real. Yet, it can still interfere with the time-reversal even part of $\eta^{(q)}$, which is purely imaginary only far from resonance. In fact, sufficiently close to the resonance (within $\sim\Gamma$), the complex resonant denominators of the above SHG expressions for $\eta^{(e)}$, $\eta^{(m)}$ and $\eta^{(q)}$ scramble the previous imaginary/real separation based on the time-reversal properties of the matrix elements in the numerator. In fact, whatever is the numerator $N=a+ib$ (with $a$ and $b$ real), we get:

\begin{align}
& \frac{a+ib}{(\omega_{mi}-2\omega+i\Gamma_m)(\omega_{ni}-\omega+i\Gamma_n)} = \nonumber \\
& = \big\{ [a((\omega_{mi}-2\omega)(\omega_{ni}-\omega)-\Gamma_m\Gamma_n)\nonumber \\
& +b(\Gamma_n(\omega_{mi}-2\omega)+\Gamma_m(\omega_{ni}-\omega))] \nonumber \\
& +i[-a(\Gamma_n(\omega_{mi}-2\omega)+\Gamma_m(\omega_{ni}-\omega)) \nonumber \\
& +b((\omega_{mi}-2\omega)(\omega_{ni}-\omega)-\Gamma_m\Gamma_n)] \big \}\nonumber \\
&\big \{((\omega_{mi}-2\omega)(\omega_{ni}-\omega)-\Gamma_m\Gamma_n)^2 \nonumber \\
& +(\Gamma_n(\omega_{mi}-2\omega)+\Gamma_m(\omega_{ni}-\omega))^2\big \}^{-1} 
\label{gamma}
\end{align}

So, $a$ and $b$ matrix elements interfere, unless we are in one of the two extreme situations: 

1) out of resonance (i.e., $\omega_{mi}-2\omega\gg \Gamma_m$ and $\omega_{ni}-\omega\gg \Gamma_n$), so that $\Gamma$ is negligible and $a$ and $b$ in Eq.~(\ref{gamma}) do not interfere any more; 

2) in the case of a single, well separated resonance (within $\sim\Gamma$), exactly at resonance (i.e., $\omega_{mi}-2\omega\ll \Gamma_m$ and $\omega_{ni}-\omega\ll \Gamma_n$), so that the whole expression reduces to $-\frac{a+ib}{\Gamma_m\Gamma_n}$, and again $a$ and $b$ do not interfere.

\subsection{Identification of the SHG order parameters}
 
Here we give the explicit spherical and cartesian components of some of the polarization tensors that couple with the order parameters identified in Section III. 
We first list all the tensors of E1-E1-E1 origin ($\chi^{(e)}$), associated with the 2$^\prime$/m magnetic group and then those of E1-E1-M1 origin ($\chi^{(m)}$), associated with 2$^\prime$/m$^\prime$. For $\chi^{(e)}$, we have that the polarization dependence is determined by: {\it a)} two first-rank tensors, $\bar{O}^{(1)}$ and $\tilde{O}^{(1)}$, both coupled to an order parameter with the symmetry of a toroidal dipole, {\it b)} a second-rank $\tilde{O}^{(2)}$, coupled to an order parameter with the symmetry of a magnetic quadrupole and {\it c)} a third-rank $\tilde{O}^{(3)}$, coupled to an order parameter with the symmetry of a magnetic toroidal octupole. Their explicit expression  can be obtained from the scalar product in Eq.~(\ref{ordpar_eee}), that can be recoupled in spherical tensors as: $\epsilon^o_{\alpha} \epsilon^i_{\beta} \epsilon^i_{\gamma} {\tilde \chi}^{(e)}_{\alpha\beta\gamma}=\bar{O}^{(1)}\cdot\bar{\chi}^{(1)}+\sum_{i=1}^3 \tilde{O}^{(i)}\cdot \tilde{\chi}^{(i)}$. If, for a simpler comparison with Eq.~(\ref{azim}), we express the spherical polarization tensors in cartesian components, we have:

\begin{align}
\bar{O}_{\alpha} & = \epsilon_{\alpha}^o \vec{\epsilon}^i\cdot \vec{\epsilon}^i 
\label{tordip1cart}
\end{align}
here and below ${\alpha}$ is any of $x$, $y$ or $z$.

\begin{align}
\tilde{O}_{\alpha} & = \frac{1}{\sqrt{15}} {\epsilon}^{o}_{\alpha} \vec{\epsilon}^i\cdot \vec{\epsilon}^i - \frac{3}{\sqrt{15}} \epsilon_{\alpha}^i \vec{\epsilon}^o \cdot \vec{\epsilon}^i
\label{tordip2cart}
\end{align}

\begin{align} \label{magquadcart}
\tilde{O}_{3z^2-r^2} & =  \epsilon_z^i (\epsilon_{x}^o \epsilon_y^i - \epsilon_{y}^o \epsilon_{x}^i )  \\
\tilde{O}_{x^2-y^2} & =\frac{1}{\sqrt{3}} \big[ \epsilon_x^i (\epsilon_{y}^o \epsilon_z^i - \epsilon_{z}^o \epsilon_{y}^i ) - \epsilon_y^i (\epsilon_{z}^o \epsilon_x^i - \epsilon_{x}^o \epsilon_{z}^i )\big] \nonumber \\
\tilde{O}_{xy} & =\frac{1}{\sqrt{3}} \big[ \epsilon_x^i (\epsilon_{z}^o \epsilon_x^i - \epsilon_{x}^o \epsilon_{z}^i ) + \epsilon_y^i (\epsilon_{y}^o \epsilon_z^i - \epsilon_{z}^o \epsilon_{y}^i )\big] \nonumber \\
\tilde{O}_{xz} & =\frac{1}{\sqrt{3}} \big[ \epsilon_z^i (\epsilon_{y}^o \epsilon_z^i - \epsilon_{z}^o \epsilon_{y}^i ) + \epsilon_x^i (\epsilon_{x}^o \epsilon_y^i - \epsilon_{y}^o \epsilon_{x}^i )\big] \nonumber \\
\tilde{O}_{yz} & = \frac{1}{\sqrt{3}} \big[ \epsilon_z^i (\epsilon_{z}^o \epsilon_x^i - \epsilon_{x}^o \epsilon_{z}^i ) + \epsilon_y^i (\epsilon_{x}^o \epsilon_y^i - \epsilon_{y}^o \epsilon_{x}^i )\big] \nonumber
\end{align}

\begin{align} \label{toroctcart}
\tilde{O}_{y(3x^2-y^2)} & = \frac{1}{2} \big[ {\epsilon}^{o}_y (\epsilon^i_x\epsilon^i_x - \epsilon^i_y\epsilon^i_y ) +2 {\epsilon}^{o}_x \epsilon^i_y \epsilon^i_x \big] \\
\tilde{O}_{x(x^2-3y^2)} & = \frac{1}{2} \big[ {\epsilon}^{o}_x (\epsilon^i_x\epsilon^i_x - \epsilon^i_y\epsilon^i_y ) +2 {\epsilon}^{o}_y \epsilon^i_y \epsilon^i_x \big]  \nonumber \\
\tilde{O}_{z(x^2-y^2)} & = \frac{1}{\sqrt{6}} \big[ 2\epsilon^i_z ({\epsilon}^{o}_x\epsilon^i_x - {\epsilon}^{o}_y\epsilon^i_y) + {\epsilon}^{o}_z (\epsilon^i_x\epsilon^i_x - \epsilon^i_y\epsilon^i_y )  \big] \nonumber \\
\tilde{O}_{xyz} &  = \sqrt{\frac{2}{3}} \big[ {\epsilon}^{o}_z \epsilon^i_x\epsilon^i_y + {\epsilon}^{o}_x \epsilon^i_z\epsilon^i_y + {\epsilon}^{o}_y \epsilon^i_x\epsilon^i_z  \big]  \nonumber \\
\tilde{O}_{xz^2} & = \frac{1}{2\sqrt{15}} \big[ 8{\epsilon}^{o}_z \epsilon^i_x\epsilon^i_z +4{\epsilon}^{o}_x \epsilon^i_z\epsilon^i_z \nonumber \\ 
& -3 {\epsilon}^{o}_x \epsilon^i_x\epsilon^i_x  - {\epsilon}^{o}_x \epsilon^i_y\epsilon^i_y -2{\epsilon}^{o}_y \epsilon^i_x\epsilon^i_y \big] \nonumber \\
\tilde{O}_{yz^2} & = \frac{1}{2\sqrt{15}} \big[ 8{\epsilon}^{o}_z \epsilon^i_y\epsilon^i_z +4{\epsilon}^{o}_y \epsilon^i_z\epsilon^i_z  \nonumber \\ 
& -3 {\epsilon}^{o}_y \epsilon^i_y\epsilon^i_y - {\epsilon}^{o}_y \epsilon^i_x\epsilon^i_x -2{\epsilon}^{o}_x \epsilon^i_y\epsilon^i_x \big] \nonumber \\
\tilde{O}_{z^3} & = \frac{1}{\sqrt{10}} {\epsilon}^{o}_z (3\epsilon^{i}_z\epsilon^{i}_z - \vec{\epsilon}^{i} \cdot \vec{\epsilon}^{i}) -2\epsilon^{i}_z \big({\epsilon}^{o}_x \epsilon^{i}_x + {\epsilon}^{o}_y \epsilon^{i}_y \big) \nonumber 
\end{align}

Here, for example, $\tilde{O}_{yz}$ couples to the corresponding susceptibility $\tilde{\chi}_{yz}$, with the symmetry of a magnetic quadrupole. Eqs.~(\ref{tordip1cart}) to (\ref{toroctcart}) allow us to associate with each multipole component a well-determined azimuthal scan which constitutes a quantitative basis for our statements in Section III. For example, using the above equations with Eq.~(\ref{azim}), we remark that first-rank tensors, with the symmetry of a toroidal dipole, contribute to the SHG signal with just a $\sin\psi$ ($\cos\psi$) dependence. However, in order to explain the SHG azimuthal scans, $\sin^3\psi$ ($\cos^3\psi$) terms are necessary. This implies that, as noted in Section III, we cannot neglect the signal determined by the magnetic toroidal octupole, Eq.~(\ref{toroctcart}). We remark also that the azimuthal-scan technique employed for SHG by Ref.~\onlinecite{shg} proves to be a very powerful tool to extract the relative weight of each multipole order parameter, in full analogy with the RXS case \cite{dima,love}.

We can look at an analogous treatment of the $\chi^{(m)}$ terms of the 2$^\prime$/m$^\prime$ group, though the algebra is slightly more involved, given the number of terms to be treated in the E1-E1-M1 case (see Fig.~\ref{fig3}).
Consider the case of the E1-E1-M1 transition amplitude, $A_{SHG}^{(m)}$, associated with $\chi^{(m)}$, in full analogy with what was done above for $\chi^{(e)}$. In this case, from Eq.~(\ref{scatSHG2}) we get:

\begin{align}
& A_{SHG}^{(m)} = \sum_{l,n} \sum_{i,j=1}^3 \Delta_{l,n}^{(i)}\chi^{(m,i,j)}_{\alpha\beta\gamma}O^{(m,i,j)}_{\alpha\beta\gamma} \nonumber \\
& = O^{(m,1,1)}_{\alpha\beta\gamma} \big(\sum_{l,n} \Delta_{l,n}^{(1)}\chi^{(m,1,1)}_{\alpha\beta\gamma}
+ \Delta_{l,n}^{(2)}\chi^{(m,2,2)}_{\gamma\alpha\beta} + \Delta_{l,n}^{(3)}\chi^{(m,3,3)}_{\beta\gamma\alpha} \big) \nonumber \\
& + O^{(m,2,1)}_{\alpha\beta\gamma} \big(\sum_{l,n} \Delta_{l,n}^{(1)}(\chi^{(m,2,2)}_{\beta\alpha\gamma} + \chi^{(m,3,3)}_{\beta\gamma\alpha}) \nonumber \\
&  + \Delta_{l,n}^{(2)}(\chi^{(m,1,1)}_{\alpha\beta\gamma} + \chi^{(m,3,3)}_{\gamma\beta\alpha}) + \Delta_{l,n}^{(3)}(\chi^{(m,2,2)}_{\alpha\gamma\beta} + \chi^{(m,1,1)}_{\gamma\alpha\beta}) \big) \nonumber \\
& \equiv O^{(m,1,1)}_{\alpha\beta\gamma} {\tilde {\tilde \chi}}^{(m)}_{\alpha\beta\gamma} + O^{(m,2,1)}_{\alpha\beta\gamma} \underline{\tilde{\tilde{\chi}}}^{(m)}_{\alpha\beta\gamma} 
\label{ordpar_eem}
\end{align}

In the second step, we used the equalities (Eqs.~(A24) to (A32)): $O^{(m,1,1)}_{\alpha\beta\gamma}=O^{(m,2,2)}_{\gamma\alpha\beta}=O^{(m,3,3)}_{\beta\gamma\alpha}$ and $O^{(m,2,1)}_{\alpha\beta\gamma}=O^{(m,3,2)}_{\gamma\alpha\beta}=O^{(m,1,3)}_{\beta\gamma\alpha}=O^{(m,3,1)}_{\alpha\gamma\beta}=O^{(m,1,2)}_{\beta\alpha\gamma}=O^{(m,2,3)}_{\gamma\beta\alpha}$. In the third step, we used relations (A6) to (A14). Finally, the last line defines the tensors ${\tilde {\tilde \chi}}^{(m)}_{\alpha\beta\gamma}$ and $\underline{\tilde{\tilde{\chi}}}^{(m)}_{\alpha\beta\gamma}$.

We can now analyze, as for $\chi^{(e)}$ before, the properties of the polarization tensors: $O^{(m,1,1)}_{\alpha\beta\gamma} \equiv (\vec{\epsilon}^{o}\times \vec{k}^o)_{\alpha}\epsilon^i_{\beta}\epsilon^i_{\gamma}$ (Eq.~(A24)) and $O^{(m,2,1)}_{\alpha\beta\gamma} \equiv (\vec{\epsilon}^{i}\times \vec{k}^i)_{\alpha}\epsilon^o_{\beta}\epsilon^i_{\gamma}$ (Eq.~(A25)). As above, we can write the two terms in the last line of Eq.~(\ref{ordpar_eem}) as a scalar product of spherical tensors: $(\epsilon^{o}\times k^o)_{\alpha}\epsilon^i_{\beta}\epsilon^i_{\gamma} {\tilde {\tilde \chi}}^{(m)}_{\alpha\beta\gamma} + (\epsilon^{i}\times k^i)_{\alpha}\epsilon^o_{\beta}\epsilon^i_{\gamma} \underline{\tilde{\tilde{\chi}}}^{(m)}_{\alpha\beta\gamma}=\sum_{i=1}^3 {\tilde {\tilde{O}}}^{(i)}\cdot {\tilde {\tilde{\chi}}}^{(i)} + \sum_{i=0}^3 \tilde{\tilde{P}}^{(i)}\cdot \underline{\tilde{\tilde{\chi}}}^{(i)}$.

Here we analyze in detail the transformation properties, under rotation, of the polarization spherical tensor, that can be formally derived in a simple way from Eqs.~(\ref{tordip1cart}), (\ref{tordip2cart}), (\ref{magquadcart}) and (\ref{toroctcart}) by the replacement: ${\epsilon}^{o}_{\alpha} \rightarrow (\vec{\epsilon}^{o}\times \vec{k}^o)_{\alpha}$ for ${\tilde {\tilde{O}}}^{(i)}$ and by the replacements: ${\epsilon}^{o}_{\alpha} \rightarrow (\vec{\epsilon}^{i}\times \vec{k}^i)_{\alpha}$ and ${\epsilon}^{i}_{\beta} {\epsilon}^{i}_{\gamma} \rightarrow ({\epsilon}^{i}_{\beta} {\epsilon}^{o}_{\gamma}+{\epsilon}^{o}_{\beta} {\epsilon}^{i}_{\gamma})/2$ for $\tilde{\tilde{P}}^{(i)}$. In this way, we double the number of tensors that we had in the $\chi^{(e)}$ case and obtain the cartesian components of eight tensors (that we call ${\bar {\bar{O}}}^{(1)}$, ${\tilde {\tilde{O}}}^{(1)}$, ${\tilde {\tilde{O}}}^{(2)}$, ${\tilde {\tilde{O}}}^{(3)}$, ${\bar {\bar{P}}}^{(1)}$, ${\tilde {\tilde{P}}}^{(1)}$, ${\tilde {\tilde{P}}}^{(2)}$, ${\tilde {\tilde{P}}}^{(3)}$). Three further allowed polarization tensors are obtained from $O^{(m,2,1)}_{\alpha\beta\gamma} \equiv (\vec{\epsilon}^{i}\times \vec{k}^i)_{\alpha}\epsilon^o_{\beta}\epsilon^i_{\gamma}$, (Eq.~(A25)), when we couple $\epsilon^o_{\beta}\epsilon^i_{\gamma}$ antisymmetrically so as to have $\vec{\epsilon}^o \times \vec{\epsilon}^i$ (this term was obviously zero in the previous cases with $\epsilon^i_{\beta}\epsilon^i_{\gamma}$). The three tensors are then obtained by the coupling of the two vectors $\vec{\epsilon}^o \times \vec{\epsilon}^i$ and $\vec{\epsilon}^{i}\times \vec{k}^i$: ${\breve{O}}^{(0)}$ (a scalar, with a constant azimuthal dependence), ${\breve{O}}^{(1)}$ (a vector), and ${\breve{O}}^{(2)}$ (a symmetric traceless second-rank tensor). These three tensors only contribute in SP, PS and PP geometry, as in SS geometry $\vec{\epsilon}^o \times \vec{\epsilon}^i$ is zero. This also implies that the scalar contribution is not allowed in SS geometry for any of the E1-E1-M1 terms (${\breve{O}}^{(0)}$ is the only scalar term). In detail, 

\begin{align}
{\breve{O}}^{(0)} = (\vec{\epsilon}^o \times \vec{\epsilon}^i) \cdot (\vec{\epsilon}^{i}\times \vec{k}^i)
\label{scalar}
\end{align}

\begin{align}
{\breve{O}}^{(1)}_{\alpha} = \big[(\vec{\epsilon}^o \times \vec{\epsilon}^i) \times (\vec{\epsilon}^{i}\times \vec{k}^i)\big]_{\alpha}
\label{vector}
\end{align}
again here and below ${\alpha}$ is any of $x$, $y$, $z$.

\begin{align}
& {\breve{O}}^{(2)}_{3z^2-r^2} = \frac{1}{6}\big [2(\vec{\epsilon}^o \times \vec{\epsilon}^i)_z (\vec{\epsilon}^{i}\times \vec{k}^i)_z \nonumber \\
& -(\vec{\epsilon}^o \times \vec{\epsilon}^i)_x (\vec{\epsilon}^{i}\times \vec{k}^i)_x - (\vec{\epsilon}^o \times \vec{\epsilon}^i)_y (\vec{\epsilon}^{i}\times \vec{k}^i)_y\big] \nonumber \\
&{\breve{O}}^{(2)}_{x^2-y^2} = \frac{1}{2}\big [(\vec{\epsilon}^o \times \vec{\epsilon}^i)_x (\vec{\epsilon}^{i}\times \vec{k}^i)_x - (\vec{\epsilon}^o \times \vec{\epsilon}^i)_y (\vec{\epsilon}^{i}\times \vec{k}^i)_y\big] \nonumber \\
&{\breve{O}}^{(2)}_{yz} = \frac{1}{2}\big [(\vec{\epsilon}^o \times \vec{\epsilon}^i)_y (\vec{\epsilon}^{i}\times \vec{k}^i)_z +(\vec{\epsilon}^o \times \vec{\epsilon}^i)_z (\vec{\epsilon}^{i}\times \vec{k}^i)_y\big] \nonumber \\
&{\breve{O}}^{(2)}_{xz} =\frac{1}{2}\big [(\vec{\epsilon}^o \times \vec{\epsilon}^i)_x (\vec{\epsilon}^{i}\times \vec{k}^i)_z +(\vec{\epsilon}^o \times \vec{\epsilon}^i)_z (\vec{\epsilon}^{i}\times \vec{k}^i)_x\big] \nonumber \\
&{\breve{O}}^{(2)}_{xy} =\frac{1}{2}\big [(\vec{\epsilon}^o \times \vec{\epsilon}^i)_x (\vec{\epsilon}^{i}\times \vec{k}^i)_y +(\vec{\epsilon}^o \times \vec{\epsilon}^i)_y (\vec{\epsilon}^{i}\times \vec{k}^i)_x\big] \nonumber \\
\label{traceless}
\end{align}

Now we list the cartesian components of the former eight tensors. They are, for ${\bar {\bar{O}}}^{(1)}$, ${\tilde {\tilde{O}}}^{(1)}$, ${\tilde {\tilde{O}}}^{(2)}$, ${\tilde {\tilde{O}}}^{(3)}$:

\begin{align}
\bar{\bar{O}}_{\alpha} & = (\vec{\epsilon}^{o}\times \vec{k}^o)_{\alpha} \vec{\epsilon}^i\cdot \vec{\epsilon}^i 
\label{magdip1cart}
\end{align}

\begin{align}
\tilde{\tilde{O}}_{\alpha} = \frac{1}{\sqrt{15}} (\vec{\epsilon}^{o}\times \vec{k}^o)_{\alpha} \vec{\epsilon}^i\cdot \vec{\epsilon}^i - \frac{3}{\sqrt{15}} \epsilon_{\alpha}^i \big[ (\vec{\epsilon}^{o}\times \vec{k}^o) \cdot \vec{\epsilon}^i \big] 
\label{magdip2cart}
\end{align}

\begin{align}
\tilde{\tilde{O}}_{3z^2-r^2} & =  \epsilon_z^i ((\vec{\epsilon}^{o}\times \vec{k}^o)_{x} \epsilon_y^i - (\vec{\epsilon}^{o}\times \vec{k}^o)_{y} \epsilon_{x}^i ) \nonumber \\
\tilde{\tilde{O}}_{x^2-y^2} & =\frac{1}{\sqrt{3}} \big[ \epsilon_x^i ((\vec{\epsilon}^{o}\times \vec{k}^o)_{y} \epsilon_z^i - (\vec{\epsilon}^{o}\times \vec{k}^o)_{z} \epsilon_{y}^i ) \nonumber \\
&- \epsilon_y^i ((\vec{\epsilon}^{o}\times \vec{k}^o)_{z} \epsilon_x^i - (\vec{\epsilon}^{o}\times \vec{k}^o)_{x} \epsilon_{z}^i )\big] \nonumber \\
\tilde{\tilde{O}}_{xy} & =\frac{1}{\sqrt{3}} \big[ \epsilon_x^i ((\vec{\epsilon}^{o}\times \vec{k}^o)_{z} \epsilon_x^i - (\vec{\epsilon}^{o}\times \vec{k}^o)_{x} \epsilon_{z}^i ) \nonumber \\
&+ \epsilon_y^i ((\vec{\epsilon}^{o}\times \vec{k}^o)_{y} \epsilon_z^i - (\vec{\epsilon}^{o}\times \vec{k}^o)_{z} \epsilon_{y}^i )\big] \nonumber \\
\tilde{\tilde{O}}_{xz} & =\frac{1}{\sqrt{3}} \big[ \epsilon_z^i ((\vec{\epsilon}^{o}\times \vec{k}^o)_{y} \epsilon_z^i - (\vec{\epsilon}^{o}\times \vec{k}^o)_{z} \epsilon_{y}^i ) \nonumber \\
& + \epsilon_x^i ((\vec{\epsilon}^{o}\times \vec{k}^o)_{x} \epsilon_y^i - (\vec{\epsilon}^{o}\times \vec{k}^o)_{y} \epsilon_{x}^i )\big] \nonumber \\
\tilde{\tilde{O}}_{yz} & = \frac{1}{\sqrt{3}} \big[ \epsilon_z^i ((\vec{\epsilon}^{o}\times \vec{k}^o)_{z} \epsilon_x^i - (\vec{\epsilon}^{o}\times \vec{k}^o)_{x} \epsilon_{z}^i ) \nonumber \\
& + \epsilon_y^i ((\vec{\epsilon}^{o}\times \vec{k}^o)_{x} \epsilon_y^i - (\vec{\epsilon}^{o}\times \vec{k}^o)_{y} \epsilon_{x}^i )\big]
\label{magtorquadcart}
\end{align}

\begin{align}
& \tilde{\tilde{O}}_{y(3x^2-y^2)} \nonumber \\
& = \frac{1}{2} \big[ (\vec{\epsilon}^{o}\times \vec{k}^o)_y (\epsilon^i_x\epsilon^i_x - \epsilon^i_y\epsilon^i_y ) +2 (\vec{\epsilon}^{o}\times \vec{k}^o)_x \epsilon^i_y \epsilon^i_x \big] \nonumber \\
& \tilde{\tilde{O}}_{x(x^2-3y^2)} \nonumber \\
& =  \frac{1}{2} \big[ (\vec{\epsilon}^{o}\times \vec{k}^o)_x (\epsilon^i_x\epsilon^i_x - \epsilon^i_y\epsilon^i_y ) +2 (\vec{\epsilon}^{o}\times \vec{k}^o)_y \epsilon^i_y \epsilon^i_x \big]  \nonumber \\
& \tilde{\tilde{O}}_{z(x^2-y^2)} = \frac{1}{\sqrt{6}} \big[ 2\epsilon^i_z ((\vec{\epsilon}^{o}\times \vec{k}^o)_x\epsilon^i_x     \nonumber \\
& - (\vec{\epsilon}^{o}\times \vec{k}^o)_y\epsilon^i_y) + (\vec{\epsilon}^{o}\times \vec{k}^o)_z (\epsilon^i_x\epsilon^i_x - \epsilon^i_y\epsilon^i_y )  \big] \nonumber \\
& \tilde{\tilde{O}}_{xyz} \nonumber \\
&  = \sqrt{\frac{2}{3}} \big[ (\vec{\epsilon}^{o}\times \vec{k}^o)_z \epsilon^i_x\epsilon^i_y + (\vec{\epsilon}^{o}\times \vec{k}^o)_x \epsilon^i_z\epsilon^i_y + (\vec{\epsilon}^{o}\times \vec{k}^o)_y \epsilon^i_x\epsilon^i_z  \big]  \nonumber \\
& \tilde{\tilde{O}}_{xz^2} = \frac{1}{2\sqrt{15}} \big[ 8(\vec{\epsilon}^{o}\times \vec{k}^o)_z \epsilon^i_x\epsilon^i_z +4(\vec{\epsilon}^{o}\times \vec{k}^o)_x \epsilon^i_z\epsilon^i_z \nonumber \\ 
& -3 (\vec{\epsilon}^{o}\times \vec{k}^o)_x \epsilon^i_x\epsilon^i_x  - (\vec{\epsilon}^{o}\times \vec{k}^o)_x \epsilon^i_y\epsilon^i_y -2(\vec{\epsilon}^{o}\times \vec{k}^o)_y \epsilon^i_x\epsilon^i_y \big] \nonumber \\
& \tilde{\tilde{O}}_{yz^2} = \frac{1}{2\sqrt{15}} \big[ 8(\vec{\epsilon}^{o}\times \vec{k}^o)_z \epsilon^i_y\epsilon^i_z +4(\vec{\epsilon}^{o}\times \vec{k}^o)_y \epsilon^i_z\epsilon^i_z  \nonumber \\ 
& -3 (\vec{\epsilon}^{o}\times \vec{k}^o)_y \epsilon^i_y\epsilon^i_y - (\vec{\epsilon}^{o}\times \vec{k}^o)_y \epsilon^i_x\epsilon^i_x -2(\vec{\epsilon}^{o}\times \vec{k}^o)_x \epsilon^i_y\epsilon^i_x \big] \nonumber \\
& \tilde{\tilde{O}}_{z^3} = \frac{1}{\sqrt{10}} (\vec{\epsilon}^{o}\times \vec{k}^o)_z (3\epsilon^{i}_z\epsilon^{i}_z - \vec{\epsilon}^{i} \cdot \vec{\epsilon}^{i}) \nonumber \\
& -2\epsilon^{i}_z \big((\vec{\epsilon}^{o}\times \vec{k}^o)_x \epsilon^{i}_x + (\vec{\epsilon}^{o}\times \vec{k}^o)_y \epsilon^{i}_y \big) 
\label{magoctcart}
\end{align}

To finish, we list the terms coming from the second polarization term: $O^{(m,2,1)}_{\alpha\beta\gamma} \equiv (\vec{\epsilon}^{i}\times \vec{k}^i)_{\alpha}\epsilon^o_{\beta}\epsilon^i_{\gamma}$ (Eq.~(A25)). In this case, as stated above, there is no symmetry between the two polarizations associated with the electric-dipole transitions, $\epsilon^o_{\beta}\epsilon^i_{\gamma}$. This implies that, besides the zeroth and second-rank tensors, analogous to the previous case, there is also the possibility of an antisymmetric coupling of $\epsilon^o_{\beta}$ and $\epsilon^i_{\gamma}$, listed above. All three (zeroth, first and second-rank) tensors must then be coupled to the last vector, $(\vec{\epsilon}^{i}\times \vec{k}^i)$. We have:

\begin{align}
\bar{ \bar{P}}_{\alpha} & = (\vec{\epsilon}^{i}\times \vec{k}^i)_{\alpha} \vec{\epsilon}^o\cdot \vec{\epsilon}^i 
\label{magdipcart1}
\end{align}

\begin{align}
\tilde{ \tilde{P}}_{\alpha} & = \frac{1}{\sqrt{15}} (\vec{\epsilon}^{i}\times \vec{k}^i)_{\alpha} \vec{\epsilon}^o\cdot \vec{\epsilon}^i \nonumber \\
& - \frac{3}{2\sqrt{15}} \big(\epsilon_{\alpha}^i \big[ (\vec{\epsilon}^{i}\times \vec{k}^i) \cdot \vec{\epsilon}^o \big] + \epsilon_{\alpha}^o \big[ (\vec{\epsilon}^{i}\times \vec{k}^i) \cdot \vec{\epsilon}^i \big] \big)
\label{magdipcart2}
\end{align}

\begin{align}
\tilde{\tilde{P}}_{3z^2-r^2} & = \frac{1}{2}\big[(\vec{\epsilon}^{i}\times \vec{k}^i)_{x}  (\epsilon_z^i \epsilon_y^o+\epsilon_z^o \epsilon_y^i)
\nonumber \\
& - (\vec{\epsilon}^{i}\times \vec{k}^i)_{y}(\epsilon_z^i \epsilon_{x}^o+\epsilon_z^o \epsilon_{x}^i)  \big]\nonumber \\
\tilde{\tilde{P}}_{x^2-y^2} & =\frac{1}{2\sqrt{3}} \big[ (\vec{\epsilon}^{i}\times \vec{k}^i)_{y} (\epsilon_x^i\epsilon_z^o+\epsilon_x^o\epsilon_z^i) \nonumber \\
& + (\vec{\epsilon}^{i}\times \vec{k}^i)_{x} (\epsilon_{z}^i \epsilon_y^o+\epsilon_{z}^o \epsilon_y^i)
-2 (\vec{\epsilon}^{i}\times \vec{k}^i)_{z} (\epsilon_{y}^i\epsilon_x^o+\epsilon_x^i \epsilon_y^o) \big] \nonumber \\
\tilde{\tilde{P}}_{xy} & =\frac{1}{2\sqrt{3}} \big[2(\vec{\epsilon}^{i}\times \vec{k}^i)_{z}(\epsilon_x^i\epsilon_x^o -\epsilon_y^i\epsilon_y^o)  \nonumber \\
& - (\vec{\epsilon}^{i}\times \vec{k}^i)_{x} (\epsilon_{z}^i \epsilon_x^o+\epsilon_{z}^o \epsilon_x^i)+ (\vec{\epsilon}^{i}\times \vec{k}^i)_{y} (\epsilon_y^i \epsilon_z^o+\epsilon_y^o \epsilon_z^i) \big] \nonumber \\
\tilde{\tilde{P}}_{xz} & =\frac{1}{2\sqrt{3}} \big[2 (\vec{\epsilon}^{i}\times \vec{k}^i)_{y}(\epsilon_z^i \epsilon_z^o-\epsilon_x^i \epsilon_x^o) \nonumber \\
& - (\vec{\epsilon}^{i}\times \vec{k}^i)_{z} (\epsilon_{y}^o\epsilon_z^i+\epsilon_{y}^i\epsilon_z^o)  + (\vec{\epsilon}^{i}\times \vec{k}^i)_{x} (\epsilon_y^i\epsilon_x^o + \epsilon_y^o\epsilon_x^i )  \big] \nonumber \\
\tilde{\tilde{P}}_{yz} & = \frac{1}{2\sqrt{3}} \big[ 2 (\vec{\epsilon}^{i}\times \vec{k}^i)_{x}(\epsilon_y^i \epsilon_y^o-\epsilon_z^i \epsilon_z^o) \nonumber \\
& - (\vec{\epsilon}^{i}\times \vec{k}^i)_{y} (\epsilon_{y}^o\epsilon_x^i+\epsilon_{y}^i\epsilon_x^o)  + (\vec{\epsilon}^{i}\times \vec{k}^i)_{z} (\epsilon_z^i\epsilon_x^o + \epsilon_z^o\epsilon_x^i )\big]
\label{magtorquadcart2}
\end{align}

\begin{align}
& \tilde{\tilde{P}}_{y(3x^2-y^2)} = \frac{1}{2} \big[ (\vec{\epsilon}^{i}\times \vec{k}^i)_y (\epsilon^i_x\epsilon^o_x - \epsilon^i_y\epsilon^o_y ) \nonumber \\
&  + (\vec{\epsilon}^{i}\times \vec{k}^i)_x (\epsilon^i_y \epsilon^o_x+\epsilon^o_y \epsilon^i_x ) \big] \nonumber \\
& \tilde{\tilde{P}}_{x(x^2-3y^2)} =  \frac{1}{2} \big[ (\vec{\epsilon}^{i}\times \vec{k}^i)_x (\epsilon^i_x\epsilon^o_x - \epsilon^i_y\epsilon^o_y )  \nonumber \\
&  + (\vec{\epsilon}^{i}\times \vec{k}^i)_y (\epsilon^i_y \epsilon^o_x+\epsilon^o_y \epsilon^i_x ) \big]  \nonumber \\
& \tilde{\tilde{P}}_{z(x^2-y^2)} = \frac{1}{\sqrt{6}} \big[ (\vec{\epsilon}^{i}\times \vec{k}^i)_x (\epsilon^i_z \epsilon^o_x +  \epsilon^o_z \epsilon^i_x)  \nonumber \\
& - (\vec{\epsilon}^{i}\times \vec{k}^i)_y (\epsilon^i_z \epsilon^o_y+ \epsilon^o_z \epsilon^i_y)+ (\vec{\epsilon}^{i}\times \vec{k}^i)_z (\epsilon^i_x\epsilon^o_x - \epsilon^i_y\epsilon^o_y )  \big] \nonumber \\
& \tilde{\tilde{P}}_{xyz}  = \sqrt{\frac{2}{3}} \big[ (\vec{\epsilon}^{i}\times \vec{k}^i)_z (\epsilon^i_x\epsilon^o_y +\epsilon^o_x\epsilon^i_y)\nonumber \\
&  + (\vec{\epsilon}^{i}\times \vec{k}^i)_x (\epsilon^i_z\epsilon^o_y+\epsilon^o_z\epsilon^i_y) + (\vec{\epsilon}^{i}\times \vec{k}^i)_y (\epsilon^i_x\epsilon^o_z +\epsilon^o_x\epsilon^i_z) \big]  \nonumber \\
& \tilde{\tilde{P}}_{xz^2} = \frac{1}{2\sqrt{15}} \big[ 4(\vec{\epsilon}^{i}\times \vec{k}^i)_z (\epsilon^i_x\epsilon^o_z+\epsilon^o_x\epsilon^i_z) +4(\vec{\epsilon}^{i}\times \vec{k}^i)_x \epsilon^i_z\epsilon^o_z \nonumber \\ 
& -3 (\vec{\epsilon}^{i}\times \vec{k}^i)_x \epsilon^i_x\epsilon^o_x  - (\vec{\epsilon}^{i}\times \vec{k}^i)_x \epsilon^i_y\epsilon^o_y -(\vec{\epsilon}^{i}\times \vec{k}^i)_y (\epsilon^i_x\epsilon^o_y +\epsilon^o_x\epsilon^i_y)\big] \nonumber \\
& \tilde{\tilde{P}}_{yz^2} = \frac{1}{2\sqrt{15}} \big[ 4(\vec{\epsilon}^{i}\times \vec{k}^i)_z (\epsilon^i_y\epsilon^o_z+\epsilon^o_y\epsilon^i_z) +4(\vec{\epsilon}^{i}\times \vec{k}^i)_y \epsilon^i_z\epsilon^o_z  \nonumber \\ 
& -3 (\vec{\epsilon}^{i}\times \vec{k}^i)_y \epsilon^i_y\epsilon^o_y - (\vec{\epsilon}^{i}\times \vec{k}^i)_y \epsilon^i_x\epsilon^o_x -(\vec{\epsilon}^{i}\times \vec{k}^i)_x (\epsilon^i_y\epsilon^o_x+\epsilon^o_y\epsilon^i_x) \big] \nonumber \\
& \tilde{\tilde{P}}_{z^3} = \frac{1}{\sqrt{10}} (\vec{\epsilon}^{i}\times \vec{k}^i)_z (3\epsilon^{i}_z\epsilon^{o}_z - \vec{\epsilon}^{i} \cdot \vec{\epsilon}^{o}) \nonumber \\
& -\big((\vec{\epsilon}^{i}\times \vec{k}^i)_x (\epsilon^{i}_x\epsilon^{o}_z+\epsilon^{o}_x\epsilon^{i}_z)  + (\vec{\epsilon}^{i}\times \vec{k}^i)_y (\epsilon^{o}_y\epsilon^{i}_z +\epsilon^{i}_y\epsilon^{o}_z) \big) 
\label{magoctcart2}
\end{align}

\subsection{Magnetic subgroups of Sr$_2$IrO$_4$}

To follow the conclusions of Section III.B in more detail, we make use of the following table for the symmetry behavior of the $\vec{P}$ and $\vec{M}$ components under the 2/m1$^\prime$ magnetic group (in this subsection, we use the notation $P_\alpha = e r_\alpha$ and $M_\alpha = \mu_B (L_\alpha + 2 S_\alpha)$, which is more often employed in the SHG community \cite{fiebig,muto}):
\begin{table}[ht!]
	\caption{Symmetries of $\vec{P}$ and $\vec{M}$ in the 2/m1$^\prime$ magnetic group}
	\centering
	\begin{ruledtabular}
		\begin{tabular}{cccccccc}
			$\hat{E}$ & $\hat{C}_{2z}$ & $\hat{m}_z$ & $\hat{I}$ & $\hat{T}$ & $\hat{T}\hat{C}_{2z}$ & $\hat{T}\hat{m}_z$ & $\hat{T}\hat{I}$ \\
			\colrule
			P$_z$ & P$_z$ & -P$_z$ & -P$_z$ & P$_z^*$ & P$_z^*$ & -P$_z^*$ & -P$_z^*$ \\
			P$_x$ & -P$_x$ & P$_x$ & -P$_x$ & P$_x^*$ & -P$_x^*$ & P$_x^*$ & -P$_x^*$ \\ 
			P$_y$ & -P$_y$ & P$_y$ & -P$_y$ & P$_y^*$ & -P$_y^*$ & P$_y^*$ & -P$_y^*$ \\ 
			M$_z$ & M$_z$ & M$_z$ & M$_z$ & -M$_z^*$ & -M$_z^*$ & -M$_z^*$ & -M$_z^*$ \\ 
			M$_x$ & -M$_x$ & -M$_x$ & M$_x$ & -M$_x^*$ & M$_x^*$ & M$_x^*$ & -M$_x^*$ \\ 
			M$_y$ & -M$_y$ & -M$_y$ & M$_y$ & -M$_y^*$ & M$_y^*$ & M$_y^*$ & -M$_y^*$ \\ 
		\end{tabular}
	\end{ruledtabular}
	\label{table1}
\end{table}

From Table III, we can extract the allowed tensors for 2/m1$^\prime$ and each of its subgroups. 
For 2/m1$^\prime$, the total signal needs to be invariant under the sum over all its symmetries. If the sum is calculated for each of the above components, P$_z$, P$_x$, P$_y$, M$_z$, M$_x$, M$_y$, we get, respectively, 0, 0, 0, 8$i\Im$M$_z$, 0, 0. This result simply expresses the fact that no matrix element for P$_\alpha$ is allowed in the above group (as a consequence of inversion symmetry that forbids matrix elements of polar vectors). Therefore $\chi^{(e)}$ is zero in 2/m1$^\prime$. However, this is not the case for the imaginary part of $\chi^{(m)}$, time-reversal even, because of the 8$i\Im$M$_z$ term. Such a term can interfere with the $\chi^{(q)}$ signal.
However, the only components of $\chi^{(m)}$ that are different from zero are $\chi^{(m)}_{zzz}$, $\chi^{(m)}_{xxz}$, $\chi^{(m)}_{yyz}$, and $\chi^{(m)}_{xyz}$. Therefore, from Eq.~(\ref{azim}), this tensor is associated with a $2\psi$ azimuthal dependence, in disagreement with the experimental data. We can therefore disregard such a magnetic group and look for all possible symmetry reductions.

A cautionary note is necessary: in principle, we should study the symmetry behavior of third-rank tensors with 27 cartesian components. We can simplify it and just study the sum of each line of Table \ref{table1} as done above for the following reasons: a) the square of all symmetry operations of Table \ref{table1} is +1; b) the couples ($P_x$, $P_y$) and ($M_x$, $M_y$) have the same behavior and c) the products $P_{x(y)}P_z$ and $M_{x(y)}M_z$ are always zero. Putting all this together implies that studying the behavior of $\sum_{\rm symmetries} A_iB_jC_k$ (where $A$, $B$ and $C$ are any of $P$ and $M$) is the same as studying $\sum_{\rm symmetries} C_k$ alone and then `add' the non-zero $A_iB_j$ part only to the non-zero $\sum_{\rm symmetries} C_k$-terms, as in the previous case.
 
Turning to subgroups of 2/m1$^\prime$, we have 7 subgroups with 4 symmetry elements, 2/m, 2$^\prime$/m, 2/m$^\prime$, 2$^\prime$/m$^\prime$, 21$^\prime$, m1$^\prime$, $\overline{\rm 1}$1$^\prime$. We can repeat the above analysis for each subgroup (keeping the order P$_z$, P$_x$, P$_y$, M$_z$, M$_x$, M$_y$) and get:
\begin{itemize}

\item 2/m: The sum over the symmetry elements gives 0, 0, 0, 4M$_z$, 0, 0. No $\chi^{(e)}$ is allowed. We have both the real and imaginary parts of $\chi^{(m)}$, but again, as for the full 2/m1$^\prime$ group, no odd dependence on the azimuthal angle $\psi$. Therefore, this magnetic subgroup is excluded. 

\item 2$^\prime$/m: The sum over the symmetry elements gives 0, 4$i\Im$P$_x$, 4$i\Im$P$_y$, 4$i\Im$M$_z$, 0, 0. This implies a signal from the time-reversal odd part of $\chi^{(e)}$ when an odd number of $x$, $y$ components is considered, meaning an even number of $z$ components. As correctly recognized in Ref.~\onlinecite{shg}, this can explain the interference with the $\chi^{(q)}$ signal, being $\psi$-odd. The $\chi^{(m)}$ signal has the same behavior as for the 2/m1$^\prime$ group, a $2\psi$-azimuthal dependence. Therefore it is not responsible for the signal. We remark that this magnetic symmetry is the one of the $-+-+$ pattern studied in the previous section. We shall analyze this case more below.

\item 2/m$^\prime$: The sum over the symmetry elements gives 4$i\Im$P$_z$, 0, 0, 4$i\Im$M$_z$, 0, 0. Both $\chi^{(e)}$ and $\chi^{(m)}$ can contribute, but none have the correct odd-$\psi$ dependence.

\item 2$^\prime$/m$^\prime$: The sum over the symmetry elements gives 0, 0, 0, 4$i\Im$M$_z$, 4$\Re$M$_x$, 4$\Re$M$_y$. No $\chi^{(e)}$ is allowed and the imaginary, time-reversal even $\chi^{(m)}$ tensor has no odd-$\psi$ terms. However, the real, time-reversal odd $\chi^{(m)}$ tensor has the desired odd-$\psi$ terms (e.g., $\chi^{(m)}_{zzx}$) and it can interfere with the imaginary part of $\chi^{(q)}$ (through the damping factor $i\Gamma$). Its order parameter is either a magnetic dipole or a magnetic octupole. Interestingly, this magnetic group corresponds to the $++++$ state. It is further discussed below. 

\item m1$^\prime$: The sum over the symmetry elements gives 0, 4$\Re$P$_x$, 4$\Re$P$_y$, 4$i\Im$M$_z$, 0, 0. The imaginary, time-reversal even $\chi^{(m)}$ tensor does not have odd-$\psi$ terms, because of the odd number of $z$ components and does not contribute to the signal. An odd-$\psi$ dependence for $\chi^{(e)}$ is possible (even number of $z$ components), but only for its real, time-reversal even part. Again, the real part of $\chi^{(e)}$ can interfere with the imaginary part of $\chi^{(q)}$ because of the damping factor $i\Gamma$. The order parameter associated with this magnetic group has the symmetry of a time-reversal even electric polarization (or octupole). On the basis of the experimental evidence, it appears as highly implausible that it determines the SHG signal because such an order parameter implies the displacement of atoms, so as to break the global inversion, and this would be detectable by other means.
Moreover, the order parameter is time-reversal even and should have contributed also in the high-temperature phase, as no crystal distortion is detected in passing from the high-temperature phase to the low-temperature, magnetic phase. This is against the experimental evidence.

\item 21$^\prime$: The sum over the symmetry elements gives 4$\Re$P$_z$, 0, 0, 4$i\Im$M$_z$, 0, 0. As for the 2/m$^\prime$ group, both $\chi^{(e)}$ (real part, time-reversal even) and $\chi^{(m)}$ can contribute, but none have the correct odd-$\psi$ dependence.

\item $\overline{\rm 1}$1$^\prime$: The sum over the symmetry elements gives 0, 0, 0, 4$i\Im$M$_z$, 4$i\Im$M$_x$, 4$i\Im$M$_y$. No $\chi^{(e)}$ is allowed. The imaginary, time-reversal even $\chi^{(m)}$ tensor is allowed with an even number of $z$ components and can have the right odd-$\psi$ dependence. It can interfere with the imaginary part of $\chi^{(q)}$, but, as for the m1$^\prime$ group, it seems implausible since its associated order parameter is time-reversal even (symmetry of an axial toroidal dipole or of an electric quadrupole): 
either it should have been different from zero also in the high-temperature, tetragonal phase, against the experimental evidence, or it should be a secondary order-parameter, induced by the ordered magnetic moments. In the latter case, however, magnetic moments should break the $2^\prime$ symmetry by tilting along the $c$-axis, against the experimental evidence as well.

\end{itemize}

\section{Iridium ground state in a tetragonal crystal field}

In this Appendix, we list our own definitions for the atomic-like ground state of Sr$_2$IrO$_4$, highlighting the differences with other authors when needed.
Notice that the method that maps $t_{2g}$ orbitals to effective $p$ orbitals with angular momentum $L_{\rm eff}=-1$ is conceived to work within the $t_{2g}$ subspace, so care must be taken when generalizing to transition matrix elements involving other subspaces, as in the case of the L$_2$ and L$_3$ edges. As our main objective is to write down the L$_{2,3}$-edge cross-section, we shall not work with the effective angular momentum, but with the real ones. Below, we give the formulas to pass from one representation to the other, in order to compare with the existing literature.

\subsection{Basic equations for an Ir ion in a tetragonal crystal field}

We define the `cartesian' angular part of the $5d$ orbitals in terms of spherical harmonics $Y_{2m}\equiv d_m$ as follows: 

$d_{3z^2-r^2}=d_0$

$d_{x^2-y^2}=(d_2+d_{-2})/\sqrt{2}$

$d_{xy}=-i(d_2-d_{-2})/\sqrt{2}$

$d_{xz}=-(d_1-d_{-1})/\sqrt{2}$

$d_{yz}=i(d_1+d_{-1})/\sqrt{2}$

\noindent From this, we get the inverse transformations as: 

$d_0=d_{3z^2-r^2}$

$d_1=-(d_{xz}+id_{yz})/\sqrt{2}$

$d_{-1}=(d_{xz}-id_{yz})/\sqrt{2}$

$d_2=(d_{x^2-y^2}+id_{xy})/\sqrt{2}$

$d_{-2}=(d_{x^2-y^2}-id_{xy})/\sqrt{2}$

\noindent Given its importance in the analysis of the L$_2$ and L$_3$ edges of Ir, we write down also the $J=5/2$ and $J=3/2$ states of the $5d$ electrons in terms of spherical harmonics and spin functions ($\uparrow$, $\downarrow$):

$|\frac{5}{2},\frac{5}{2}\rangle=|d_{2\uparrow}\rangle$

$|\frac{5}{2},\frac{3}{2}\rangle=(|d_{2\downarrow}\rangle+2|d_{1\uparrow}\rangle)/\sqrt{5}$

$|\frac{5}{2},\frac{1}{2}\rangle=(\sqrt{2}|d_{1\downarrow}\rangle+\sqrt{3}|d_{0\uparrow}\rangle)/\sqrt{5} $

$|\frac{5}{2},-\frac{1}{2}\rangle=(\sqrt{3}|d_{0\downarrow}\rangle+\sqrt{2}|d_{-1\uparrow}\rangle)/\sqrt{5} $

$|\frac{5}{2},-\frac{3}{2}\rangle=(2|d_{-1\downarrow}\rangle+|d_{-2\uparrow}\rangle)/\sqrt{5} $

$|\frac{5}{2},-\frac{5}{2}\rangle=|d_{-2\downarrow}\rangle$

$|\frac{3}{2},\frac{3}{2}\rangle=(2|d_{2\downarrow}\rangle-|d_{1\uparrow}\rangle)/\sqrt{5} $

$|\frac{3}{2},\frac{1}{2}\rangle=(\sqrt{3}|d_{1\downarrow}\rangle-\sqrt{2}|d_{0\uparrow}\rangle)/\sqrt{5} $

$|\frac{3}{2},-\frac{1}{2}\rangle=(\sqrt{2}|d_{0\downarrow}\rangle-\sqrt{3}|d_{-1\uparrow}\rangle)/\sqrt{5} $

$|\frac{3}{2},-\frac{3}{2}\rangle=(|d_{-1\downarrow}\rangle-2|d_{-2\uparrow}\rangle)/\sqrt{5} $

\noindent It should be reminded that a direct transition from the L$_2$ edge, characterized by $J_{2p}=1/2$, to the $|j=\frac{5}{2},j_z\rangle$ subspace is dipole forbidden. Therefore the decomposition of the half-filled Kramers doublet  of Ir in Sr$_2$IrO$_4$ as $|j=\frac{5}{2},j_z\rangle$ states will directly inform us about whether this transition is allowed or not at the L$_2$ edge. 
In order to do this, we write down the inverse formulas to get $d_m$ states as a function of $|j,j_z\rangle$:

\begin{align}
|d_{2\uparrow}\rangle= |\frac{5}{2},\frac{5}{2}\rangle   \nonumber\\
|d_{2\downarrow}\rangle = (|\frac{5}{2},\frac{3}{2}\rangle+2|\frac{3}{2},\frac{3}{2}\rangle)/\sqrt{5} \nonumber \\
|d_{1\uparrow}\rangle= (2|\frac{5}{2},\frac{3}{2}\rangle-|\frac{3}{2},\frac{3}{2}\rangle)/\sqrt{5} \nonumber \\
|d_{1\downarrow}\rangle = (\sqrt{2}|\frac{5}{2},\frac{1}{2}\rangle+\sqrt{3}|\frac{3}{2},\frac{1}{2}\rangle)/\sqrt{5} \nonumber\\
|d_{0\uparrow}\rangle= (\sqrt{3}|\frac{5}{2},\frac{1}{2}\rangle-\sqrt{2}|\frac{3}{2},\frac{1}{2}\rangle)/\sqrt{5}  \nonumber\\
|d_{0\downarrow}\rangle = (\sqrt{3}|\frac{5}{2},-\frac{1}{2}\rangle+\sqrt{2}|\frac{3}{2},-\frac{1}{2}\rangle)/\sqrt{5} \nonumber\\
|d_{-1\uparrow}\rangle= (\sqrt{2}|\frac{5}{2},-\frac{1}{2}\rangle-\sqrt{3}|\frac{3}{2},-\frac{1}{2}\rangle)/\sqrt{5}  \nonumber\\
|d_{-1\downarrow}\rangle = (2|\frac{5}{2},-\frac{3}{2}\rangle+|\frac{3}{2},-\frac{3}{2}\rangle)/\sqrt{5} \nonumber \\
|d_{-2\uparrow}\rangle= (|\frac{5}{2},-\frac{3}{2}\rangle-2|\frac{3}{2},-\frac{3}{2}\rangle)/\sqrt{5} \nonumber \\
|d_{-2\downarrow}\rangle = |\frac{5}{2},-\frac{5}{2}\rangle   
\label{jj3}
\end{align}

These expressions can be directly related to the `cartesian' states ($d_{xy}$, $d_{xz}$, etc.) through the relations given above. However, it is more convenient to keep them in this form and rather to express the Kramers doublet (Eq.~(\ref{kramers})) in the spherical basis $d_m$, as detailed below.
It should be noted in passing that Fig.~1(e) of Ref.~\onlinecite{kim1} is somewhat misleading in one aspect: the $|j=\frac{3}{2},j_z\rangle$ atomic (spherical) states do not branch just to the $t_{2g}$ ones in an octahedral environment (called $J_{\rm eff}=3/2$ in that figure), but to both $t_{2g}$ and $e_g$. The same is true for the $|j=\frac{5}{2},j_z\rangle$ states. However, it is true that the Kramers doublet in an octahedral crystal field (i.e., $J_{\rm eff}=1/2$, in the absence of a tetragonal distortion), is entirely due to the $|j=\frac{5}{2},j_z\rangle$ states, as we shall see below. This is expressed in Fig. \ref{fig6}.
  
As done in a number of publications on iridates \cite{moretti}, and as detailed in Abragam and Bleaney \cite{abragam}, the Ir$^{4+}$ state when restricted to the $t_{2g}$ subspace, and in the presence of a weak tetragonal crystal field in the $z$ direction, can be determined by the following Hamiltonian: 
\begin{align}
H=\lambda \vec{L}\cdot\vec{S} + \Delta_t (1-L_z^2)
\label{ham}
\end{align}

Here $\lambda$ is the spin-orbit coupling and $\Delta_t$ represents the tetragonal distortion. 
But, to facilitate calculations at the L$_2$ and L$_3$ edges, we consider this Hamiltonian in the $L=2$ basis instead of the $L_{\rm eff}=-1$ basis. For comparison with earlier work, we report below the relation between the two bases.
Following Balhausen \cite{balhausen}, we define the effective representation $L_{\rm eff}=-1$ through the three states $\phi_1$, $\phi_0$ and $\phi_{-1}$, corresponding to effective angular momentum $L_{\rm eff}^z=1,0,-1$, respectively. They are defined as:
$\phi_1\equiv d_{-1} = (d_{xz}-id_{yz})/\sqrt{2}$,
$\phi_0 \equiv (d_2-d_{-2})/\sqrt{2} = id_{xy}$,
$\phi_{-1} \equiv -d_{1} = (d_{xz}+id_{yz})/\sqrt{2}$.

\noindent The inverse transformations are:
$d_{xz}= (\phi_1+\phi_{-1})/\sqrt{2}$,
$d_{xy}= -i\phi_0$,
$d_{yz}= i(\phi_1-\phi_{-1})/\sqrt{2}$.

\noindent From the diagonalization of the previous Hamiltonian, we find three Kramers doublets \cite{perkins}, one bonding, one non-bonding and one antibonding. For what follows, we are interested only in the half-filled doublet that can be written as follows:

\begin{align}
|\psi_{+}\rangle= \frac{1}{\sqrt{N}} \left(d_{1\uparrow}+R\frac{(d_{2\downarrow}-d_{-2\downarrow})}{\sqrt{2}} \right)   \nonumber\\
|\psi_{-}\rangle =\frac{1}{\sqrt{N}} \left(d_{-1\downarrow}-R\frac{(d_{2\uparrow}-d_{-2\uparrow})}{\sqrt{2}} \right)
\label{kramers}
\end{align}
where the coefficients $R(\eta)$ and $N(\eta)$
depend solely on the ratio $\eta=\Delta_t/\lambda$ between the tetragonal crystal field, $\Delta_t$, and the spin-orbit coupling, $\lambda$:

$R(\eta)=-\frac{1}{\sqrt{2}}\left(1-\frac{1}{2}(1+2\eta+\sqrt{9-4\eta+4\eta^2}) \right)$

$N(\eta)=1+ \frac{1}{2}\left(1-\frac{1}{2}(1+2\eta+\sqrt{9-4\eta+4\eta^2})\right)^2$

\noindent Notice that in the octahedral limit, $\eta=0$, $R(\eta=0)=\frac{1}{\sqrt{2}}$ and $N(\eta=0)=\frac{3}{2}$.
In order to understand the behavior at the L$_2$ and L$_3$ edges, it is instructive to rewrite this Kramers doublet in terms of the $|j,j_z\rangle$ basis. This is done in Eqs.~(\ref{kramers2a}) and (\ref{kramers2b}) in Section IV. Here we report the expressions of Eqs.~(\ref{psireal}) and (\ref{psirealpm}) in the $|j,j_z\rangle$ basis:

\begin{align} 
|\psi_{\rm any}\rangle = \frac{1}{\sqrt{5N}} \left[ \cos(\beta)\left( (2+\frac{R}{\sqrt{2}})|\frac{5}{2},\frac{3}{2}\rangle - R\sqrt{\frac{5}{2}}|\frac{5}{2},-\frac{5}{2}\rangle \right. \right.  \nonumber\\
 \left. + (\sqrt{2}R-1)|\frac{3}{2},\frac{3}{2}\rangle \right) +\sin(\beta) e^{-i\gamma} \left( - (\sqrt{2}R-1)|\frac{3}{2},-\frac{3}{2}\rangle \right. \nonumber\\
 \left. \left. +(2+\frac{R}{\sqrt{2}})|\frac{5}{2},-\frac{3}{2}\rangle - R\sqrt{\frac{5}{2}}|\frac{5}{2},\frac{5}{2}\rangle  \right)\right] 
\label{psirealj}
\end{align}

and 

\begin{align}
|\psi_{\rm real}^+\rangle= \frac{1}{\sqrt{10N}} \left[ \left( (2+\frac{R}{\sqrt{2}})|\frac{5}{2},\frac{3}{2}\rangle  -R\sqrt{\frac{5}{2}}|\frac{5}{2},-\frac{5}{2}\rangle  \right. \right. \nonumber \\
\left. + (\sqrt{2}R-1)|\frac{3}{2},\frac{3}{2}\rangle \right) + \frac{1}{\sqrt{2}} (1-i) \left( - (\sqrt{2}R-1)|\frac{3}{2},-\frac{3}{2}\rangle \right. \nonumber\\
 \left. \left. +(2+\frac{R}{\sqrt{2}})|\frac{5}{2},-\frac{3}{2}\rangle  - R\sqrt{\frac{5}{2}}|\frac{5}{2},\frac{5}{2}\rangle  \right)\right]  \\
|\psi_{\rm real}^-\rangle= \frac{1}{\sqrt{10N}} \left[ \left( (2+\frac{R}{\sqrt{2}})|\frac{5}{2},\frac{3}{2}\rangle -R\sqrt{\frac{5}{2}}|\frac{5}{2},-\frac{5}{2}\rangle \right. \right. \nonumber \\
\left. + (\sqrt{2}R-1)|\frac{3}{2},\frac{3}{2}\rangle \right) - \frac{1}{\sqrt{2}} (1-i) \left( - (\sqrt{2}R-1)|\frac{3}{2},-\frac{3}{2}\rangle \right. \nonumber\\
 \left. \left. +(2+\frac{R}{\sqrt{2}})|\frac{5}{2},-\frac{3}{2}\rangle  - R\sqrt{\frac{5}{2}}|\frac{5}{2},\frac{5}{2}\rangle  \right)\right]  
\label{psirealjpm}
\end{align}

\subsection{Details of the calculations at the L$_2$ and L$_3$ edges}

In order to perform calculations at the L$_2$ and L$_3$ edges, we move to a spherical basis, so that the Gaunt coefficients are more directly evaluated. Eq.~(\ref{psireal}) can be written as:

\begin{align}
|\psi_{\rm any}\rangle= \frac{1}{\sqrt{N}} \left[ \cos(\beta)\left(d_{1\uparrow}+R\frac{(d_{2\downarrow}-d_{-2\downarrow})}{\sqrt{2}}\right) \right.   \nonumber\\
\left. +\sin(\beta) e^{-i\gamma} \left(d_{-1\downarrow}-R\frac{(d_{2\uparrow}-d_{-2\uparrow})}{\sqrt{2}}\right)\right]
\label{psireal2}
\end{align}

We need to evaluate matrix elements that appear in the calculation of both RXS and XAS, of the kind

$\sum_{j_z} \langle\psi_{\rm any}|r_\alpha ||\frac{1}{2},j_z\rangle\langle\frac{1}{2},j_z|| r_\beta |\psi_{\rm any}\rangle \\$

\noindent for the L$_2$ edge and of the kind 

$\sum_{j_z} \langle\psi_{\rm any}|r_\alpha ||\frac{3}{2},j_z\rangle\langle\frac{3}{2},j_z|| r_\beta |\psi_{\rm any}\rangle\\$

\noindent for the L$_3$ edge, reminding that $\alpha,\beta$ refer to any of the three components $x,y,z$. In order to clearly identify the core-hole states $j,j_z$ with respect to the $5d$ $j,j_z$ of Eq.~(\ref{jj3}), we used a $||$ convention for their bras and kets. As the transition operator $r_\alpha$ only changes the orbital angular momentum, it is more convenient to rewrite $||j,j_z\rangle$ states in terms of their $2p$ orbital counterparts.
We have: 

$||\frac{1}{2},\frac{1}{2}\rangle=(\sqrt{2}|p_{1\downarrow}\rangle-|p_{0\uparrow}\rangle)/\sqrt{3} $

$||\frac{1}{2},-\frac{1}{2}\rangle=(|p_{0\downarrow}\rangle-\sqrt{2}|p_{-1\uparrow}\rangle)/\sqrt{3} $

$||\frac{3}{2},\frac{3}{2}\rangle=|p_{1\uparrow}\rangle$

$||\frac{3}{2},\frac{1}{2}\rangle=(|p_{1\downarrow}\rangle+\sqrt{2}|p_{0\uparrow}\rangle)/\sqrt{3} $

$||\frac{3}{2},-\frac{1}{2}\rangle=(\sqrt{2}|p_{0\downarrow}\rangle+|p_{-1\uparrow}\rangle)/\sqrt{3} $

$||\frac{3}{2},-\frac{3}{2}\rangle=|p_{-1\downarrow}\rangle $

Finally, the last step before performing the calculation is to remember the expression for $r_\alpha$ in terms of spherical harmonics: $z =cr Y_{10}$; $x =cr (Y_{1,-1}-Y_{11})/\sqrt{2}$; $y =cr i(Y_{1,-1}+Y_{11})/\sqrt{2}$. Here $c=\sqrt{4\pi/3}$ is a normalization constant (not important for the following, as it can be absorbed into the radial part, $r$). 

Having all the coefficients, the transition matrix elements are now easily calculated after noting that the spin is not changed in the transition (so, only equal-spin states are coupled during the x-ray excitation), and given the expressions for the only two Gaunt coefficients that appear in the calculation:

$\langle Y_{2m}| Y_{1m_1} |Y_{1m_2}\rangle = \frac{\sqrt{3}}{\sqrt{10\pi}} (1,m_1;1,m_2|2,m)$

\noindent and

$\langle Y_{1m}| Y_{1m_1} |Y_{2m_2}\rangle = -\frac{1}{\sqrt{2\pi}} (1,m_1;2,m_2|1,m)$

\noindent where $(1,m_1;2,m_2|1,m)$ represents the Clebsch-Gordan coefficient for the vector sum of $(1,m_1)$ and $(2,m_2)$ to give $(1,m)$. Notice that in the following, we shall drop the radial matrix elements, $|\langle R_{2p}(r)| r |R_{5d}(r)\rangle|^2$, as they are common to both the $L_2$ and $L_3$ edges,
if we neglect relativistic corrections.
At L$_2$, we obtain the following values for the matrix elements:

$\langle\frac{1}{2},\frac{1}{2}|| Y_{11} |\psi_{\rm any}\rangle = 0$

$\langle\frac{1}{2},\frac{1}{2}|| Y_{1,-1} |\psi_{\rm any}\rangle = \frac{1}{\sqrt{10\pi N}}\cos(\beta) \left(\frac{1}{\sqrt{2}} -R\right)$

$\langle\frac{1}{2},\frac{1}{2}|| Y_{10} |\psi_{\rm any}\rangle = 0$

$\langle\frac{1}{2},-\frac{1}{2}|| Y_{11} |\psi_{\rm any}\rangle = \frac{1}{\sqrt{10\pi N}}\sin(\beta) e^{-i\gamma} \left(R-\frac{1}{\sqrt{2}}\right)$

$\langle\frac{1}{2},-\frac{1}{2}|| Y_{1,-1} |\psi_{\rm any}\rangle = 0$

$\langle\frac{1}{2},-\frac{1}{2}|| Y_{10} |\psi_{\rm any}\rangle = 0$

\noindent These matrix elements are sufficient to derive the full scattering matrix, noting that $\langle j,j_z|| Y_{1m} |\psi_{\rm any}\rangle = -(\langle\psi_{\rm any}| Y_{1,-m} ||j,j_z\rangle)^*$, because of the spherical harmonics phase rule $Y_{1,m}=-Y^*_{1,-m}$.

From this, we get the following expressions for the total matrix $L^{(2)}_{\alpha\beta}$ ($\alpha,\beta=x,y,z$) at the L$_2$ edge (with a common constant ${\tilde C}$):

\begin{align}
& L^{(2)}_{xx}=\frac{{\tilde C}}{20\pi N}(R-\frac{1}{\sqrt{2}})^2 \nonumber \\
& L^{(2)}_{yy}=\frac{{\tilde C}}{20\pi N}(R-\frac{1}{\sqrt{2}})^2 \nonumber \\
& L^{(2)}_{zz}=0 \nonumber \\
& L^{(2)}_{xy}=\frac{i{\tilde C}}{20\pi N}(R-\frac{1}{\sqrt{2}})^2 \cos(2\beta) \nonumber \\
& L^{(2)}_{yx}=-\frac{i{\tilde C}}{20\pi N}(R-\frac{1}{\sqrt{2}})^2 \cos(2\beta) \nonumber \\
& L^{(2)}_{xz}=L^{(2)}_{zx}=L^{(2)}_{yz}=L^{(2)}_{zy}=0  
\label{L2edge} 
\end{align}

\noindent Interestingly, as already highlighted after Eq.~(\ref{kramers2b}), we get that in the octahedral limit, $R(\eta=0)=\frac{1}{\sqrt{2}}$, all the matrix elements at the L$_2$ edge are zero, not just the magnetic ones. In particular, the XAS signal should be zero in this limit. Therefore, if an XAS signal is confirmed at the energy of this edge, this would show that $R$ deviates from $\frac{1}{\sqrt{2}}$, as discussed in Section IV.
Notice however that, differently from magnetic RXS, XAS can also see the other empty states that are higher in energy, in particular the $e_g$ states, that are accessible from the L$_2$ edge because they have a sizable component in the $J=3/2$ subspace as seen above. Therefore, the XAS signal should be reanalyzed with a better energy sensitivity, so as to clearly disentangle the unoccupied $t_{2g}$ states from the $e_g$ ones, as discussed in Section V. If such a signal is clearly detected at the edge itself, this is a definitive proof that the half-filled Kramers doublet deviates significantly from the octahedral limit. If such a signal is not detected, to the contrary, this is clear proof that the doublet is composed purely of $J=5/2$ states, as is the case of $|\psi_{\rm any}\rangle$ above (Eq.~(\ref{psirealj})) in the octahedral limit ($R(\eta=0)=\frac{1}{\sqrt{2}}$). In the latter case, this also implies the absence of a signal in the magnetic RXS at the L$_2$ edge.
In the former case, instead, absence of the magnetic signal at the L$_2$ edge can also be explained by an in-plane magnetic moment, that makes $\cos(2\beta)=0$,
since the magnetic signal originates from the off-diagonal matrix element, $xy$.
It should be noted, however, that this latter case would not permit one to explain the lack of a RXS signal that is also seen when the magnetic moment is along $c$, as in Mn-doped \cite{calderMn} and Ru-doped \cite{calderRu} samples.

At L$_3$, we obtain the following values for the matrix elements:

$\langle\frac{3}{2},\frac{3}{2}|| Y_{11} |\psi_{\rm any}\rangle = 0$

$\langle\frac{3}{2},\frac{3}{2}|| Y_{1,-1} |\psi_{\rm any}\rangle = \sqrt{\frac{3}{20\pi N}} R \sin(\beta) e^{-i\gamma}$

$\langle\frac{3}{2},\frac{3}{2}|| Y_{10} |\psi_{\rm any}\rangle = \sqrt{\frac{3}{20\pi N}} \cos(\beta)$

$\langle\frac{3}{2},\frac{1}{2}|| Y_{11} |\psi_{\rm any}\rangle = 0$

$\langle\frac{3}{2},\frac{1}{2}|| Y_{1,-1} |\psi_{\rm any}\rangle = -\frac{1}{\sqrt{20\pi N}}\cos(\beta) (\sqrt{2} +R)$

$\langle\frac{3}{2},\frac{1}{2}|| Y_{10} |\psi_{\rm any}\rangle = 0$

$\langle\frac{3}{2},-\frac{1}{2}|| Y_{11} |\psi_{\rm any}\rangle = -\frac{1}{\sqrt{20\pi N}}\sin(\beta) e^{-i\gamma} (\sqrt{2} +R)$

$\langle\frac{3}{2},-\frac{1}{2}|| Y_{1,-1} |\psi_{\rm any}\rangle = 0$

$\langle\frac{3}{2},-\frac{1}{2}|| Y_{10} |\psi_{\rm any}\rangle = 0$

$\langle\frac{3}{2},-\frac{3}{2}|| Y_{11} |\psi_{\rm any}\rangle = \sqrt{\frac{3}{20\pi N}} R \cos(\beta)$

$\langle\frac{3}{2},-\frac{3}{2}|| Y_{1,-1} |\psi_{\rm any}\rangle = 0$

$\langle\frac{3}{2},-\frac{3}{2}|| Y_{10} |\psi_{\rm any}\rangle = \sqrt{\frac{3}{20\pi N}} \sin(\beta) e^{-i\gamma}$

From this, we get the following expressions for the total matrix $L^{(3)}_{\alpha\beta}$ ($\alpha,\beta=x,y,z$) at the $L_3$ edge (with a common constant ${\bar C}$, differing from that at the L$_2$ edge):

\begin{align} 
& L^{(3)}_{xx}=\frac{{\bar C}}{20\pi N}(2R^2+\sqrt{2}R+1)  \nonumber \\
& L^{(3)}_{yy}=\frac{{\bar C}}{20\pi N}(2R^2+\sqrt{2}R+1) \nonumber \\
& L^{(3)}_{zz}=-\frac{3{\bar C}}{20\pi N} \nonumber \\
& L^{(3)}_{xy}=\frac{i{\bar C}}{40\pi N}(R^2+2\sqrt{2}R-1) \cos(2\beta) \nonumber \\
& L^{(3)}_{yx}=-\frac{i{\bar C}}{40\pi N}(R^2+2\sqrt{2}R-1) \cos(2\beta) \nonumber \\
& L^{(3)}_{xz}=\frac{3\sqrt{2}i{\bar C}}{40\pi N} R \sin(2\beta) \sin(\gamma) \nonumber \\
& L^{(3)}_{zx}=-\frac{3\sqrt{2}i{\bar C}}{40\pi N} R \sin(2\beta) \sin(\gamma) \nonumber \\
& L^{(3)}_{yz}=-\frac{3\sqrt{2}i{\bar C}}{40\pi N} R \sin(2\beta) \cos(\gamma) \nonumber \\
& L^{(3)}_{zy}=\frac{3\sqrt{2}i{\bar C}}{40\pi N} R \sin(2\beta) \cos(\gamma) 
\label{L3edge} 
\end{align}

\noindent We see that at least one off-diagonal matrix element always differs from zero, whatever $R$ and $\beta$ are. This implies that the magnetic RXS signal is always different from zero, as explained in Section IV.

\subsection{Direction of the magnetic moment}

The evaluation of $<M_z> = <L_z + 2 S_z>$ is straightforward using Eq.~(\ref{kramers}) and Eq.~(\ref{psireal2}):

\begin{align}
 & \langle M_z \rangle_{\rm any} = \cos^2(\beta) \langle \psi_+ | M_z|\psi_{+}\rangle + \sin^2(\beta) \langle \psi_- | M_z|\psi_{-}\rangle = \nonumber \\
 & \cos^2(\beta) \left( \frac{1}{N} + \frac{1}{N} - \frac{R^2}{N} \right) + \sin^2(\beta) \left( -\frac{1}{N} - \frac{1}{N} + \frac{R^2}{N} \right) \nonumber \\
 & =(\cos^2(\beta) - \sin^2(\beta)) \frac{2-R^2}{N} = \cos(2\beta) \frac{2-R^2}{N}
\label{mz}
\end{align}

Notice that $\langle M_z \rangle_{\rm any} =0$, i.e., the moment lies in the $ab$-plane, when $\cos(2\beta)=0$ or $R=\sqrt{2}$. The latter case corresponds to $\Delta_t = \lambda$, that we exclude for Sr$_2$IrO$_4$.
The former condition coincides with the one that makes the L$_2$ magnetic RXS zero (in keeping with Ref.~\onlinecite{lovesey} and Ref.~\onlinecite{moretti}).

If we use the two partners of the Kramers doublet, Eq.~(\ref{psirealpm}), it is straightforward to verify, using the usual definitions for $L_x = (L_+ + L_-)/2$ and $L_y = (L_+ - L_-)/(2i)$, that the eigenstates of $L_x$ and $S_x$ correspond to $\beta = \pi/4 + k\pi/2$, with $\gamma=0$ (positive $x$) or $\gamma=\pi$ (negative $x$) and that the eigenstates of $L_y$ and $S_y$ correspond to $\beta = \pi/4 + k\pi/2$, with $\gamma=\pi/2$ (positive $y$) or $\gamma=3\pi/2$ (negative $y$).

Finally, using $L_x=(L_+ + L_-)/2$ and $L_y=(L_+ - L_-)/(2i)$ and analogously for the spins $S_x$ and $S_y$, it is possible to show that the condition $\langle M_x \rangle_{\rm any} =\langle M_y \rangle_{\rm any}$ implies $\gamma=\pi/4$ (modulo $n\pi$).



\begin{thebibliography}{99}
\bibitem{footye}
The magnetic pattern presented in Fig.~2c of Ref.~\onlinecite{ye15} appears to be that for $-+-+$.  It is actually
$++++$ since the authors shift registry by $z=\frac{1}{4}$ when converting to I4$_1$/a notation.
But, since they only measure at (1,0,L), they cannot differentiate between $-+-+$ and $++++$.
The authors thank Prof.~Hsieh for raising this point, and Dr.~Ye for this clarification.
\bibitem{twin}
Sometimes, a signal is also seen when $h$ and $k$ in the reflection are inverted.  This is due to
the AF moments aligning along $b$ rather than $a$ (noting that the high temperature phase is tetragonal).
For a large beam size (like neutrons), both twin domains are often seen.
\bibitem{nota2}
Order of magnitude wise, $\chi^{(e)}$ is bigger than $\chi^{(q)}$ and $\chi^{(m)}$ by a factor $\lambda/a$ (here $\lambda$ is the wavelength and $a$ is a typical interatomic space). This, however, is true for its real, time-reversal even part, not for the imaginary, magnetic part. The latter, defined in Appendix A, can be comparable to $\chi^{(m)}$, as calculated in Ref.~\onlinecite{muto}. This is further confirmed in the present case if the interference with $\chi^{(q)}$, which is of the same order as $\chi^{(m)}$, turns out to be determined by $\chi^{(e)}$.
\bibitem{fluence}
From $\sigma_1$ and $\epsilon_1$, one can estimate an optical penetration depth for
Sr$_2$IrO$_4$ of 700 \AA~at 1.55 eV.
From this, one can derive that for a 1 mJ/cm$^2$ fluence, the number of photons per Ir is $\sim$0.055.
\bibitem{nota7}
In this section and in Appendix B, $x$, $y$ and $z$ refer to the local octahedron axes that define t$_{2g}$ and e$_g$ orbitals and not the $a$, $b$ and $c$ crystal axes, as in the previous sections.
\bibitem{nota4}
As clear from the choice of t$_{2g}$ orbitals, the $x$, $y$ and $z$ axes are directed towards the surrounding oxygens. The angle $\gamma$ is defined in this basis. Of course, because of the -33$^\circ$ rotation of the $x$-axis of the octahedron compared to the $a$-axis, the magnetic moment at Ir$_1$ is oriented 12$^\circ$ to the $a$-axis.
\bibitem{notakim}
We note that Fig.~1(e) of Ref.~\onlinecite{kim1} is somewhat misleading, because it suggests that $e_g$ orbitals have a purely $j=5/2$ character and the $J_{\rm eff}=3/2$ manifold a purely $j=3/2$ character.
\bibitem{input}
We used FDMNES in an atomic input mode, with the input corresponding to S=1/2 per Ir ion.  More exact results could be achieved by
using a band structure input instead.
\bibitem{nota8}
Partial fluorescence yield was employed in Ref.~\onlinecite{clancy} for the Ir L$_3$ edge, but the XAS peak was still 4 eV broad.
\bibitem{xmcd}
Concerning the interpretation of the XMCD at the L$_2$ edge advanced in Ref.~\onlinecite{haskel}, suggesting a zero contribution for all energies for the purely octahedral J$_{\rm eff}=1/2$ Kramers empty partner, we notice that it is apparently based on the incorrect assumption that the $e_g$ states totally belong to the $j=5/2$ manifold. This assumption can be traced back to Fig.~1 of Ref.~\onlinecite{angeles}. Our Eq.~(\ref{jj3}) shows instead that e$_g$ states have a sizable $j=3/2$ contribution and this can lead to an XMCD signal at the L$_2$ edge by spin polarization of the $e_g$ states. Therefore, the XMCD detected around the energy of the $e_g$ states \cite{haskel}  does not contradict the main message of Section IV.
\end{thebibliography}
\end{document}